\documentclass[aps,pra,superscriptaddress,showpacs,twocolumn,longbibliography]{revtex4-1}

\usepackage{graphicx}
\usepackage{amsmath}
\usepackage{amsfonts}
\usepackage{amssymb}
\usepackage{color}
\usepackage[usenames,dvipsnames]{xcolor}
\usepackage[colorlinks=true,linkcolor=blue,urlcolor=blue,citecolor=blue]{hyperref}
\usepackage{mathtools}
\usepackage[normalem]{ulem}
\usepackage{afterpage}

\newcommand{\qq}{\mathbf{q}}
\newcommand{\QQ}{\mathsf{Q}}
\newcommand{\RR}{\mathbf{R}}

\newcommand{\kk}{\mathbf{k}}

\newcommand{\up}{\uparrow}
\newcommand{\dn}{\downarrow}
\newcommand{\GG}{\mathcal{G}}
\newcommand{\FF}{\mathcal{F}}

\newcommand{\uphii}{\overline{\varphi}}
\newcommand{\uhGG}{\overline{\hat{\GG}}}

\newcommand{\Odrv}{\Omega_\mathrm{drv}}

\newcommand{\hGG}{\hat{\GG}}
\newcommand{\hFF}{\hat{\FF}}

\newcommand{\hsx}{\hat{\sigma}_x}

\newcommand{\hsz}{\hat{\sigma}_z}
\newcommand{\hSIG}{\hat{\Sigma}}

\newcommand{\AM}{\mathcal{A}}
\newcommand{\BM}{\mathcal{B}}

\newcommand{\epss}{\varepsilon}
\newcommand{\phii}{\varphi}
\newcommand{\dd}{\mathrm{d}}
\newcommand{\CC}{\mathcal{C}}
\newcommand{\DD}{\mathcal{D}}
\newcommand{\DDD}{\mathfrak{D}}

\newcommand{\KKK}{\mathsf{K}}
\newcommand{\FFF}{\mathsf{F}}

\newcommand{\PSI}{\Psi^{\phantom{\dagger}}}
\newcommand{\PSID}{\Psi^{\dagger}}
\newcommand{\GAM}{\mathsf{\Gamma}}

\newcommand{\AAA}{\mathsf{A}}

\newcommand{\low}[1]{{#1}^{\phantom{\dagger}}}
\newcommand{\remark}[1]{\noindent {\bf #1:}}

\newcommand{\eqfigscl}[2]{\vcenter{\hbox{\includegraphics[scale=#1]{#2}}}}

\newcommand{\eq}[1]{Eq.~(\ref{#1})}

\makeatletter
\def\@maketitle{%
  \newpage
  \null
  \vskip 2em%
  \begin{center}%
  \let \footnote \thanks
    {\Large\bfseries \@title \par}%
    \vskip 1.5em%
    {\normalsize
      \lineskip .5em%
      \begin{tabular}[t]{c}%
        \@author
      \end{tabular}\par}%
    \vskip 1em%
    {\normalsize \@date}%
  \end{center}%
  \par
  \vskip 1.5em}
\makeatother

\makeatletter
\newsavebox{\@brx}
\newcommand{\llangle}[1][]{\savebox{\@brx}{\(\m@th{#1\langle}\)}%
  \mathopen{\copy\@brx\kern-0.5\wd\@brx\usebox{\@brx}}}
\newcommand{\rrangle}[1][]{\savebox{\@brx}{\(\m@th{#1\rangle}\)}%
  \mathclose{\copy\@brx\kern-0.5\wd\@brx\usebox{\@brx}}}
\makeatother

\makeatletter
\DeclareRobustCommand{\cev}[1]{%
  \mathpalette\do@cev{#1}%
}
\newcommand{\do@cev}[2]{%
  \fix@cev{#1}{+}%
  \reflectbox{$\m@th#1\vec{\reflectbox{$\fix@cev{#1}{-}\m@th#1#2\fix@cev{#1}{+}$}}$}%
  \fix@cev{#1}{-}%
}
\newcommand{\fix@cev}[2]{%
  \ifx#1\displaystyle
    \mkern#23mu
  \else
    \ifx#1\textstyle
      \mkern#23mu
    \else
      \ifx#1\scriptstyle
        \mkern#22mu
      \else
        \mkern#22mu
      \fi
    \fi
  \fi
}
\makeatother

\allowdisplaybreaks

\begin{document}

\title{The theory of parametrically amplified electron-phonon superconductivity}

\author{Mehrtash Babadi}%
\affiliation{Institute for Quantum Information and Matter, Caltech, Pasadena, CA 91125, USA}%
\affiliation{Broad Institute of MIT and Harvard, Cambridge, MA 02138, USA}%
\author{Michael Knap}%
\affiliation{Department of Physics, Walter Schottky Institute, and Institute for Advanced Study, Technical University of Munich, 85748 Garching, Germany}
\author{Ivar Martin}%
\affiliation{Materials Science Division, Argonne National Laboratory, Argonne, Illinois 60439, USA}%
\author{Gil Refael}%
\affiliation{Institute for Quantum Information and Matter, Caltech, Pasadena, CA 91125, USA}%
\author{Eugene Demler}%
\affiliation{Department of Physics, Harvard University, Cambridge, MA 02138, USA}%

\date{\today}

\begin{abstract}
Ultrafast optical manipulation of ordered phases in strongly correlated materials is a topic of significant theoretical, experimental, and technological interest. Inspired by a recent experiment on light-induced superconductivity in fullerenes~[Mitrano {\em et al.}, Nature {\bf 530}, 2016], we develop a comprehensive theory of light-induced superconductivity in driven electron-phonon systems with lattice nonlinearities. In analogy with the operation of parametric amplifiers, we show how the interplay between the external drive and lattice nonlinearities lead to significantly enhanced effective electron-phonon couplings. We provide a detailed and unbiased study of the nonequilibrium dynamics of the driven system using the real-time Green's function technique. To this end, we develop a Floquet generalization of the Migdal-Eliashberg theory and derive a numerically tractable set of quantum Floquet-Boltzmann kinetic equations for the coupled electron-phonon system. We study the role of parametric phonon generation and electronic heating in destroying the transient superconducting state. Finally, we predict the transient formation of electronic Floquet bands in time- and angle-resolved photoemission spectroscopy experiments as a consequence of the proposed mechanism.
\end{abstract}

\maketitle


\section{Introduction}

In the recent years, the rapid progress of the field of ultrafast pump-probe spectroscopy experiments has enabled an unprecedented exploration of many-body quantum dynamics in far-from-equilibrium states (for reviews, see Refs.~\cite{basov2011electrodynamics,giannetti2016ultrafast}). The application of strong ultrafast laser pulses can dramatically alter the equilibrium state, outreach the linear response regime, and enable the induction of novel ordered states and stimulation of phase transitions via transient nonequilibrium states.

One of the main motivations behind these experiments is to shed light on the interplay between competing orders in strongly-correlated superconductors, along with the tantalizing outlook of stimulating the superconducting transition at temperatures above the critical temperature $T_c$. An early experimental evidence proving the possibility of stimulating superconductivity via external fields is the Wyatt-Dayem effect~\cite{wyatt1966microwave,dayem1967behavior}, where microwave radiation of superconducting micro-bridges in the MHz to GHz frequency range was found to increase $T_c$ by a few percents. This  effect was explained theoretically by Eliashberg~\cite{eliashberg1970film} on the basis of the nonequilibrium shift of the quasiparticle occupation to high energies. Subsequent theoretical work~\cite{chang1978nonequilibrium} and experiments in double-barrier tunnel junctions and strips~\cite{pals1979measurements,blamire1991extreme,heslinga1993enhancement,komissinski1996superconductivity} found a much larger effect up to several times larger than the equilibrium $T_c$. An experimental proposal for investigating this effect using ultracold fermionic atoms has also been given~\cite{robertson2009nonequilibrium}.

Recently, Mitrano {\em et al.}~\cite{mitrano2016possible} have reported a large increase in carrier mobility and the opening of an optical gap upon stimulating the intercalated fullerene superconductor $K_3 C_{60}$ with a femto-second mid-infrared light pulse in the frequency range $80$--$200$~meV ($19$--$48$~THz). These effects persist for several pico-seconds after pumping, and remarkably for initial temperatures up to $T_i \sim 100$~K, much higher than the equilibrium $T_c \sim 20$~K, providing a compelling evidence for a possibly light-induced superconducting state. The experimentally observed resonance with several $C_{60}$ vibrational modes suggests that the underlying mechanism for enhanced Cooper pairing in this experiment stems from lattice distortions and is distinct from the Wyatt-Dayem effect.

\begin{figure}[t!]
\includegraphics[width=\linewidth]{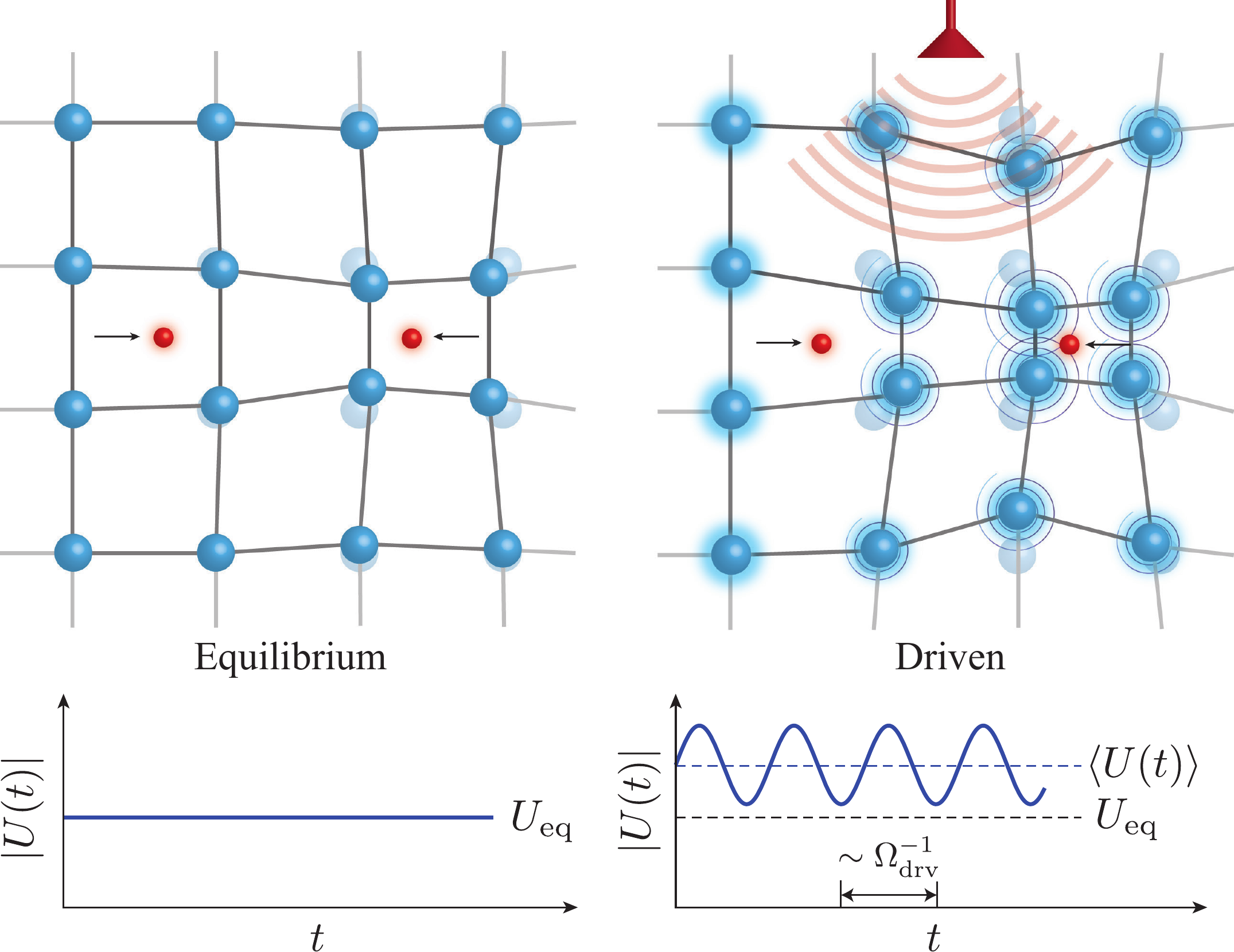}
\caption{{\bf Parametric amplification of the phonon response.} (left) Phonon-mediated electron attraction in the absence of external drive, (right) The external drive and lattice nonlinearities parametrically amplify lattice distortions which in turn mediate stronger attraction between the electrons.}
\label{fig:schem}
\end{figure}

The application of a strong pump pulse alters the initial equilibrium state in various ways and is a complex function of the strength of the drive, strength of coupling to different degrees of freedom and energetic proximity to resonances. The accurate theoretical modeling of light-stimulated superconductivity in $K_3 C_{60}$ is exacerbated by the structural complexity of $K_3 C_{60}$, including the three-fold degeneracy of the conduction $t_{1u}$ bands and their coupling to eight $H_g$ intra-molecular Jahn-Teller phonons~\cite{gunnarsson2004narrow}, strong electron-phonon coupling $\lambda \sim 0.5$--$1$, narrow conduction band $\omega_\mathrm{ph}/W_\mathrm{el} \sim 0.1$--$0.25$ ($\omega_\mathrm{ph}$ is the typical phonon energy scale and $W_\mathrm{el}$ is the conduction bandwidth), and strong Coulomb interaction $U_c/W_\mathrm{el} \sim 1.5$--$2.5$~\cite{gunnarsson1997superconductivity}. A reliable material-specific analysis must combine state-of-the-art {\em ab initio} modeling including nonlinear interactions and beyond-Migdal vertex corrections~\cite{grimaldi1995nonadiabatic} into the framework of nonequilibrium field theory. We do not pursue this formidable goal here; rather, {\em inspired} by the experiment and embracing the Occam's razor tradition, we explore a simplified model with fairly generic electron-phonon interaction which retains the essential features observed in experiments with light-stimulated superconductivity in $K_3C_{60}$.

At the simplest level, the pumping pulse with frequency $\Odrv \sim 100$ meV strongly drives near-resonant infra-red (IR) active lattice vibrational mode, such as $T_{1u}$ modes in fullerenes. As a first approximation, one may average out fast oscillations at the scale of $\Odrv^{-1}$. The presence of lattice anharmonicities and nonlinear coupling between vibrational modes results in the expansion and contraction of time-averaged lattice constants and electronic orbital configurations. The time-averaged electron-phonon coupling constants and electronic density of states are consequently renormalized. This approach is adopted in Ref.~\cite{mitrano2016possible} where an {\em ab initio} analysis in the static ``frozen-phonon'' approximation is performed and it is shown that time-averaged lattice deformations exhibit a favorable trend toward increasing $T_c$.\\

In this paper, we aim to show that the implications of a strong periodic drive and its interplay with lattice nonlinearities reaches beyond statically renormalized model parameters. In fact, we will show that the mechanism which yields the most striking enhancement of electron-phonon coupling is purely dynamical in nature and is not described by time-averaged Hamiltonians.

The phonon-mediated electron-electron attraction $U$ is usually understood using second-order perturbation theory: an electron distorts the lattice and the other electron is attracted to the lattice distortion, see Fig.~\ref{fig:schem} (left). In other words, this attractive potential is proportional to the retarded phonon response function. We will show that the enhancement of superconductivity in a driven nonlinear lattice is conceptually similar to the operation of a parametric amplifier circuit: the ``nonlinear capacitor'' is realized by the lattice nonlinearity, the ``ac pump source'' is realized by an excited lattice vibrational mode, the ``input signal'' is the phonon excitation caused by a momentum kick from an electron, and the ``output signal'' is the parametrically amplified phonon response observed by the other electron. In essence, lattice nonlinearities {\em convert} the coherent motion of the driven mode into a source of parametric drive for the phonon that couples to conduction electrons. When this drive is near parametric resonances, the retarded response will be significantly amplified, leading to a much stronger electron-electron attraction. Parametric driving also induces strong temporal oscillations in the effective electron-electron attraction, allowing it to visit very large values during each cycle. We will show that such temporal oscillations can significantly enhance $T_c$ even if the time-averaged attraction remains constant, see Figs.~\ref{fig:schem} (right) and~\ref{fig:T_c_analytic}.

A rigorous quantitative analysis of this simple mechanism and its consequences in a realistic electron-phonon model goes beyond the amplifier analogy as one must take into account several competing effects. Most importantly, the nearly resonant drive also results in parametric generation of high-energy phonons that dissipate their excess energy to electrons, leading to to higher scattering rates and heating. It is not {\em a priori} clear which subset of these phenomena prevails, even for short times, without resorting to an unbiased and rigorous framework. Ultimately, we find that without an external cooling mechanism, Cooper pairing may only be enhanced for a short time similar to the experiments, and the normal state takes over as high-energy phonon excitations equilibrate their energy with electrons.\\

Our goal in this paper is two-fold. First and foremost, we wish to present a transparent and physical analysis of the role of parametric resonances of the lattice in enhancing electron-phonon interactions and stimulating Cooper pair formation. The major part of this goal is achieved in the first part of the paper using perturbation theory, BCS theory, classical dynamics, along with a number of common-sense simplifications. Secondly, we aim to develop a rigorous theoretical formalism for analyzing the nonequilibrium dynamical nature of light-stimulated superconductivity experiments; a formalism that takes into account the detailed driven-dissipative evolution of phonons and heating of electrons while being flexible enough to include material-specific properties and paving the way for future investigations. To this end, we develop an extension of the Migdal-Eliashberg theory~\cite{migdal1958interaction,eliashberg1960interactions} to periodically driven electron-phonon systems with lattice nonlinearities, and utilize it to substantiate the results of the first part as an immediate application.

The conventional Migdal-Eliashberg theory is a cornerstone of the modern theory of superconductivity, both for qualitative understandings and accurate {\em ab initio} calculations. The existing attempts at the real-time extension of the Migdal-Eliashberg theory are known to be intractably difficult to work with due to the complicated temporal structure of the equations~\cite{bennemann2008superconductivity,sentef2016theory}. Here, we combine ideas from effective actions, Floquet theory, dynamical mean-field theory, and quantum kinetic theory to develop a formalism that is well-suited for numerical and analytical studies of periodically driven systems. The Floquet quantum kinetic formalism trades fast drive-induced oscillations of nonequilibrium propagators with slowly-varying Floquet components, and memory convolution integrals with algebraic products along with derivative corrections~\cite{martinez2003floquet,tsuji2008correlated,aoki2014nonequilibrium,genske2015floquet}. These controlled approximations effectively reduce the two-time Kadanoff-Baym integro-differential equations~\cite{kadanoff1962quantum} to (implicit) ordinary differential equations which are much easier to solve numerically. The extension of the quantum kinetic formalism to periodically driven systems has been considered before in Ref.~\cite{genske2015floquet} in a different context and in the Boltzmann ``quasiparticle'' approximation. The latter is obtained by neglecting off-shell processes~\cite{danielewicz1984quantum,berges2004introduction}. We do not adopt this approximation here. As we pointed out earlier, suppression of electronic and phononic quasiparticle coherence is an important factor in the analysis of transient superconductivity. Hence, a detailed study of the changes in the spectral functions of Floquet quasiparticles will be an important ingredient of our theory.\\

\begin{figure}[th!]
\includegraphics[width=\linewidth]{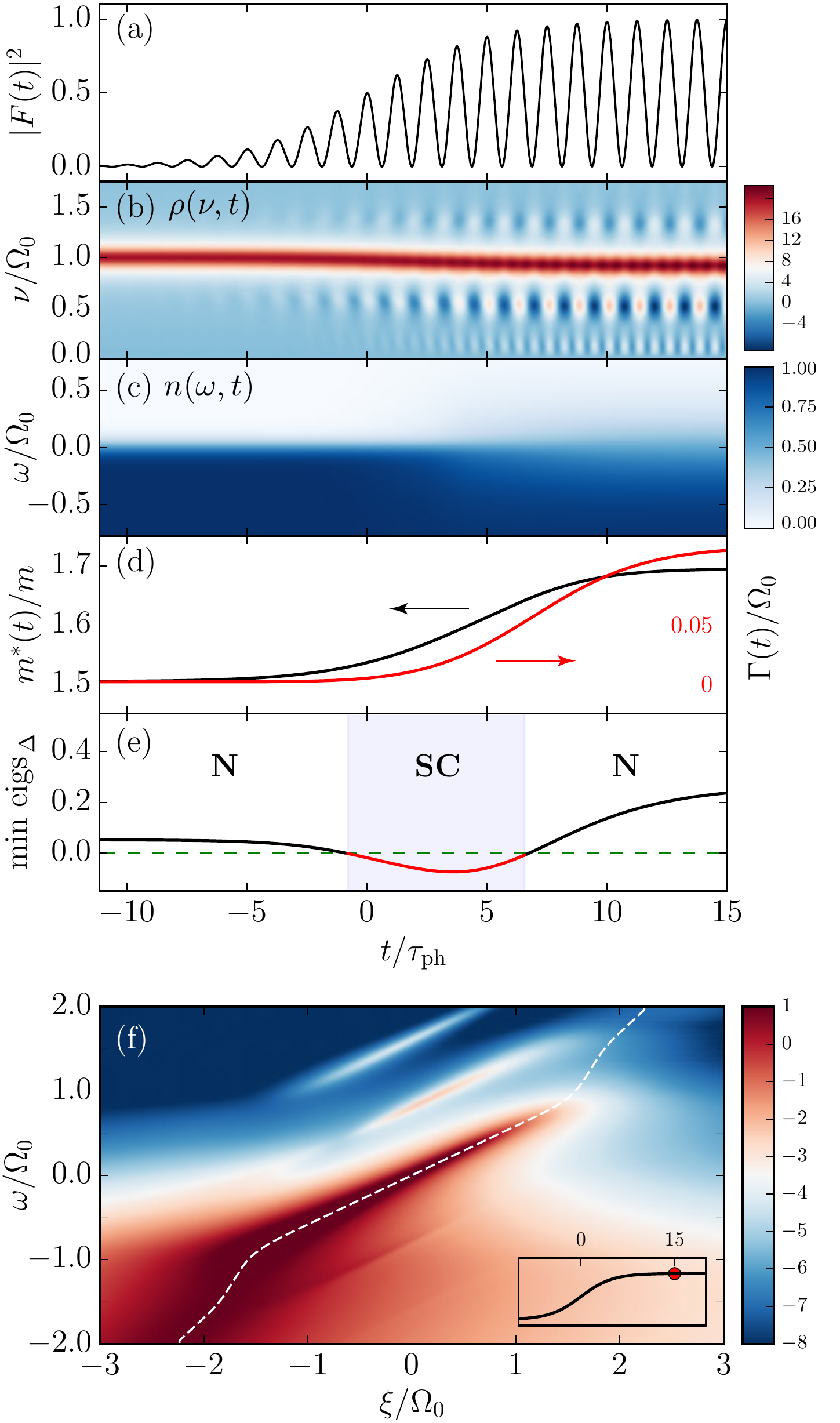}
\caption{{\bf Non-equilibrium evolution of the driven electron-phonon system obtain using the Floquet-Migdal-Eliashberg formalism.} (a) intensity of the external drive, (b) phonon spectral function $\rho(\nu, t)$, showing the redshift of the phonon peak and along with emergent oscillatory features, (c) electron distribution $n(\omega, t)$ showing the smearing of the Fermi surface as the electrons heat up, (d) electron effective mass (black, left axis) and damping (red, right axis), (e) lowest eigenvalue of the Floquet-Migdal-Eliashberg gap functional, where N and SC correspond to normal conducting and superconducting (instability) intervals, (f) predicted time-resolved ARPES signal in the log scale as a function of electron frequency $\omega$ and kinetic energy $\xi$ at $t = 15\,\tau_\mathrm{ph}$, showing the formation of electronic Floquet bands. The lattice nonlinearity is cubic-type with $\kappa_3 = 0.1\,\Omega_0$, and the drive frequency and amplitudes are $\Odrv = 0.4\,\Omega_0$ and $\mathcal{A} = 0.75$ (refer to Sec.~\ref{sec:FME_results} for additional details).} 
\label{fig:summary}
\end{figure}

The experimental observation of the light-induced superconducting state in $K_3 C_{60}$~\cite{mitrano2016possible} has inspired several theoretical works. Sentef {\em et al.}~\cite{sentef2016theory} have studied the transient dynamics of the superconducting gap following a change in the coupling constants. In an earlier work, we outlined the role of parametric driving in enhancing the electron-phonon coupling~\cite{knap2015dynamical} and analyzed the problem using a Floquet extension of the BCS theory. Komnik {\em et al.}~\cite{komnik2016bcs} have recently worked out a similar BCS framework using a more concise analytical approach. More recently, Kennes {\em et al.}~\cite{kennes2016electronic} have suggested nonlinear electron-phonon couplings as another plausible source of enhancing $T_c$ in a highly-pumped state. Last but not least, Kim {\em et al.}~\cite{kim2016enhancing} have suggested light-induced changes in the screened Coulomb matrix elements as a factor for enhancing superconductivity in intercalated fullerenes. We would like to mention that none of these works, except for Ref.~\cite{knap2015dynamical}, have studied the role of undesirable competing factors within their respective models. The electrons are always assumed to remain in the initial thermal state and heating is ignored. Given that superconductivity in fullerenes is mediated by high-frequency optical phonons, the issue of heating is a crucial aspect of the phenomenology even for short times. Another important goal of our paper to provide a first complete and unbiased analysis of the competition between processes that enhance and suppress Cooper pairing.

Finally, we would like to emphasize that the parametric amplification of electron-phonon coupling is not limited to enhancing Cooper pairing and is expected to find similar applications to other systems. For example, the same framework can be employed to study the recently observed enhancement of electron-phonon coupling in periodically distorted graphene~\cite{gierz2016enhanced} and driven opto-mechanical cavities~\cite{xu2015quantum,lemonde2016enhanced,stehlik2016double,zeytinoglu2016engineering}.

\subsection{Organization of the paper}

This paper is organized as follows. We describe the model in Sec.~\ref{sec:model} as the first step, and present its analysis in two separate stages. Before delving into the detailed formalism, we give a more intuitive account using perturbation theory, classical dynamics, and the BCS theory to demonstrate the idea of parametric amplification in nonlinear lattices and its implications in Sec.~\ref{sec:param}. Many of the relevant details such as feedback to electrons, heating, and competing factors are left to the second stage.

Sec.~\ref{sec:FME_main} and its multiple subsections are dedicated to developing the formalism of Floquet-Migdal-Eliashberg quantum kinetics. In particular, a pairing instability criterion is derived in Sec.~\ref{sec:FME_instab} that generalizes the result of Scalapino, Schrieffer, and Wilkins~\cite{scalapino1966strong} to quasi-steady Floquet states. As a first application of the formalism, we study the stationary solutions of the driven-dissipative state of phonons while neglecting the heating of electrons. This allows us to gain insight about the parameter regimes of maximal electron-phonon coupling enhancement, and to study the role of individual factors in enhancing and suppressing Cooper pairing. We move on the fully non-equilibrium scenario in Sec.~\ref{sec:full_picture} where we discuss the dynamics of the coupled electron-phonon system and show that a window of transient superconducting instability can exist even if the heating of electrons are taken into account. Finally, we use our theory to make additional experimental predictions in Sec.~\ref{sec:arpes}, in particular, the dynamical formation of Floquet conduction bands which can be probed using time- and angle-resolved photo-emission spectroscopy (tr-ARPES).

Some of the technical details, in particular those pertaining to numerical methods, have been moved to the appendices. The appendices also include an extensive discussion of the role of electrons in generating phonon nonlinearities (see Sec.~\ref{sec:nonlin}). In particular, we show that the magnitude of electron-mediated phonon nonlinearities increases near parametric resonances and can make a significant contribution to intrinsic lattice nonlinearities.

Finally, Fig.~\ref{fig:summary} shows a summary of the non-equilibrium dynamics obtained from the Floquet-Migdal-Eliashberg formalism; refer to the figure caption for details.

\section{The model}\label{sec:model}
We start with a general model for conduction electrons and a single phonon branch, along with an external drive that couples to the uniform lattice displacement, a local nonlinear lattice potential, and a linear electron-phonon coupling. The Lagrangian for this system is given as:
\begin{align}\label{eq:lag}
&\mathcal{L}[\phii,\Psi](t) = \sum_{\kk} \Psi^\dagger_{\kk}\left(i\partial_t\mathbb{I} - \xi_\kk\hat{\sigma}_3\right)\Psi^{\phantom{\dagger}}_{\kk}\nonumber\\
&- \frac{1}{2}\sum_\qq \frac{1}{2\omega_\qq} \, \low{\phii}_\qq \left(\partial_t^2 + \omega_\qq^2\right)\low{\phii}_{-\qq} - \sum_{j\in\,\mathrm{lattice}} \mathcal{V}^\mathrm{ph}(\phii_j)\nonumber\\
&-\frac{1}{\sqrt{N}}\sum_{\kk,\kk'} \low{g}_{\kk,\kk'} \, \low{\phii}_{\kk-\kk'}\,\PSID_{\kk'}\hat{\sigma}_3\PSI_{\kk}\nonumber\\
&+ \frac{\Lambda}{2}\,|F(t)|^2\,\sum_{j \in \mathrm{lattice}} \phii_j
\end{align}
Here, $\PSI_{\kk} = (c_{\kk\up}, c_{-\kk\dn}^\dagger)^T$ is the Nambu spinor of the conduction electrons, $\xi_\kk$ is the electron dispersion, $\phii_\qq \equiv b^\dagger_\qq + b^{\phantom{\dagger}}_{-\qq} = \sum_{j}e^{-i\qq\cdot \RR_j}\phii_j/\sqrt{N}$ is the lattice displacement operator, $\omega_\qq$ is the phonon dispersion, and $\low{g}_{\kk,\kk'}$ is the linear electron-phonon coupling constant. Furthermore, $\mathcal{V}^\mathrm{ph}(\phii)$ is the local lattice anharmonic potential which, for low-amplitude deformations, can be modeled as:
\begin{equation}\label{eq:Vnonlin}
\mathcal{V}^\mathrm{ph}(\phii) = -\frac{\kappa_3}{3!}\,\phii^3 -\frac{\kappa_4}{4!}\,\phii^4,
\end{equation}
We assume $\kappa_4>0$ since the lattice potential generically softens for large deformations. The sign of $\kappa_3$ is inconsequential due to symmetries. We neglect Coulomb interaction to simplify the analysis. We will briefly comment on its effect later on and argue that it does not play a consequential role in the phenomenon that is the case here. Finally, $F(t)$ is the external classical drive that couples to the uniform $\mathbf{q}=0$ lattice displacement with strength $\Lambda/2$.

\subsection{The origin of the drive term}
The generic model we introduced in \eq{eq:lag} is compatible with several scenarios suggested for modeling the role of the drive in pump-probe experiments of different materials. If $\phii$ describes a polarizable (IR-active) phonon, $F(t)$ can be directly identified with the external electric field, in which case, the coupling strength $\Lambda$ will be proportional to the polarizability of $\phii$. On the other hand, if $\phii$ is a non-polarizable (Raman-active) phonon, even though the incident light does not directly influence it through dipole coupling, the classical drive term can still be obtained via nonlinear coupling to a driven ``proxy'' IR-active phonon. 
The leading order nonlinear IR/Raman coupling allowed by symmetries is the cubic $\propto \phii_\mathrm{IR}^2\,\phii$ interaction. In this case, we can identify $F(t) \sim \langle \phii_{\mathrm{IR}}(t) \rangle$ as the coherent oscillations of the driven IR-active mode and $\Lambda$ as the strength of the cubic coupling to the Raman phonon $\phii$. Regardless of the origin of the classical drive, we assume:
\begin{equation}\label{eq:F_def}
F(t) = F_\mathrm{env}(t) \cos(\Omega_\mathrm{drv}t),
\end{equation}
where $\Omega_\mathrm{drv}$ is the principal frequency of the classical drive and $F_\mathrm{env}(t)$ is its slowly-varying envelope. Note that the classical drive couples to $\phii$ in intensity, $|F(t)|^2$, such that the effective principal drive frequency is $2\Odrv$.

If the drive term originates from nonlinear coupling to an IR-active phonon, $\Omega_\mathrm{drv}$ may no longer be identified with the frequency of the incident light after the pump pulse is ramped down; rather, the pump pulse coherently drives the proxy IR-active phonon out of its equilibrium position and subsequently, the coupled IR-active and Raman phonons oscillate together at a frequency predominantly determined by the IR-active mode. The proposed model still applies to this case with the appropriate choice of $\Odrv$.

\subsection{Different routes to parametric driving}
For the purposes of this work, the necessary ingredient of the model is a mechanism to achieve parametric driving of the $\phii$ phonon, i.e. a route for achieving the effective substitution $\omega_\qq^2 \rightarrow \omega_\qq^2[1 + 2\alpha_\qq \cos(2\Odrv t)]$, where $\alpha_\qq$ is the effective parametric driving amplitude. In the model proposed by Eq.~\eqref{sec:model}, this is achieved from the interplay between the nonlinearities of the $\phii$ phonon and its own coherent displacement, as further explained in the next section. There exists, however, a multitude of other physically realizable routes that all lead to parametric driving. This situation closely resembles the multitude of architectures proposed for building electronic parametric amplifiers over the years using elements such as variable capacitance diodes, nonlinear inductors, and Josephson junctions. The common theme remains the same: the interplay between pumping and nonlinear elements.

The strong pumping of a material with a complex crystal structure will induce coherent oscillations in a few primary modes. These oscillations trickle down to several other modes as a result of nonlinear couplings. Thus, {\em every} symmetry-allowed mode will be parametrically driven to a degree with strong enough pumping. With this understanding, the model proposed here is only one out of the numerous possible other routes to achieve parametric driving. For instance, the symmetry-allowed quartic coupling in fulleride superconductors $\sim \phii_\mathrm{IR}^2 \phii^2$ directly translates the coherent motion of $\phii_\mathrm{IR}$ to a parametric drive for $\phii$. Here, $\phii_\mathrm{IR}$ is one of the IR-active modes of $C_{60}$ such as $\mathrm{T}_{1u}(1-4)$, and $\phii$ is a Raman-active mode such as $\mathrm{Hg}(7-8)$ that couples strongly to conduction electrons~\cite{mitrano2016possible,varma1991superconductivity}. Even though achieving parametric driving is material-specific, it leads to the same qualitative physics. The present paper mainly deals with the universal consequences of parametric driving.

\section{Parametric amplification of phonon-mediated electron-electron attraction: a first look}\label{sec:param}
Our goal in this section is to demonstrate the resonant amplification of the electron-phonon coupling in the presence of the drive. For the time being, we neglect the complex epiphenomena such as the nonequilibrium evolution of electrons, phonon dissipation and retardation, and the feedback between electrons and phonons. Instead, we resort to a perturbative treatment and elementary methods in order to elucidate the main ideas. We will revisit the problem again in later section and provide a comprehensive account using the nonequilibrium Migdal-Eliashberg theory. The latter treatment is naturally more cumbersome than the physical account given in this section. The present analysis serves a guideline to identify and interpret the results of the upcoming detailed analysis.

As a first step, we assume that the lattice nonlinearity $\mathcal{V}(\phii)$ and the electron-phonon coupling $g_{\kk,\kk'}$ are both weak compared to the drive, such that we can study the coherent motion of the lattice in isolation. The classical equation of motion for $\langle \hat{\phii}_0(t) \rangle$ (the $\qq=0$ mode) is easily found as:
\begin{equation}
\partial_t^2 \langle \hat{\phii}_0(t) \rangle + \omega_0^2\,\langle \hat{\phii}_0(t) \rangle = \frac{\Lambda\omega_0}{2}\,\sqrt{N}\,F^2_\mathrm{env}(t)\,\cos^2(\Omega_\mathrm{drv}t),
\end{equation}
The normalization constant $\sqrt{N}$ results from the definition of the Fourier operators given earlier, i.e. $\langle \hat{\phii}_{0} \rangle = \sqrt{N}\,\langle \hat{\phii}_{j} \rangle$ where $\langle \hat{\phii} \rangle$ is the coordinate of an arbitrary single ion $j$. We assume that the temporal variation scale of $F_\mathrm{env}(t)$ is much longer than the drive period. Thus, for an adiabatically ramped up $F_\mathrm{env}(t)$, we find:
\begin{multline}\label{eq:phi0_approx_sol}
\langle \hat{\phii}_0(t) \rangle \approx \frac{\Lambda\,\sqrt{N}}{4\omega_0}\, F_\mathrm{env}^2(t) + \\
\frac{\Lambda\,\sqrt{N}\omega_0}{4(\omega_0^2-4\Omega_\mathrm{drv}^2)}\, F_\mathrm{env}^2(t) \, \cos(2\Omega_\mathrm{drv} t).
\end{multline}
Near the resonance $\Omega_\mathrm{drv} = \omega_0/2$, the oscillatory term dominates the dc term in amplitude. The precise value of the prefactors of the dc and ac terms are not important for the present discussion and in a more realistic setting, both get corrections from phonon damping, nonlinearities, etc. Quite generally though, we have $\langle \hat{\phii}_0(t) \rangle  \approx \sqrt{N} \, \phii_0(t) + \sqrt{N} \, \phii_1(t) \cos(2\Omega_\mathrm{drv}t)$ where $\phii_0(t)$ and $\phii_1(t)$ are slowly-varying functions of time. With this understanding, we drop the time labels from $\phii_0$ and $\phii_1$ hereafter and treat than as given quasi-steady parameters.

The local lattice nonlinearity terms couple the coherent uniform motion of the lattice to $\pm \qq$ modes. For instance, the leading order correction resulting from the cubic nonlinearity $\sim \phii^3$ is found by replacing one of the operators with $\langle \hat{\phii}_0(t) \rangle$. Momentum conservation implies opposite momenta for the remaining two operators:
\begin{multline}\label{eq:cubic_effective}
-\frac{\kappa_3}{3!}\sum_{j \in \mathrm{lattice}} \hat{\phii}_j^3 \rightarrow\\
-\frac{\kappa_3}{2} \left[\phii_0 + \phii_1\,\cos(2\Omega_\mathrm{drv} t)\right] \sum_{\qq \neq 0} \hat{\phii}_{\qq}\,\hat{\phii}_{-\qq}.
\end{multline}
Likewise, the leading order contribution from the quartic nonlinearity is found by replacing two of the operators with $\qq=0$, which yields:
\begin{multline}\label{eq:quartic_effective}
-\frac{\kappa_4}{4!}\sum_{j \in \mathrm{lattice}} \hat{\phii}_j^4 \rightarrow\\
-\frac{\kappa_4}{4} \left[\phii_0 + \phii_1\,\cos(2\Omega_\mathrm{drv} t)\right]^2 \sum_{\qq \neq 0} \hat{\phii}_{\qq}\,\hat{\phii}_{-\qq}.
\end{multline}
The dc terms result in the renormalization of the phonon frequency, e.g. $\omega_\qq^2 \rightarrow \omega_\qq^2 - \kappa_3 \phii_0/2$ for the cubic nonlinearity, and $\omega_\qq^2 \rightarrow \omega_\qq^2 - \kappa_4 (\phii_0^2/4 + \phii_1/8)$. Such corrections are precisely the time-averaged renormalized lattice properties that we discussed earlier in the introduction and can enhance or suppress the effective electron-phonon coupling on their own account.

\begin{figure}
\includegraphics[width=\linewidth]{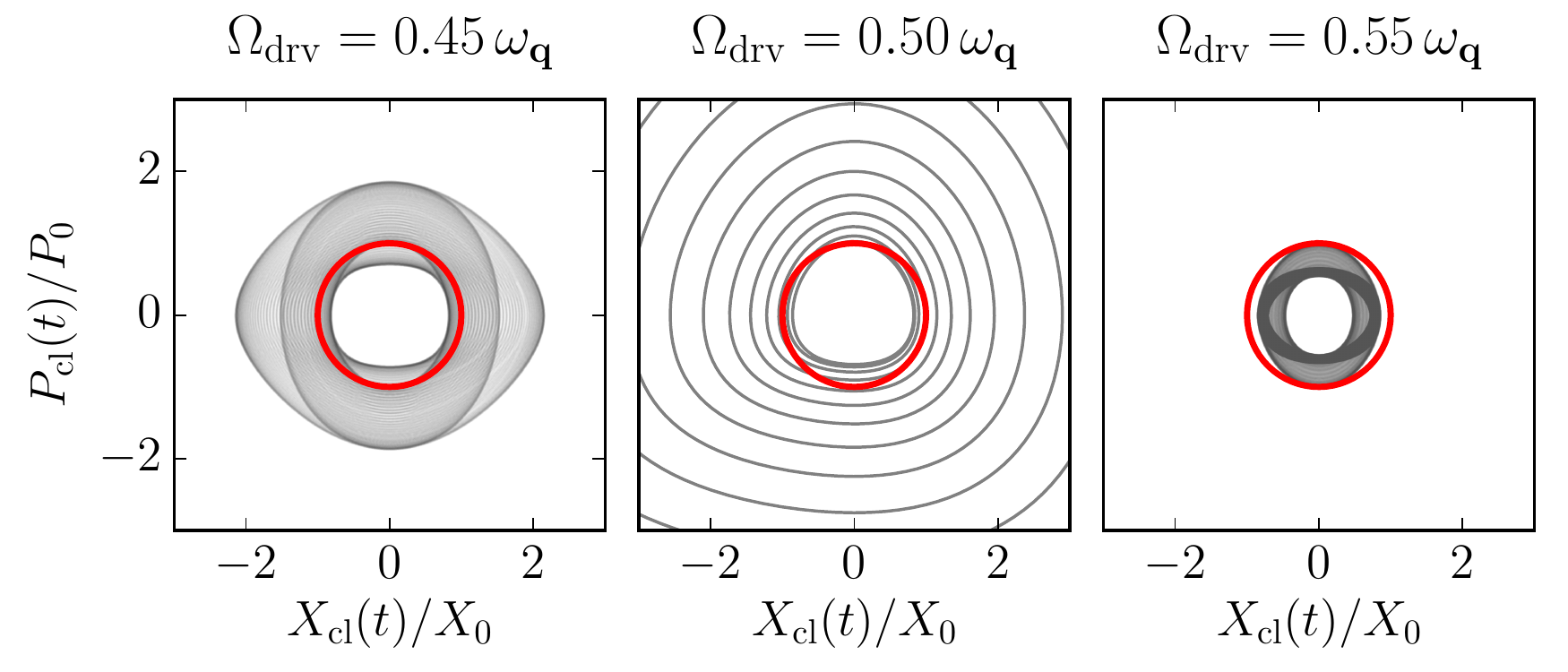}
\caption{{\bf Illustration of parametric amplification from classical phase-space trajectories.} The classical phase-space trajectories corresponds to a parametrically driven oscillator in response to a momentum jump with magnitude $P_0$ at $t=0$ for $\Omega_\mathrm{drv}$ below resonance (left), on resonance (middle), and above resonance (right). Here, $X_0 \equiv P_0/(2M\omega_\qq)$ is a normalization constant and the Mathieu parameter is set to $\alpha = 0.2$. The red circle is the periodic trajectory in the absence of the drive ($\alpha = 0$). Note the significantly amplified response below resonance $\Omega_\mathrm{drv} < \omega_\qq/2$, the diverging response on resonance $\Omega_\mathrm{drv} = \omega_\qq/2$, and suppressed response above resonance $\Omega_\mathrm{drv} > \omega_\qq/2$.}
\label{fig:trajectories}
\end{figure}

As we will show soon, the most intriguing effect is purely dynamical and stems from the ac term. We neglect the dc corrections for simplicity hereafter. At the present order in perturbation theory, only $\pm \qq$ opposite momentum pairs couple and thus we may focus on a single momentum pair without the loss of generality. We also only consider the cubic nonlinearity. It will soon become apparent that both types of nonlinearity give rise to the same resonant amplification phenomenon. The Hamiltonian is given as $\hat{H}_{\pm\qq}(t) = \hat{H}_\mathrm{e} + \hat{H}_\mathrm{p,\pm\qq}(t) + \hat{H}_\mathrm{ep,\pm\qq}$, where $\hat{H}_e = \sum_{\kk,\sigma} \xi_\kk \, c^\dagger_{\kk,\sigma} c^{\phantom{\dagger}}_{\kk,\sigma}$ and:
\begin{align}
H_\mathrm{p,\pm\qq}(t) &= \frac{\hbar \omega_\qq}{2} \, \hat{\phii}_\qq \, \hat{\phii}_{-\qq} + 2\,\hbar \omega_\qq \, \hat{\pi}_{\qq} \, \hat{\pi}_{-\qq}\nonumber\\
&\qquad\qquad -\kappa_3\, \phii_1 \, \cos(2\Odrv t) \, \hat{\phii}_{\qq} \, \hat{\phii}_{-\qq}\nonumber\\
H_\mathrm{ep,\pm\qq} &= g_\qq \, \hat{\phii}_{\qq}\,\hat{\rho}_{-\qq} + g_{-\qq} \, \hat{\phii}_{-\qq}\,\hat{\rho}_{\qq},
\end{align}
where $\hat{\pi}_{\qq} = (b^{\phantom{\dagger}}_{-\qq} - b^\dagger_{\qq})/(2i)$ is the conjugate momentum to $\hat{\phii}_\qq$~\footnote{Our definition of $\hat{\phi}_\qq$ and $\hat{\pi}_\qq$ deviates from the standard literature, resulting in unbalanced prefactors of the kinetic and potential terms in the harmonic oscillator Hamiltonian.}, and $g_\qq$ is the linear electron-phonon coupling constant (we have assumed $g_{\kk,\kk'} \approx g_{\kk - \kk'}$).

Finally, $\rho_\qq = \sum_{\kk,\sigma} c^{\dagger}_{\kk+\qq, \sigma} c^{\phantom{\dagger}}_{\kk, \sigma}$ is the electron charge density operator. The analysis can be further simplified by performing a canonical change of variables to standing wave phonon operators:
\begin{align}
\hat{Q}_+ &= \sqrt{\frac{\hbar}{4M\omega_\qq}}\left(\hat{\phii}_\qq + \hat{\phii}_{-\qq}\right),\nonumber\\
\hat{Q}_- &= \sqrt{\frac{\hbar}{4M\omega_\qq}}\left(\hat{i\phii}_\qq - i\hat{\phii}_{-\qq}\right),
\end{align}
and their corresponding conjugate momenta:
\begin{align}
\hat{P}_+ &= \sqrt{\hbar M \omega_\qq}\left(\hat{\pi}_\qq + \hat{\pi}_{-\qq}\right),\nonumber\\
\hat{P}_- &= \sqrt{\hbar M \omega_\qq}\left(i\hat{\pi}_\qq - i\hat{\pi}_{-\qq}\right).
\end{align}
Here, $M$ is the ion mass. It is readily verified that $[\hat{Q}_+, \hat{P}_+] = [\hat{Q}_-, \hat{P}_-] = i\hbar$ while all other commutators vanish. The Hamiltonian can be easily written in terms of $\hat{Q}_\pm$ and $\hat{P}_\pm$ operators:
\begin{subequations}
\begin{align}
\label{eq:H_driven_HO}
H_{\mathrm{p},\qq} &= \sum_{\alpha=\pm}\left(\frac{\hat{P}_\alpha^2}{2M} + \frac{1}{2}\,M\Omega_\qq^2(t)\, \hat{Q}_\alpha^2\right),\\
H_{\mathrm{ep},\qq} &= \tilde{g}_\qq\sum_{\alpha=\pm}\hat{Q}_\alpha \hat{\rho}_\alpha,
\end{align}
\end{subequations}
where $\tilde{g}_\qq = g_\qq \sqrt{2M\omega_\qq/\hbar}$, $\hat{\rho}_+ = (\hat{\rho}_{\qq} + \hat{\rho}_{-\qq})/\sqrt{2}$, $\hat{\rho}_- = i(\hat{\rho}_{\qq} - \hat{\rho}_{-\qq})/\sqrt{2}$, and:
\begin{align}\label{eq:alpha_def}
\Omega_\qq^2(t) &= \omega_\qq^2\left[1 + 2\alpha\cos(2\Odrv t)\right],\nonumber\\
\alpha &= - \kappa_3 \omega_\qq \, \phii_1 / (\hbar\omega_\qq)
\end{align}
The Hamiltonian is the sum of two decoupled harmonic oscillators ($\alpha=\pm$) each with linear coupling to a standing electronic charge density wave. Since the $\alpha=\pm$ modes undergo a similar evolution, we will drop the $\pm$ index hereafter and focus on a one of the modes. In the absence of electron-phonon coupling, the problem reduces to a parametrically driven harmonic oscillator. Corrections arising from electron-phonon coupling can be be studied order by order using time-dependent perturbation theory in the weak-coupling limit $g_\qq \nu(0) \ll 1$. This can be done, e.g. via a perturbative expansion of the unitary evolution operator $\hat{U}(t) = \mathcal{T}\exp\left[-i\hbar^{-1} \int_0^t \dd t' \, \hat{H}_{\pm\qq}(t')\right]$ in the powers of $\tilde{g}_\qq$~\cite{fujita2010quantum}. The leading order correction to the action is $\mathcal{O}(\tilde{g}_\qq^2)$ and is easily found as:
\begin{equation}
\Delta S^{(2)}(t) = -\frac{|\tilde{g}_\qq|^2}{\hbar} \int_{-\infty}^t \dd t' \, e^{-\epsilon(t-t')} \, \mathcal{D}^R_{QQ}(t,t') \, \hat{\rho}(t) \, \hat{\rho}(t'),
\end{equation}
where $\mathcal{D}^R_{QQ}(t,t') = - i\theta(t-t')\langle [\hat{Q}(t) , \hat{Q}(t')]\rangle$ is the retarded phonon correlator and $\epsilon$ is an infinitesimal. It is well-known that this correction implies an attractive interaction between the electrons in the long wave-length regime $|\xi_{\kk \pm \qq}- \xi_\kk| \ll \omega_\qq$. In this regime of interest, the phase winding of the electron charge density excitation is much slower than the phonon time scale. Thus, to simplify the discussion further, we simply neglect the relative phase winding of the electron charge density waves and set $\hat{\rho}(t') \rightarrow \hat{\rho}(t)$ from the outset. Note that this coincides with the usual $\varepsilon_\kk = \varepsilon_{\kk'} = \varepsilon_F$ approximation in the BCS treatment. This results in the following simple expression for the phonon-mediated attractive interaction, $\Delta S^{(2)}(t) = U(t) \, \hat{\rho}(t) \, \hat{\rho}(t)$ where:
\begin{equation}\label{eq:U_pert}
U(t) = \frac{|\tilde{g}_\qq|^2}{\hbar} \int_{-\infty}^t \dd t' \, e^{-\epsilon(t-t')} \, \mathcal{D}^R_{QQ}(t,t').
\end{equation}
Since the Hamiltonian is time-dependent, $U(t)$ is expected to be time-dependent as well. In particular, for the periodic Hamiltonian given in Eq.~\eqref{eq:H_driven_HO}, $U(t)$ further admits a Fourier expansion $U(t) = \sum_{n=-\infty}^{+\infty} U_n \, e^{-2in\Omega_\mathrm{drv} t}$ in the harmonics of $2\Odrv$. In plain words, $U(t)$ is proportional to the {\em mean displacement} of the oscillator up to time $t$ in response to a momentum boost at all prior times.

The retarded phonon propagator $\DD_{QQ}^R(t,t')$ is most easily calculated using Heisenberg equations for $\hat{Q}(t)$ and $\hat{P}(t)$:
\begin{subequations}
\begin{align}
\frac{\dd \hat{Q}(t)}{\dd t} &= \frac{\hat{P}(t)}{M},\\
\frac{\dd \hat{P}(t)}{\dd t} &= -M\omega_\qq^2 \big[1 + 2\alpha \cos(2\Omega_\mathrm{drv}t)\big]\hat{Q}(t),
\end{align}
\end{subequations}
The Heisenberg equations exactly coincide with the classical equations of motion due to the Ehrenfest's theorem. The formal solution of these equations can be expressed in terms of Mathieu functions. For given Heisenberg operators at time $t'$, we find:
\begin{align}\label{eq:QP_dyn}
\hat{Q}(t) &= \mathfrak{M}_{QQ}(t,t') \, \hat{Q}(t') - \mathfrak{M}_{QP}(t,t')\,\frac{\hat{P}(t')}{M\Odrv},\nonumber\\
\hat{P}(t) &=  \mathfrak{M}_{PQ}(t,t')\, M\Odrv\,\hat{Q}(t') + \mathfrak{M}_{PP}\,\hat{P}(t'),
\end{align}
The explicit expressions for the $\mathfrak{M}$-functions are given in Eq.~\eqref{eq:mathieu} in terms of even and odd Mathieu functions. In the limit $\alpha \rightarrow 0$ (no drive), we have $\lim_{\alpha \rightarrow 0} \mathfrak{M}_{QQ}(t,t') = \lim_{\alpha \rightarrow 0} \mathfrak{M}_{PP}(t,t') = \cos[\omega_\qq(t-t')]$, and $(\omega_\qq/\Odrv)\,\lim_{\alpha \rightarrow 0} \mathfrak{M}_{QP}(t,t') = -(\Odrv/\omega_\qq)\,\lim_{\alpha \rightarrow 0} \mathfrak{M}_{PQ}(t,t') = -\sin[\omega_\qq(t-t')]$, thus reducing Eq.~\eqref{eq:QP_dyn} the usual non-driven harmonic oscillator evolution. Furthermore, $\mathfrak{M}_{QQ}(t',t') = \mathfrak{M}_{PP}(t',t') = 1$ and $\mathfrak{M}_{PQ}(t',t') = \mathfrak{M}_{QP}(t',t') = 0$, as it is also required from the initial condition. The retarded correlator is immediately calculated using Eq.~\eqref{eq:QP_dyn}, giving an exceedingly simple result:
\begin{align}\label{eq:D_ret_mathieu}
\DD^R_{QQ}(t,t') &= -i\,\theta(t-t')\langle[\hat{Q}(t'), \hat{P}(t')]\rangle\, \times \frac{-\mathfrak{M}_{QP}(t,t')}{M\Odrv}\nonumber\\
&=-\frac{\hbar}{M\Odrv}\,\theta(t-t')\,\mathfrak{M}_{QP}(t,t').
\end{align}
This result presents several important features:
\begin{itemize}
  \item{The retarded phonon propagator $\DD^R_{QQ}(t,t')$, and consequently $U(t)$ (see Eq.~\ref{eq:U_pert}), are {\em fully determined} by $\mathfrak{M}_{QP}(t,t')$. The latter is in turn fully determined by the Heisenberg equations, and is independent of the initial wavefunction of phonons. In other words, for the Hamiltonian given in Eq.~\eqref{eq:H_driven_HO}, one obtains the same effective interaction $U(t)$ in all equilibrium and nonequilibrium phonon states. This is a direct consequence of the linearity of the harmonic oscillator evolution. As a corollary, this result immediately shows the peculiar cancellation of Bose enhancement factors between phonon absorption and emission processes in the conventional textbook diagrammatic calculation of $U$ in equilibrium~\cite{schrieffer1983theory}.}
  \item{$\DD_{QQ}^R(t,t') \propto \mathfrak{M}_{QP}(t,t')$ admits a simple classical interpretation: it coincides with the {\em displacement} of a classical driven oscillator at time $t$ in response to a {\em momentum} jump at time $t'$, i.e. $\DD_{QQ}^R(t,t')~=~\hbar \, \delta Q_\mathrm{cl}(t)/ \delta P_\mathrm{cl}(t')$. This result can be obtained independently and more directly using quantum phase-space methods~\cite{polkovnikov2010phase}.}
  \item{The origin of phonon-mediated attraction is purely quantum mechanical. Had $\hat{Q}$ been a classical operator, it would commute with itself at different times and we would obtain $\lim_{\hbar \rightarrow 0}\DD^R_{QQ}(t,t') = 0$. Nonetheless, this analysis provides a classical analogy as a response to a momentum jump (see the previous remark).}
\end{itemize}
The second remark implies that classical trajectories of a parametrically driven oscillator following a momentum jump encode the necessary and sufficient information to calculate $U(t)$. As a first example, let us consider the non-driven limit. In this case, we find $\lim_{\alpha \rightarrow 0} \mathfrak{M}_{QP}(t,t') = -(\Odrv/\omega_\qq)\,\sin \omega_\qq (t-t')$
\begin{equation}
\lim_{\alpha \rightarrow 0} \DD^R_{QQ}(t,t') = \hbar\,\theta(t-t')\,\frac{\sin \omega_\qq(t-t')}{M\omega_\qq}.
\end{equation}
The above result indeed corresponds the $QP-$response of the classical harmonic oscillator up to a factor of $\hbar$. Plugging this result in Eq.~\eqref{eq:U_pert}, we find $U = -|\tilde{g}_\qq|^2/(M\omega_\qq^2) = -2|g_\qq|^2/(\hbar \omega_\qq)$ which is the usual time-independent equilibrium result~\cite{mahan2013many}. Note that regularizing prefactor $e^{-\epsilon(t-t')}$ is crucial for obtaining this result, without which the $t'$-integration is ill-defined. In a more realistic model with finite phonon damping, regularization is unnecessary.

\begin{figure}
\includegraphics[width=\linewidth]{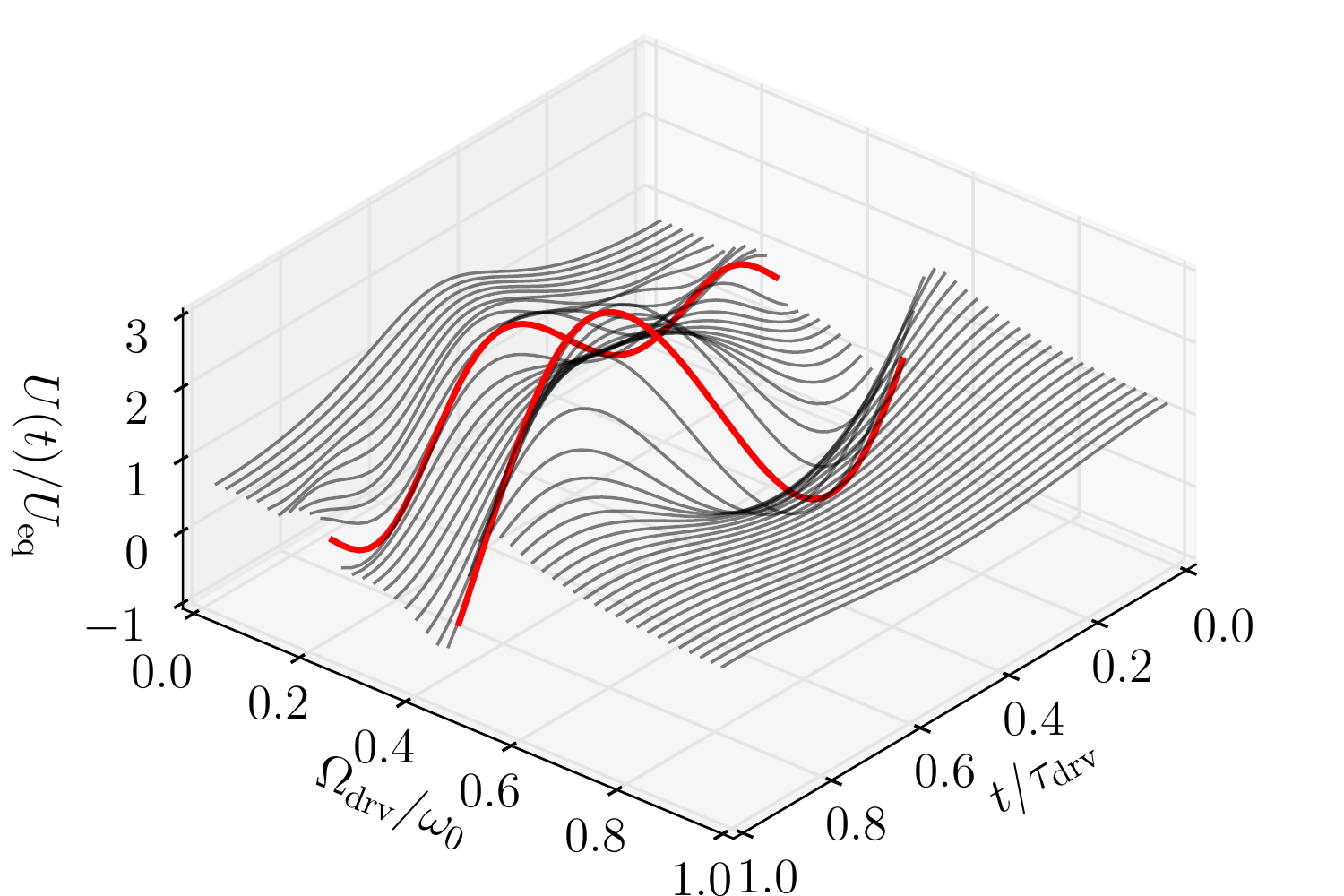}
\caption{{\bf The effective electron-electron interaction $U(t)$ as a function of drive frequency $\Omega_\mathrm{drv}$ and time.} Here, $\tau_\mathrm{drv} = \pi/\Omega_\mathrm{drv}$ and $U_\mathrm{eq} = - 2|g_\qq|^2/(\hbar\omega_\qq)$ is attraction strength in the absence of the drive. The Mathieu parameter is $\alpha = 0.2$ and we have set the damping rate to $\epsilon=0.1 \omega_\qq$. The red thick lines show $U(t)$ on the first two resonances $\Odrv/\omega_\qq = 1/2, 1/4$.} 
\label{fig:semiclassical_U}
\end{figure}

\begin{figure}
\includegraphics[width=\linewidth]{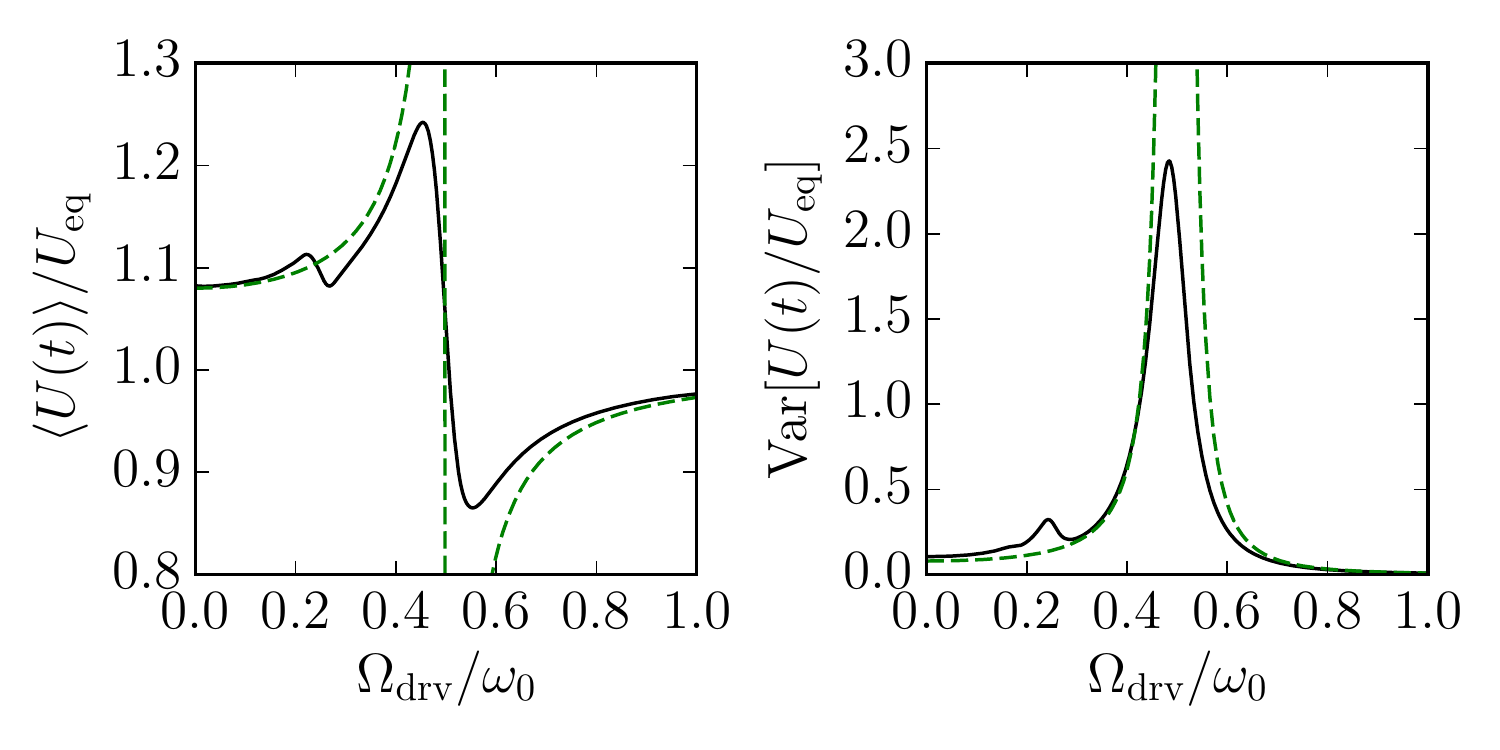}
\caption{{\bf Time average and variance of the effective electron-electron interaction.} Solid lines are numerical results obtained from the solutions of the Mathieu equation. Dashed lines correspond to the perturbative results given in \eq{eq:U_meanvar}. The Mathieu parameter is set to $\alpha=0.2$.} 
\label{fig:semiclassical_stats}
\end{figure}

Fig.~\ref{fig:trajectories} shows the classical trajectories in response to a sudden momentum jump at $t=0$ in the presence of a finite parametric drive. The red circular orbit shows the response in the absence of the drive. It is noticed that the classical trajectory diverge on resonance $\Odrv = \omega_\qq/2$ (middle panel), implying an {\em infinitely enhanced} response. Intuitively, the lattice becomes critically unstable in the presence of a resonant drive of finite amplitude such that smallest perturbation causes an infinitely large deformation. Just below the resonance (left panel), the trajectories are noticeably expanded in the phase space though remain bounded. This corresponds to a moderately enhanced $U(t)$. Finally, the response is attenuated just above the resonance (right panel), which is the expected asymmetric behavior near a parametric resonance.

The time-dependent effective attraction $U(t)$ can be obtained numerically using Eqs.~\eqref{eq:D_ret_mathieu} and~\eqref{eq:U_pert}. The results are shown in Fig.~\ref{fig:semiclassical_U}. According to the previously mentioned periodicity $U(t + \pi/\Odrv) = U(t)$, we have only shown one full period in $t$. The immediately noticeable feature is the large temporal variations of $U(t)$ near parametric resonances, which is a manifestation of the wild variations of classical trajectories in the phase-space in this regime. The perturbation series for $U(t)$ can be worked out using Eqs.~\eqref{eq:U_pert},~\eqref{eq:D_ret_mathieu}, and the expressions given in Appendix~\ref{sec:mathieu}. We quote the final result here:
\begin{multline}\label{eq:U_t_pert}
\frac{U(t)}{U_\mathrm{eq}} = 1 - \frac{2\alpha \,\omega_\qq^2\,\cos(2\Odrv t)}{\omega_\qq^2 - 4\Odrv^2}\\
+ \frac{2\alpha^2\,\omega_\qq^2\left[\omega_\qq^2 - 16\Odrv^2 + \omega_\qq^2\cos(4\Odrv t)\right]}{(\omega_\qq^2 - 16 \Odrv^2)(\omega_\qq^2 - 4\Odrv^2)} + \mathcal{O}(\alpha^4)
\end{multline}
Note the parametric resonances at $\Odrv = \omega_\qq/2$ in the first order term and $\Odrv = \omega_\qq/4$ in the second order term, as well as the appearance of higher order harmonics of $2\Odrv$.

We remark that the parametric resonances of the Mathieu oscillator, $\Odrv = \omega_\qq/n$, $n \in \mathbb{N}$, do not necessarily imply a corresponding resonance in the effective attraction $U(t)$. As remarked after Eq.~\eqref{eq:U_pert}, $U(t)$ can be interpreted as the mean displacement of the oscillator in response to a momentum boost. The mean displacement can behave properly even for divergent trajectories. For example, the leading $n=1$ parametric resonance at $\Odrv = \omega_\qq$ leaves $U(t)$ regular while leading to a divergent response at the same time. This can be easily noticed in the perturbation analysis: Eq.~\eqref{eq:traj_pert} shows a resonance at $\Odrv = \omega_\qq$ while $U(t)$ remains regular, see Eq.~\eqref{eq:U_t_pert} and Fig.~\ref{fig:semiclassical_stats}.

The time average and variance of $U(t)/U_\mathrm{eq}$ can be found readily using the above result:
\begin{subequations}\label{eq:U_meanvar}
\begin{align}\label{eq:U_mean}
\left\langle\frac{U(t)}{U_\mathrm{eq}}\right\rangle &= 1 + \frac{2\alpha^2\,\omega_\qq^2}{\omega_\qq^2 - 4\Odrv^2} + \mathcal{O}(\alpha^4),\\
\label{eq:U_var}
\mathrm{Var}\left[\frac{U(t)}{U_\mathrm{eq}}\right] &= \frac{2 \, \alpha^2 \,\omega_\qq^4}{(\omega_\qq^2 -4\Odrv^2)^2} + \mathcal{O}(\alpha^4).
\end{align} 
\end{subequations}
The leading order correction to $\langle U(t) \rangle$ is $\mathcal{O}(\alpha^2)$. Fig.~\ref{fig:semiclassical_stats} shows the average and variance of $U(t)$ within a period. The dashed lines show the above perturbative results. As expected from the study of classical trajectories (see Fig.~\ref{fig:trajectories}), $\langle U(t) \rangle$ is enhanced for frequencies below the resonance and is suppressed above the resonance. Perhaps more importantly, the temporal variance of $U(t)$ is significantly increased on either side of the resonance. This can be seen e.g. from Eq.~\eqref{eq:U_var}. In the next section, we discuss the important role of enhanced temporal variations of $U(t)$ in enhancing the superconducting transition temperature.

\subsection{The superconducting transition temperature for a time-dependent effective interaction}\label{sec:floquet_BCS}
We showed that the parametric drive of the lattice results in the enhancement of the effective attraction mediated between the electrons, $U(t)$. We further showed that $U(t)$ exhibits large oscillations in $t$ near parametric resonances (see Fig.~\ref{fig:semiclassical_U}), such that $U(t)$ takes on values that are significantly higher and lower than its equilibrium value. According to the BCS theory, $T_c[U] \sim \exp[-1/(\nu(0) U)]$ where $\nu(0)$ is the electronic density of states (EDOS) at the Fermi surface. It is tempting to naively propose a replacement $U \rightarrow U(t)$ in the BCS formula and propose $T_c^\mathrm{drv} \sim \langle T_c[U(t)] \rangle$. Since $T_c[U]$ is a convex function of $U$ in the weak-coupling limit, one would then conclude $\langle T_c[U(t)] \rangle > T_c[\langle U(t) \rangle]$, i.e. temporal variations of $U(t)$ can increase $T_c$ even if $\langle U(t) \rangle$ remains constant or even decrease. Of course, this argument lacks rigor and $T_c^\mathrm{drv}$ must be found within a proper Floquet extension of the BCS theory~\cite{knap2015dynamical}. 
%
%
To this end, we assume:
\begin{equation}
U(t) \approx U_0 + U_1\,\cos(2\Odrv t),
\end{equation}
where $U_0$ and $U_1$, can be read from Eq.~\eqref{eq:U_t_pert}:
\begin{align}\label{eq:U_01}
U_0 &= -\frac{g^2}{2\omega_0}\left[1 + \frac{2\alpha^2\,\omega_\qq^2}{\omega_\qq^2 - 4\Odrv^2} + \mathcal{O}(\alpha^4)\right],\nonumber\\
U_1 &= \frac{g^2}{2\omega_0}\left[\frac{2\alpha \,\omega_\qq^2}{\omega_\qq^2 - 4\Odrv^2} + \mathcal{O}(\alpha^4)\right].
\end{align}

The superconducting gap inherits the periodicity of $U(t)$ such that $\Delta(t) = \frac{1}{N}\sum_\kk \langle \psi_{\kk,\up}(t) \, \psi_{-\kk,\dn}(t)\rangle \rightarrow \sum_{n=-\infty}^{+\infty} \Delta_n \, e^{2ni\Odrv t}$. At the onset of pairing, the Floquet BCS gap equation takes the following form~\cite{knap2015dynamical}:
\begin{equation}\label{eq:gap_eq}
\left(1 - U_0 F_n\right)\Delta_n + \frac{U_1}{2} F_n\left(\Delta_{n-1} + \Delta_{n+1}\right) = 0,
\end{equation}
where:
\begin{equation}
F_n = -\nu(0) \int_{-\omega_c/2}^{\omega_c/2} \dd\xi\,\frac{\tanh[\xi/(2T)]}{2\xi - 2n\Odrv + i0^+}.
\end{equation}
Here, $T$ is the temperature and $\omega_c$ is the UV cutoff for $U$. For an Einstein phonon with frequency $\omega_0$, we expect $\omega_c \sim \omega_0$. In principle, $T_c$ must be found such that Eq.~\eqref{eq:gap_eq} admits a non-trivial solution for $\{\Delta_n\}$ and since it is a homogeneous equation in $\{\Delta_n\}$, it reduces to the vanishing determinant condition for an (infinitely large) matrix.

A closed form solution for $T_c$ seems to be out of reach in general and one must resort to numerical methods. We attempt to find an approximate analytic solution with additional assumptions $U_0 \nu(0) \ll 1$ and $U_1 \ll U_0$. Strictly speaking, the last assumption does not generally apply to our problem since $|U_1| \geq |U_0|$ near the resonances. Our main goal here is to demonstrate how temporal variations in $U(t)$ increases $T_c$ rather than presenting a rigorous analysis; the latter is the objective of the upcoming sections. Thus, the additional simplifying assumptions must be taken with this understanding in mind.

As a first step, we observe that $\Delta_{n+1}/\Delta_n \sim \mathcal{O}(U_1/U_0) \ll 1$. Therefore, we may neglect $\Delta_n$ for $|n| \geq 2$ for small $U_1$. This reduces the infinite set of equations for $\{\Delta_n\}$ to only three equations for $\Delta_0$, $\Delta_{1}$, and $\Delta_{-1}$. Omitting $\Delta_{\pm 1}$ between the equations and assuming $\Delta_0 \neq 0$, we find the following approximate pairing condition:
\begin{equation}\label{eq:effective_gap_eq}
U_0 F_0 + \frac{U_1^2}{4} F_0 \left(\frac{F_1}{1 - U_0 F_1} + \frac{F_{-1}}{1 - U_0 F_{-1}}\right) = 1.
\end{equation}
Approximate expressions for $F_n$ can be found in the limit $T/\omega_c, T/\Odrv \ll 1$ using the Sommerfeld expansion technique. We quote the final result here:
\begin{align}
F_0 &= -\nu(0)\left[\ln(2\omega_c/T) + \gamma - \ln \pi + \mathcal{O}(T/\omega_c)\right],\nonumber\\
F_n &= -\frac{\nu(0)}{2}\bigg[-i \pi\,\mathrm{sgn}(n) + \ln\left|\frac{\omega_c^2}{n^2 \Odrv^2} - 1\right|\nonumber\\
&\qquad + \mathcal{O}(e^{-\omega_c/T_c})\bigg], \qquad (|n| > 0),
\end{align}
where $\gamma \simeq 0.577$ is the Euler-Mascheroni constant. The final result resembles the BCS formula for $T_c$, though, with $U_0$ replaced with an effective interaction $U_\mathrm{eff}$:
\begin{subequations}
\begin{equation}\label{eq:T_c_analytic}
T^\mathrm{drv}_c \simeq \frac{2e^\gamma}{\pi}\, \omega_c\, \exp[-1/(\nu(0) U_\mathrm{eff})],
\end{equation}
where:
\begin{equation}\label{eq:U_eff}
U_\mathrm{eff} \approx \left\{\begin{array}{ll}\displaystyle
|U_0| + \frac{\nu(0)\,U_1^2}{4}\,\ln\left|\frac{\omega_c^2}{\Odrv^2} - 1\right| & \quad \displaystyle\frac{\omega_c}{\Odrv} \nsim 1,\\
\\
\displaystyle
|U_0| + \frac{U_1^2}{2 |U_0|} & \displaystyle \quad \frac{\omega_c}{\Odrv} \sim 1.
\end{array}\right.
\end{equation}
\end{subequations}
We notice that $U_\mathrm{eff} > |U_0|$ in both cases. The last result provides a well-founded justification for our preliminary heuristic argument based on the convexity of $T_c[U]$.

\begin{figure}[t]
\includegraphics[width=\linewidth]{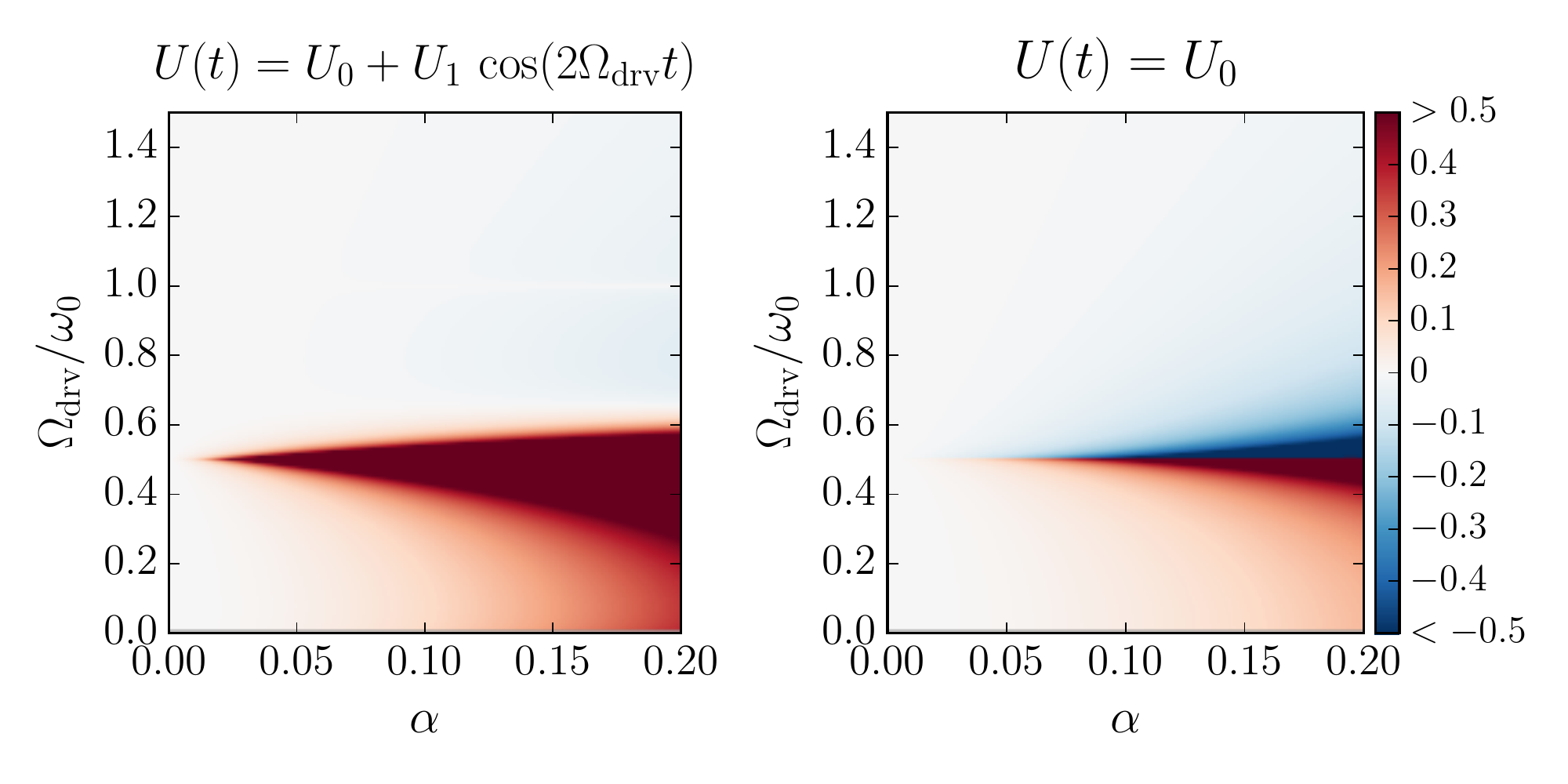}
\caption{{\bf Heat map plot of $T_c^\mathrm{drv}/T_c^\mathrm{eq} -1$ based on the analytic Floquet-BCS analysis of Sec.~\ref{sec:floquet_BCS}.} The plot on the left is the result obtained from $U(t) = U_0 + U_1\cos(2\Odrv t)$. The plot on the right is obtained by neglecting the ac component and setting $U(t) = U_0 = \mathrm{const.}$ The significant role of oscillations of $U(t)$ in enhancing $T_c$ is clearly noticeable. Note that $U_0$ and $U_1$ are both functions of $\alpha$ and $\Odrv$ as given in Eq.~\eqref{eq:U_01}, and $T_c$ is calculated using Eqs.~\eqref{eq:T_c_analytic}-\eqref{eq:U_eff}. In both cases, we have set $\nu(0) U_\mathrm{eq} = 0.5$ and $\omega_c = \omega_0$.}
\label{fig:T_c_analytic}
\end{figure}

Finally, we have plotted $T_c^\mathrm{drv}$ based on the approximate analytic results obtained in this section (Eqs.~\ref{eq:U_01},~\ref{eq:T_c_analytic},~\ref{eq:U_eff}) in Fig.~\ref{fig:T_c_analytic}. The left heat map plot shows the full result, when both dc and ac components of $U(t)$ are taken into account. It is noticed that $T_c$ is dramatically enhanced both below and above the resonance. The right plot shows the result when only the dc component, $U_0$, is kept. While $T_c^\mathrm{drv}$ is enhanced below the resonance, it is suppressed for $\Odrv > \omega_0/2$. This result can be understood by appealing to Fig.~\ref{fig:semiclassical_stats}. The time-averaged interaction $\langle U(t) \rangle = U_0$ is lower than $U_\mathrm{eq}$ for $\Odrv > \omega_0/2$, and $T_c$ is decreased accordingly. In contrast, including the ac component brings in large oscillations $\propto U_1$ which offset the loss in $U_0$ by allowing $U(t)$ to exceed $U_\mathrm{eq}$ during a fraction of each cycle. 

We will obtain a plot similar to Fig.~\ref{fig:T_c_analytic} later using a Floquet extension of the Migdal-Eliashberg theory (see Fig.~\ref{fig:T_c}), and we will find that the approximate analytic picture provided here remains remarkably accurate. We conclude this section by noting that had we included higher-order harmonic corrections in $U(t)$, we would get additional parametric resonances beside to the main one. Those would subsequently lead to additional ``tongues'' in Fig.~\ref{fig:T_c_analytic} around $\Odrv = \omega_0/4, \omega_0/8, \ldots$ (see Fig.~\ref{fig:T_c}).

\subsection{Higher-order nonlinearities, phonon damping, and parametric resonance}\label{sec:higher_order_nonlins}
Our discussion in the last two sections was based on a cubic lattice nonlinearity. We showed that the nonlinearity leads to a constant renormalization of $\omega_\qq$ and produces a $\sim \cos(2\Odrv t)$ periodic correction to $\omega_\qq$ at the leading order (see Eq.~\ref{eq:cubic_effective}). The scenario remains similar for quartic nonlinearities, save for an additional $\sim \cos(4\Odrv t)$ correction to $\omega_\qq$. This can be seen by expanding the square brackets in Eq.~\eqref{eq:quartic_effective}. In fact, for a nonlinearity $\sim \hat{\phii}^n$ ($n \geq 3$), we find $\hat{\phii}^n \approx \langle \hat{\phii}_0(t) \rangle^{n-2} \, \sum_\qq \hat{\phii}_\qq \, \hat{\phii}_{-\qq}$. Subsequently, $\langle \hat{\phii}_0(t) \rangle^{n-2}$ can be expanded in harmonics of $2\Odrv$ such that $\omega_\qq^2 \rightarrow \omega_{\qq}^2\big[1 + \sum_{n=0}^{n-2} \alpha_n\,\cos(2n\Odrv t)\big]$ for some $\{\alpha_n\}$. This analysis can be naturally extended for any smooth nonlinear potential. Since all parametric drive terms are harmonics of $2\Odrv$, the physics is expected to remain qualitatively similarly to the single harmonic case. For example, the case of double parametric drive terms is studied in Ref.~\cite{sofroniou2014dynamics}, where it is shown if the drive frequencies are integer multiples of one another, the same instability ``tongue'' patterns are obtained. Finally, let us mention in passing that with a finite phonon damping and for a fixed driving strength, the infinite cascade of ideal parametric resonances at $\Odrv = \omega_0/(2n)$ will be truncated above a certain order~\cite{gunderson1974technique}. For instance, only two resonances are noticeable in Fig.~\ref{fig:semiclassical_U} (black lines). The reason can be traced back to using a small phonon damping in the numerics.

\subsection{Intermission}
We derived a simple formula for the phonon-mediated electron-electron attraction $U(t)$ in the presence of parametric drive (Eq.~\ref{eq:U_pert}). We showed that $U(t)$ is independent of the initial phonon wavefunction and is fully determined by the Hamiltonian and the drive. We related $U(t)$ to the momentum-jump response of a classical parametric oscillator and used this classical analogy to demonstrate the parametric amplification of $U(t)$ by appealing to the classical phase-space trajectories of the parametric oscillator. Finally, we calculated an analytical formula for $T_c$ using a Floquet generalization of the BCS theory~\cite{knap2015dynamical} and demonstrated that the oscillations $U(t)$, as well as its increased time-average, lead to enhanced $T_c$.

Parametric amplification, while enhancing $T_c$ at the first glance, leads to several undesirable consequences as well. It leads to parametric phonon generation, a well-known phenomenon in the context of early universe field theory~\cite{berges2003parametric}. The generated phonons heat up the electrons and decrease their coherence. The oscillatory effective electron-electron interaction $U(t)$ also generates electron-hole excitations on its own account~\cite{knap2015dynamical}. An unbiased and consistent analysis of these competing effects is a challenging task and requires a rigorous and unified treatment. In the remainder of the paper, we will develop such a formalism and revisit the problem one more time in full detail.

\section{The Floquet-Migdal-Eliashberg quantum kinetic theory}\label{sec:FME_main}
\begin{figure}[th]
\includegraphics[width=\linewidth]{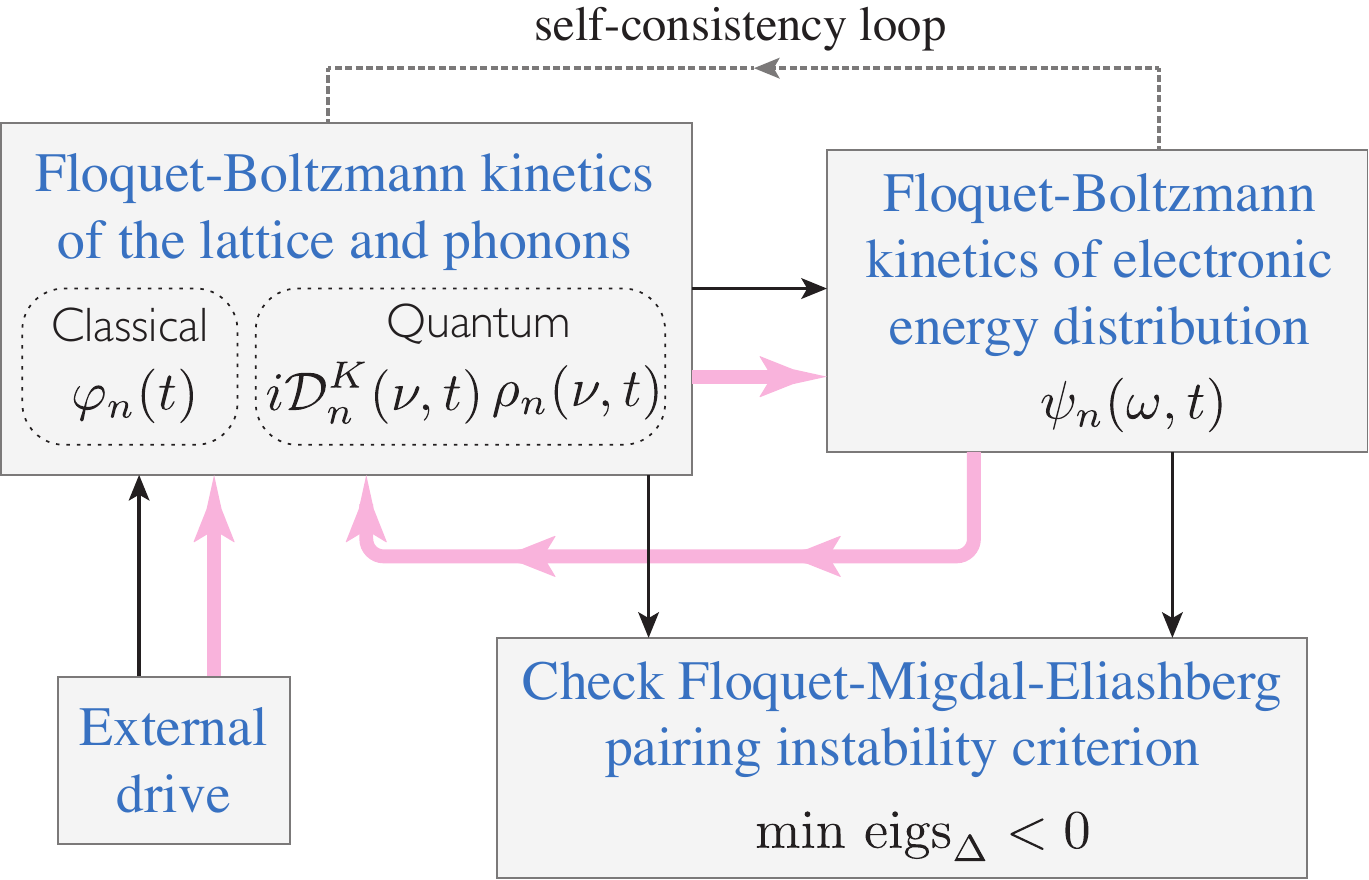}
\caption{{\bf A flowchart for the Floquet-Migdal-Eliashberg quantum kinetic formalism.} The external drive along width the initial electron propagators determine the evolution of the coherent (``classical'') lattice displacement and phonon propagators (``quantum''). Subsequently, the evolution of electronic energy distribution is calculated on the backdrop of the driven lattice. This procedure can be iterated until convergence if required. Finally, the Floquet-Migdal-Eliashberg pairing condition is assessed to determine whether the normal state exhibits a pairing instability during the evolution. The thin lines show the ``procedural flow'' of the calculations. The thick red lines show the ``heat flow'', from the external drive to phonons, then to electrons, and finally back to phonons through self-consistency.}
\label{fig:flowchart}
\end{figure}

The major steps of the forthcoming calculation is summarized graphically in Fig.~\ref{fig:flowchart}. Initially, the electron-phonon system is prepared in a normal-conducting equilibrium state with temperature $T_i>T_c$, where $T_c$ is the equilibrium critical temperature for Cooper pairing. The external drive is then ramped up and initiates the nonequilibrium quantum dynamics.

We study the coupled quantum dynamics of the lattice displacement, phonons, and electrons by deriving and numerically solving a set of quantum kinetic equations specifically tailored for investigating driven electron-phonon systems. We start this theoretical development from a two-particle irreducible effective active (2PI-EA) formulation of the Migdal-Eliashberg theory extended to nonlinear lattices in Sec.~\ref{sec:2PI_EOM}, followed by a Floquet generalization of the ensuing quantum kinetic equations in Secs.~\ref{sec:qfb}-\ref{sec:QFB_electron}.

Once the evolution of the system in the normal-conducting state is calculated, it is checked for the emergence of a Floquet superconducting instability during the evolution. This step in enabled by deriving a Floquet-Migdal-Eliashberg gap equation for the onset of pairing in Sec.~\ref{sec:FME_instab}. A summary of the numerical methods is provided in Appendix~\ref{app:numerics}.

\subsection{Two-particle irreducible effective action (2PI-EA) and real-time evolution equations}\label{sec:2PI_EOM}
We study the nonequilibrium dynamics of the driven electron-phonon system using the functional technique of 2PI effective actions (2PI-EA). A prominent feature of this approach is the guaranteed satisfaction of conservation laws and the absence of secular terms that arise in non-self-consistent perturbation theory~\cite{berges2004introduction}. Both of these features are necessary for stable and physically meaningful description of nonequilibrium quantum dynamics. In the context of our problem, the effective action is a functional of the uniform coherent lattice displacement $\phii$, the Nambu closed-time-path (CTP) electron propagator $\hGG$, and the CTP phonon propagator $\DD$. The lattice displacement field (a ``classical'' object) is coupled to the phonon propagator (a ``quantum'' object) via lattice nonlinearities and lead to the previously discussed parametric amplification effect.

We will work in the Migdal's limit $\omega_\mathrm{ph}/W_\mathrm{el} \ll 1$ ($\omega_\mathrm{ph}$ is the typical optical phonon frequency and $W_\mathrm{el}$ is the conduction electron bandwidth) where electron-phonon vertex corrections are suppressed by the powers of $\omega_\mathrm{ph}/W_\mathrm{el}$ and may be controllably neglected~\cite{migdal1958interaction,engelsberg1963coupled}. In the ideal Migdal limit $\omega_\mathrm{ph}/W_\mathrm{el} \rightarrow 0$, the 2PI-EA, $\GAM[\phii, \DD, \DD]$ truncated to the two-loop order in lattice nonlinearities is given as:
\begin{subequations}
\begin{align}\label{eq:2PI}
&\GAM[\phii, \DD, \hGG] = S_\mathrm{cl}[\phii] + \frac{i}{2}\, \mathrm{tr} \ln \DD^{-1} + \frac{i}{2}\,\mathrm{tr}[\DD_0^{-1} \DD]\nonumber\\
&\qquad -i \, \mathrm{tr} \ln \hGG^{-1} -i \, \mathrm{tr}[\hGG_0^{-1}\hGG] + \GAM_2[\phii, \DD, \hGG]\nonumber\\
&\GAM_2[\phii, \DD, \hGG] = i \sum_\kk \low{g}_{\kk,\kk} \int_\CC \dd t \, \phii(t) \, \mathrm{tr}\left[\hGG_\kk(t, t^+)\,\hat{\sigma}_3\right]\nonumber\\
&\qquad- \frac{1}{2} \sum_\qq \int_\CC \dd t_1 \, \dd t_2 \, \Pi_\qq(t_1,t_2)\, i\DD_\qq(t_2,t_1)\nonumber\\
&\qquad+ \frac{iN\kappa_4}{4}\int_\CC \dd t \, \phii^2(1) \, \DD_\ell(t,t) - \frac{N\kappa_4}{8}\int_\CC \dd t\,[\DD_\ell(t,t)]^2\nonumber\\
&\qquad+ \frac{iN\kappa_3}{2}\int_\CC \dd t \, \phii(t) \, \DD_\ell(t,t).
\end{align}
Here, $\GAM_2[\phii, \DD, \hGG]$ is the two-loop part of the action, $\phii(t) = \sqrt{N} \, \langle \phii_{\qq=0}(t) \rangle$ is the uniform coherent displacement of the lattice which is induced via coupling to the external drive $|F(t)|^2$, $\hGG_\kk(t_1, t_2) \equiv -i\,\big\langle T_\CC [\PSI_\kk(t_1) \PSID_{\kk}(t_2)]\big\rangle$ is the Nambu closed-time-path (CTP) electron propagator, and $\DD_\qq(t_1,t_2) \equiv -i\,\big\langle T_\CC [\phii_\qq(t_1) \phii_{-\qq}(t_2)]\big\rangle$ is the CTP phonon propagator. The interaction part of the effective action $\GAM_\mathrm{int}[\phii, \DD, \hGG]$ has the following diagrammatic representation:
\begin{align}\label{eq:2PI_diag}
&\GAM_\mathrm{int}[\phii,\hGG,\DD] = \,\,\eqfigscl{0.75}{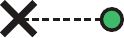} \,\,+\,\, \eqfigscl{0.75}{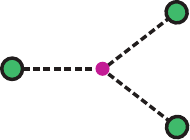} \,\,+\,\, \eqfigscl{0.75}{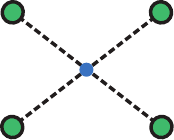}\nonumber\\
&\hspace{20pt}+\,\, \eqfigscl{0.75}{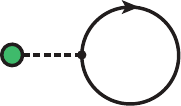} \,\,+\,\, \eqfigscl{0.75}{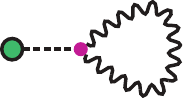} \,\,+\,\, \eqfigscl{0.75}{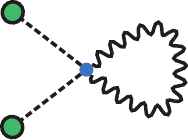} \,\,\nonumber\\
&\hspace{20pt}+\,\, \eqfigscl{0.75}{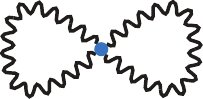} \,\,+\,\, \eqfigscl{0.75}{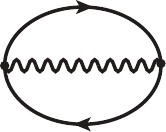}
\end{align}
The solid and wiggly lines denote $\hGG$ and $\DD$, respectively. The green dots denote $\phii$ and the cross denotes the external field. The pink, green, and black vertices denote the cubic nonlinearity, quartic nonlinearity, and the electron-phonon coupling constants, respectively.

The definition and symmetries of CTP propagators as well as the definition of various real-time components are given in Appendix~\ref{sec:props}. The bare electron and phonon propagators are given as:
\begin{align}
\hGG_{0,\kk}^{-1}(t_1,t_2) =&~(i\partial_{t_1}\mathbb{I} - \xi_\kk \hat\sigma_3)\,\delta_\CC(t_1,t_2),\nonumber\\
\DD_{0,\qq}^{-1}(t_1,t_2) =&~-\frac{1}{2\omega_\qq}\,(\partial_{t_1}^2 + \omega_\qq^2)\,\delta_\CC(t_1,t_2),
\end{align}
and the classical action of the coherent lattice displacement is given as:
\begin{align}
\label{eq:Scl}
S_\mathrm{cl}[\phii] =&~N\int \dd t \Big[-\frac{1}{4\omega_0} \, \phii(t)\,(\partial_t^2 + \omega_0^2 + \gamma_0\omega_0\partial_t)\,\phii(t)\nonumber\\
&+ \frac{\Lambda}{2} \, |F(t)|^2 \, \phii(t)+ \frac{\kappa_4}{4!}\,\phii^4(t) + \frac{\kappa_3}{3!}\,\phii^3(t) \Big].
\end{align}
We have introduced a small phenomenological Ohmic friction $\gamma_0\omega_0\partial_t$ for $\phii$ to model the effect of classical dissipation due to coupling to other lattice modes. In principle, the friction must be accompanied by a stochastic Langevin noise for consistency. The presence of a strong driving term, however, dominates over the Langevin noise in practice and allows us to neglect the latter. Finally, the electron-hole bubble $\Pi_\qq$ is defined as:
\begin{align}
\label{eq:Pidef}
\Pi_\qq(t_1,t_2) =&~\frac{1}{N}\sum_\kk |\low{g}_{\kk,\kk+\qq}|^2\nonumber\\
&\times\mathrm{tr}\left[\hGG_{\kk+\qq}(t_1,t_2)\,\hat{\sigma}_3 \, \hGG_{\kk}(t_2,t_1) \, \hat{\sigma}_3\right],
\end{align}
and the local phonon propagator $\DD_\ell$ appearing in the last three terms of $\eqref{eq:2PI}$ is defined as:
\begin{align}
\label{eq:Dlocdef}
\DD_\ell(t_1, t_2) =&~\frac{1}{N}\sum_\qq \DD_\qq(t_1, t_2).
\end{align}
\end{subequations}
We note that the external drive only appears in the classical action $S_\mathrm{cl}[\phii]$ and couples to $\phii_{\qq=0}$. The latter directly couples to finite-$\qq$ phonons via nonlinearities (see the fifth and sixth diagrams in Eq.~\ref{eq:2PI_diag}), and to electrons if $\low{g}_{\kk,\kk} \neq 0$ (SSH-type phonons, see the forth diagram in Eq.~\ref{eq:2PI_diag}).

The equations of motion are found by imposing the stationarity condition on $\GAM$ with respect to $\phii$, $\GG$, and $\DD$. For the classical displacement, we find:
\begin{multline}\label{eq:phiEOM}
\frac{1}{2\omega_0}\left(\partial_t^2 + \omega_0^2 + \gamma_0 \omega_0 \partial_t \right)\phii(t) - \frac{\kappa_4}{6}\,\phii^3(t) - \frac{\kappa_3}{2}\,\phii^2(t)\\
- \frac{\kappa_4}{2} \, \chi(t)\,\phii(t) = \frac{\Lambda}{2}\,|F(t)|^2 + \frac{\kappa_3}{2}\, \chi(t) + \eta(t),
\end{multline}
where:
\begin{subequations}
\begin{align}
\label{eq:chidef}
\chi(t) &\equiv \frac{1}{N}\sum_\qq i\DD_\qq(t,t),\\
\label{eq:etadef}
\eta(t) &\equiv \frac{i}{N}\sum_\kk \low{g}_{\kk,\kk}\,\mathrm{tr}\left[\hGG_\kk(t, t^+)\,\hat{\sigma}_3\right].
\end{align}
\end{subequations}
Here, $\chi(t)$ is the phonon tadpole and $\eta(t)$ is the electron-mediated classical force on the uniform lattice displacement. The evolution of $\DD$ is given by the following set of Kadanoff-Baym (KB) equations:
\begin{subequations}
\begin{align}
\label{eq:KBD0}
&-\frac{1}{2\omega_\qq}\big[\partial_{t_1}^2 + \omega_\qq^2\big]\,\DD_\qq(t_1,t_2) = \delta_\CC(t_1,t_2)\nonumber\\
&\qquad + V(t_1) \, \DD_\qq(t_1, t_2) + \int_\CC \dd \tau \, \Pi_\qq(t_1,\tau)\,\DD_\qq(\tau,t_2),\\
\label{eq:KBD0adj}
&-\frac{1}{2\omega_\qq}\big[\partial_{t_2}^2 + \omega_\qq^2\big]\, \DD_\qq(t_1,t_2) = \delta_\CC(t_1,t_2)\nonumber\\
&\qquad + V(t_2) \, \DD_\qq(t_1, t_2) + \int_\CC \dd \tau \, \DD_\qq(t_1,\tau)\,\Pi_\qq(\tau,t_2),
\end{align}
\end{subequations}
where:
\begin{equation}\label{eq:Vdef}
V(t) \equiv -\frac{\kappa_4}{2} \,\chi(t) - \frac{\kappa_4}{2} \, \phii^2(t) - \kappa_3\,\phii(t)
\end{equation}
is a local time-dependent potential acting on phonons. Aside from $\chi(t)$ which is a phonon self-interaction, the second and third terms are responsible for the parametric drive of phonons in connection to the analysis given in Sec.~\ref{sec:param}.

Finally, the evolution of the CTP Nambu electron propagator $\GG$ is given by the following set of KB equations:
\begin{subequations}
\begin{align}
\label{eq:KBG0}
&\big[i\partial_{t_1}\mathbb{I} - \xi_\kk \hat{\sigma}_3 - \phii(t_1) \, \low{g}_{\kk,\kk} \, \hat{\sigma}_3 \big]\hGG_\kk(t_1, t_2) =\nonumber\\
&\qquad\qquad\qquad\qquad\qquad \delta_\CC(t_1, t_2) + \int_\CC \hat{\Sigma}_\kk(t_1, \tau) \, \hGG_\kk(\tau, t_2),\\
\label{eq:KBG0adj}
&\big[i\partial_{t_2}\mathbb{I} - \xi_\kk \hat{\sigma}_3 - \phii(t_2) \, \low{g}_{\kk,\kk} \, \hat{\sigma}_3 \big]\hGG_\kk(t_1, t_2) =\nonumber\\
&\qquad\qquad\qquad\qquad\qquad \delta_\CC(t_1, t_2) + \int_\CC \hGG_\kk(t_1, \tau) \, \hat{\Sigma}_\kk(\tau, t_2),
\end{align}
\end{subequations}
where the Nambu self-energy is given as:
\begin{equation}\label{eq:ME0}
\hat{\Sigma}_\kk(t_1,t_2) = \frac{1}{N}\sum_{\kk'} |\low{g}_{\kk,\kk'}|^2 \, \hat{\sigma}_3\,\hat{\GG}_{\kk'}(t_1,t_2)\,\hat{\sigma}_3 \, i\DD_{\kk-\kk'}(t_1,t_2),
\end{equation}
which is the usual Migdal-Eliashberg self-energy in real time. Explicit equations for the retarded (R), advanced (A), and Keldysh (K) components of $\hGG$ and $\DD$ can be worked out from Eqs.~\eqref{eq:KBG0}-\eqref{eq:KBG0adj} and Eqs.~\eqref{eq:KBD0}-\eqref{eq:KBD0adj} using Langreth rules~\cite{rammer2007quantum}, respectively. 

In principle, the solution of the coupled integro-differential equations derived in this section, while being a daunting task, yields a complete and unbiased analysis. Given that our goal in the present paper is to give a transparent account of the key mechanisms that play a role in enhancing or suppressing superconductivity, we find it rather beneficial to simplify the model to the greatest possible extent without sacrificing any qualitative physics.

\subsection{The trimmed-down model}\label{sec:minmodel}
In this section, we present and discuss several simplifying approximations which we adopt in the rest of the paper. These assumptions are presented in a single section, rather than incrementally, for better clarity.\\

\noindent{\em Holstein-type electron-phonon coupling---} Depending on the nature of the electron-phonon coupling, $\low{g}_{\kk,\kk'}$ may assume different dependencies on $\kk$ and $\kk'$. For Holstein-type phonons, appropriate for describing longitudinal optical (LO) phonons, $\low{g}_{\kk,\kk'}$ only depends on the net momentum of the electron-hole excitation, i.e. $g_{\kk,\kk'} \sim g_{\kk-\kk'}$. On the other hand, for Su-Shrieffer-Hieger (SSH) type phonons, $\low{g}_{\kk,\kk'}$ will depend on the individual momenta. We restrict our analysis to Holstein-type coupling here.

For a realistic description of an electron-phonon system, some aspects of the Coulomb interaction must be incorporated into the model, in particular, the renormalized electron dispersion $\xi_\kk$ and screening of the electron-phonon coupling $g^\mathrm{(scr)}_{\mathbf{\kk} - \mathbf{\kk'}} \approx \low{g}_{\kk - \kk'} / \epss(\Omega, \kk - \kk')$~\cite{mahan2013many}. Here, $\epss$ is the dielectric function and $\Omega$ is the relevant energy scale of the dynamical screening process which is commonly set to zero. In our problem, the largest relevant frequency for dynamical screening is set by external drive and the optical phonon peak frequency, whichever is the largest. In the regime relevant to the experiments, both are an order of magnitude smaller than the typical plasma frequency $\omega_p$. For example, $\omega_p \approx 0.5 \sim 2~{\rm eV}$ for fulleride superconductors whereas the typical frequency of optical phonons is $\omega_\mathrm{ph} \approx 10 \sim 100~{\rm meV}$~\cite{gunnarsson2004narrow}. Therefore, we may safely use the static Thomas-Fermi (TF) dielectric function:
\begin{equation}\label{eq:g_scr}
g^\mathrm{(scr)}_{\kk - \kk'} \simeq \frac{\low{g}_{\kk - \kk'}}{\epss^\mathrm{TF}(\kk-\kk')} = \frac{\low{g}_{\kk - \kk'}}{1 + q_\mathrm{TF}^2/|\kk-\kk'|^2}.
\end{equation}
The screened coupling in the long wavelength $|\kk - \kk'| \ll q_\mathrm{TF}$ is almost perfectly screened. As a consequence, (1) the direct coupling of the uniform $\qq=0$ lattice displacement to conduction electrons (described by the forth diagram in Eq.~\eqref{eq:2PI_diag} and Eq.~\eqref{eq:etadef}) is vanishingly small and can be neglected; (2) the time-dependent correction $g_{\kk,\kk}\,\phii(t)$ to the electron dispersion in Eqs.~\eqref{eq:KBG0}-\eqref{eq:KBG0adj} vanishes as well.\\


\noindent{\em Non-dispersive (Einstein) optical phonons---} So far, we have assumed a general dispersion $\omega_\qq$ for the optical phonon. We neglect the phonon dispersion hereafter, i.e. $\omega_\qq \rightarrow \omega_0$, which an excellent approximation for a large class of materials, including fulleride superconductors~\cite{gunnarsson1997superconductivity}.\\

\noindent{\em Local approximation for $\Pi_\qq$---} In the absence of $\Pi_\qq$ phonon self-energy correction appearing on the right hand side of Eqs.~\eqref{eq:KBD0}-\eqref{eq:KBD0adj} and using Einstein phonons, the phonon propagator $\DD_\qq$ will have no $\qq$-dependence. Therefore, the $\qq$-dependence of $\DD_\qq$ is entirely induced by $\Pi_\qq$. The $\qq$-dependence of the latter is inherited from $\low{g}_{\kk,\kk'}$ and the electron dispersion $\xi_\kk$ (See Eq.~\ref{eq:Pidef}) and is non-universal. In an attempt to simplify the model, we propose a local approximation for $\Pi_\qq$ in the spirit of dynamical mean field theory (DMFT):
\begin{equation}
\Pi_\qq(t_1, t_2) \rightarrow \Pi_\ell(t_1, t_2) \equiv \frac{1}{N}\sum_\qq \Pi_\qq(t_1, t_2).
\end{equation}
The local approximation for $\Pi_\qq$ has an important practical advantage. It removes the $\qq$-dependence from $\DD_\qq$ and allows us to study a single momentum mode.\\

\noindent{\em Fermi-surface averaged (FSA) electron self-energy---} The Cooper pairing process in the majority of conventional superconductors only involves electrons within in a thin shell about the Fermi surface. This observation is the basis of a widely used approximation where the Migdal-Eliashberg electron self-energy $\hat{\Sigma}_\kk$ is replaced with its Fermi-surface averaged (FSA) approximation:
\begin{equation}
\hat{\Sigma}_\kk(t_1, t_2) \rightarrow \hat{\Sigma}(t_1, t_2) \equiv \llangle \hat{\Sigma}_\kk(t_1,t_2) \rrangle_\mathrm{FS}.
\end{equation}
This approximation, being akin to DMFT-type approximation, fully retains the dynamical structure of the self-energy while simplifying the momentum summations by removing the spatial structure of the self-energy. We remark that the FSA approximation is indeed an excellent approximation while adopting the previous two approximations: for a local Holstein-type electron-phonon coupling and Einstein phonons, $\hat{\Sigma}_\kk$ naturally loses is $\kk$-dependence (see Eq.~\ref{eq:ME0}).\\

\noindent{\em Flat electronic density of states (EDOS)---} We neglect the variations of the EDOS $\nu(\xi)$ and pin it to its value at the Fermi surface $\nu(0)$. This is an excellent approximation in three-dimensional systems. \\

\noindent{\em Ideal Migdal's limit---} We work in the ideal Migdal's limit $\omega_\mathrm{ph}/W_\mathrm{el} \rightarrow 0$.

\subsection{Floquet-Boltzmann quantum kinetic formalism}\label{sec:qfb}
\begin{figure}[ht!]
\includegraphics[width=\linewidth]{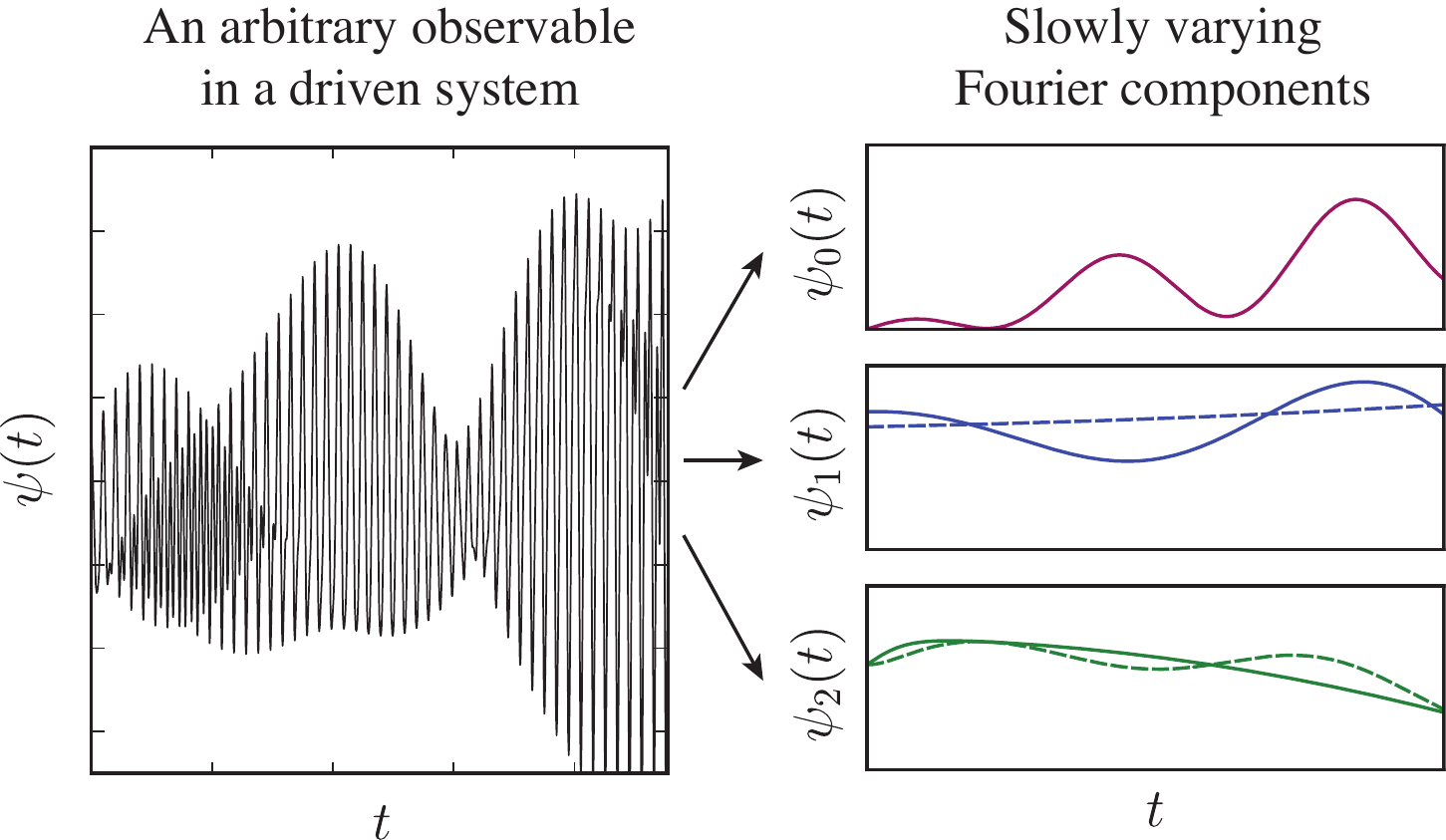}
\caption{{\bf An illustration of Floquet-Boltzmann kinetic formalism.} An arbitrary observable in a driven system is expected to have fast temporal variations on the scale of the driving frequency and a slowly varying envelope. By decomposing the observable into the harmonics of the driving frequency using short-time Fourier transforms.}
\label{fig:stft}
\end{figure}

Despite the simplifications proposed to the model in Sec.~\ref{sec:minmodel}, the solution of the KB equations and obtaining the two-time propagators remains a challenging numerical task. Provided that the external perturbing field varies on time and length scales longer than the intrinsic microscopic time and scale of the many-body system, the KB equations can be controllably reduced to one-time kinetic-type equation using the machinery of Wigner transforms and gradient expansion~\cite{kadanoff1962quantum,rammer2007quantum}. The case of driven systems is subtler though since the external drive $F(t)$ has both {\em fast} and {\em slow} components: even though the temporal variations of the amplitude $F_\mathrm{env}(t)$ occurs on long time scales (compared to the phonon period), the multiplicative oscillatory factor $\cos(\Odrv t)$ varies on the same scale as the phonon frequency in the interesting nearly-resonant regime. In this section, we show that by introducing Floquet bands via short-time Fourier transforms, we can performs gradient expansion on the amplitude of Fourier components and derive Boltzmann-type quantum kinetic equations. This procedure is schematically shown in Fig.~\ref{fig:stft}.\\

\noindent {\em Floquet-Wigner (FW) transform---} Let $\mathcal{A}(t_1, t_2)$ be an arbitrary two-time function, such as the $\hGG(t_1, t_2)$ and $\DD(t_1, t_2)$. The Wigner transform $\mathcal{A}(\omega, t)$ is formally defined as:
\begin{equation}
\mathcal{A}(\omega, t) = \int_{-\infty}^{+\infty} \dd \tau \, e^{i\omega \tau} \, \mathcal{A}(t + \tau/2, t - \tau/2).
\end{equation}
Here, $t = (t_1 + t_2)/2$ is the {\em center-of-mass (COM) time} and $\tau = t_1 - t_2$ is the relative time on which we perform a Fourier transform. The COM-time dependence vanishes at equilibrium. By continuity, we expect the presence of a slowly-varying external field to induce a similarly slowly-varying COM-dependence on $\mathcal{A}(\omega, t)$. This permits a controllable series expansion in successive COM-time derivatives of $\mathcal{A}(\omega, t)$ in the collision integrals~\cite{rammer2007quantum}. In case of periodically driven systems, however, the fast oscillatory component of the external field can induce fast harmonics on the COM-dependence of all two-time quantities. To make a connection with Fig.~\ref{fig:stft}, one must identify $\psi(t)$ as $\mathcal{A}(\omega, t)$ for a fixed value of $\omega$. In principle, we can resolve the $t$-dependence of $\mathcal{A}(\omega,t)$ into Fourier harmonics of the drive using short-time Fourier transforms:
\begin{equation}\label{eq:FW_inverse}
\mathcal{A}(\omega, t) = \sum_n \mathcal{A}_n(\omega; t) \, e^{-in\Omega t},
\end{equation}
where formally $\mathcal{A}_n(\omega; t) = \int_{-\infty}^{+\infty} \dd t' \, W(t' - t) \, e^{i n \Omega t'} \, \mathcal{A}(\omega,t')$. We express the COM-time of the harmonics as a label, i.e. $\mathcal{A}_n(\omega; t)$, to emphasize on the quasi-static nature of the Fourier amplitudes. Here, $\Omega$ is the principal frequency of the driving term. In the present context, $\Omega = 2\Odrv$ since the external drive appears as $|F(t)|^2$ in the model. $W(t)$ is a {\em window function} normalized to unity and concentrated near $t=0$ with support in the time interval $\sim (-\tau_W, +\tau_W)$ where $\Omega^{-1} \ll \tau_W \ll |F_\mathrm{env}(t)|/|F'_\mathrm{env}(t)|$. The shape of the window function is immaterial so long as this condition is satisfied~\cite{mallat1999wavelet}. We refer to the collection of $\{\mathcal{A}_n(\omega,t)\}$ as the {\em Floquet-Wigner transform} of $\mathcal{A}(t_1, t_2)$ and hereafter, we assume that the Floquet components admit a controlled series expansion in COM time derivatives.\\

\noindent {\em Floquet-Groenewold-Moyal (FGM) product formula---} The convolution integrals appearing on the right hand side of KB equations can be formally expressed as a series expansion using Groenewold-Moyal (GM) product formula~\cite{groenewold1946principles,moyal1949quantum}:
\begin{equation}\label{eq:GM}
(\AM \star \BM)(\omega, t) = \exp\left[\frac{i}{2}\left(\partial_t^\mathcal{B}\partial_\omega^\mathcal{A} - \partial_t^\mathcal{A}\partial_\omega^\mathcal{B}\right)\right]\AM(\omega,t) \, \BM(\omega,t).
\end{equation}
The left hand side represent the convolution integral of $\AM$ and $\BM$ followed by a Wigner transform. Expanding the exponentiated differential operator and truncating the series at finite orders yields an approximate expression for the convolution of $\AM$ and $\BM$ in terms of the time and frequency derivatives of their Wigner transforms. In particular, truncating the series at the first order yields the well-known {\em gradient expansion} formula which forms the basis of quantum kinetic equations~\cite{kadanoff1962quantum}. This procedure can be readily generalized to the case of Floquet-Wigner transforms. To this end, we plug the Floquet-Wigner expansion of $\AM$ and $\BM$ (as given in Eq.~\ref{eq:FW_inverse}) in Eq.~\eqref{eq:GM} and take a short-time Fourier transform of both sides. The final result is:
\begin{multline}\label{eq:FGM0}
(\mathcal{A} \star \mathcal{B})_n(\omega;t) = \int_{-\infty}^{+\infty} \dd t' \, W(t'-t) \, e^{in\Omega t'}\\
\times\exp\left[\frac{\Omega}{2}\left(n_\BM \partial_\omega^\AM - n_\AM \partial_\omega^\BM \right) + \frac{i}{2}\left(\partial_t^\mathcal{B}\partial_\omega^\mathcal{A} - \partial_t^\mathcal{A}\partial_\omega^\mathcal{B}\right)\right]\\
\times \sum_{n_\AM, n_\BM} \AM_{n_\AM}(\omega; t) \, \BM_{n_\BM}(\omega; t)\, e^{-i(n_\AM + n_\BM) \Omega t}.
\end{multline}
The COM time derivatives have been resolved into a part acting on the phasor and a part acting on the Fourier amplitudes, i.e. $\partial_{t} \rightarrow -in_{\AM/\BM}\Omega + \partial^{\AM/\BM}_t$. Performing the $t'$-integral is trivial since by construction, we can neglect the COM time variations $\{\AM_n(\omega;t\}$ and $\{\BM_n(\omega;t\}$ within the support of the window function. Using the formal Taylor's expansion formula $e^{\alpha x}f(x) = f(x+\alpha)$, we finally obtain:
\begin{multline}\label{eq:FGM1}
(\mathcal{A} \star \mathcal{B})_n(\omega,t) = \sum_{n_\AM, n_\BM}  \delta(n - n_\AM - n_\BM)\\
\times \exp\bigg[\frac{i}{2}\big(\partial_t^\mathcal{B}\partial_\omega^\mathcal{A}- \partial_t^\mathcal{A}\partial_\omega^\mathcal{B}\big)\bigg] \AM_{n_\AM, -n_\BM}(\omega;t) \, \BM_{n_\BM,n_\AM}(\omega;t),
\end{multline}
where we have defined:
\begin{equation}\label{eq:nm_def}
\AM_{n,m}(\omega;t) \equiv \AM_n(\omega - m\Omega/2; t).
\end{equation}
We refer to labels $n$ and $m$ in $\AM_{n,m}(\omega;t)$ as {\em Floquet band index} and {\em Floquet quasi-momentum}, respectively, in analogy to the Bloch band theory. Eq.~\eqref{eq:FGM1} will be referred to as {\em Floquet-Groenewold-Moyal (FGM) product formula}. Expanding the exponentiated differential operator to linear order, we obtain the {\em FGM gradient expansion formula}.\\

\subsection{Floquet-Boltzmann quantum kinetic equations for phonons}\label{sec:QFB_lattice}
The formalism outlined in the previous section can be used to obtain quantum kinetic-like (``one-time'') equations for the nonequilibrium evolution of the lattice displacement and phonon propagators in the presence of a periodic drive with a slowly varying envelope.

Starting with Eq.~\eqref{eq:phiEOM}, taking a short-time Fourier transform of the sides and neglecting the second-order derivatives, we find:
\begin{subequations}
\begin{align}\label{eq:EOM_phi}
&\left[(\gamma_0\omega_0-2in\Omega)\, \partial_t + \omega_0^2 - n^2 \Omega^2 \right]\phii_n\nonumber\\
&\qquad- \frac{\omega_0 \kappa}{3} \sum_{n_1, n_2} \phii_{n_1} \, \phii_{n_2} \, \phii_{n - n_1 - n_2}\nonumber\\
&\qquad- \omega_0 \kappa_3 \sum_{n_1} \phii_{n_1} \, \phii_{n - n_1} - \omega_0 \kappa_4 \sum_{n_1} \phii_{n_1} \, \chi_{n-n_1}\nonumber\\
&\qquad- \omega_0 \kappa_3 \, \chi_n = \omega_0^2 \,  \Phi_n + \mathcal{O}(\partial_t^2).
\end{align}
We have dropped the $t$-arguments for brevity, $\mathcal{O}(\partial_t^2)$ is a mnemonic for neglecting the second-order time derivative, and $\Phi_n$ is the Fourier amplitude for the external drive. For the drive given in Eq.~\eqref{eq:F_def}, we have:
\begin{equation}
\Phi_n(t) = \frac{\Lambda}{2\omega_0} \, F^2_\mathrm{env}(t)\left[\delta_{n,0} + \frac{1}{2}\,\delta_{n,-1} + \frac{1}{2}\,\delta_{n,1}\right],
\end{equation}
and $\Omega \equiv 2\,\Odrv$. $\chi_n$ is found using Eq.~\eqref{eq:chidef}:
\begin{equation}
\label{eq:chin}
\chi_n(t) = \frac{1}{2}\int_{-\infty}^{+\infty} \frac{\dd \omega}{2\pi} \, i\DD^K_n(\omega;t).
\end{equation}
\end{subequations}
Next, we consider the retarded component of the KB equation for $\DD$. Taking a FW transform of the sides of Eqs.~\eqref{eq:KBD0}-\eqref{eq:KBD0adj} and using the FGM gradient expansion formula in the collision convolution integrals, we find:
\begin{widetext}
\begin{subequations}
\begin{align}
\label{eq:FBDR}
\left[\DDD^{-1}(\omega + n\Omega/2) + \frac{i}{2}\,\partial_\omega \DDD^{-1}(\omega + n\Omega/2)\,\partial_t \right]\DD^R_n(\omega; t) &= \delta_{n,0} + \sum_{n_1} \left[1 - \frac{i}{2}\, \partial_\omega^\DD \, \partial_t^V \right] V_{n_1}(t)\,\DD^R_{n-n_1, n_1}(\omega; t)\nonumber\\
&\hspace{-70pt}+ \sum_{n_1} \left[1 + \frac{i}{2} \partial_\omega^\Pi \, \partial_t^\DD - \frac{i}{2} \partial_\omega^\DD \, \partial_t^\Pi\right] \Pi^{R}_{\ell; n_1, n_1 - n}(\omega; t)\, \DD^R_{n - n_1, n_1}(\omega; t) + \mathcal{O}(\partial_t^2),\\
\label{eq:FBDRadj}
\left[\DDD^{-1}(\omega - n\Omega/2) - \frac{i}{2}\,\partial_\omega \DDD^{-1}(\omega - n\Omega/2)\,\partial_t \right]\DD^R_n(\omega; t) &= \delta_{n,0} + \sum_{n_1} \left[1 + \frac{i}{2}\, \partial_\omega^\DD \, \partial_t^V \right] \DD^R_{n-n_1, -n_1}(\omega; t)\,V_{n_1}(t)\nonumber\\
&\hspace{-70pt}+ \sum_{n_1} \left[1 + \frac{i}{2} \partial_\omega^\DD \, \partial_t^\Pi - \frac{i}{2} \partial_\omega^\Pi \, \partial_t^\DD\right] \DD^R_{n_1, n_1 - n}(\omega; t)\, \Pi^{R}_{\ell; n - n_1, n_1}(\omega; t) + \mathcal{O}(\partial_t^2),
\end{align}
\end{subequations}
\end{widetext}
where $\DDD^{-1}(\omega) \equiv (\omega^2 - \omega_0^2)/(2\omega_0)$ is the bare inverse phonon propagator in the frequency time. $V(t)$ was defined earlier in Eq.~\eqref{eq:Vdef} and is the self-consistently determined potential that parametrically drives the phonons. The Fourier amplitudes of $V(t)$ are trivially found as $V_n(t) = -(\kappa_4/2)\sum_{n_1}\phii_{n_1}(t)\,\phii_{n-n_1}(t) - (\kappa_4/2)\,\chi_n(t) - \kappa_3 \, \phii_n(t)$. We have employed local approximation for $\Pi$. The phonon propagator thus can be thought of either as that of a single momentum-mode or as the momentum-summed (local) one. We do not need a separate evolution equation for the advanced propagator since it can be determined from the retarded propagator via the identity $\DD^A_n(\omega;t) = [\DD^R_{-n}(\omega;t)]^*$.

The kinetic equations for the Keldysh phonon propagator is found in a similar fashion. First, we use the Langreth rules to find an explicit KB equation for $\DD^K$, followed by a FW transform and FGM gradient expansion of the collision integral convolutions. The final result is:
\begin{widetext}
\begin{subequations}
\begin{align}
\label{eq:FBDK}
&\left[\DDD^{-1}(\omega + n\Omega/2) + \frac{i}{2}\,\partial_\omega \DDD^{-1}(\omega + n\Omega/2)\,\partial_t \right] i\DD^K_n(\omega; t) = \sum_{n_1} \left[1 - \frac{i}{2}\, \partial_\omega^\DD \, \partial_t^V \right] V_{n_1}(t)\,i\DD^K_{n-n_1, n_1}(\omega; t)\nonumber\\
&\,\,\,\,+ \sum_{n_1} \left[1 + \frac{i}{2} \partial_\omega^\Pi \, \partial_t^\DD - \frac{i}{2} \partial_\omega^\DD \, \partial_t^\Pi\right] \left[\Pi^{(\ell), R}_{n_1, n_1 - n}(\omega; t)\, i\DD^K_{n - n_1, n_1}(\omega; t) + i\Pi^{K}_{\ell; n_1, n_1 - n}(\omega; t)\, \DD^A_{n - n_1, n_1}(\omega; t)\right] + \mathcal{O}(\partial_t^2),\\
\label{eq:FBDKadj}
&\left[\DDD^{-1}(\omega - n\Omega/2) - \frac{i}{2}\,\partial_\omega \DDD^{-1}(\omega - n\Omega/2)\,\partial_t \right] i\DD^K_n(\omega; t) = \sum_{n_1} \left[1 + \frac{i}{2}\, \partial_\omega^\DD \, \partial_t^V \right] i\DD^K_{n-n_1, -n_1}(\omega; t)\,V_{n_1}(t)\nonumber\\
&\,\,\,\,+ \sum_{n_1} \left[1 + \frac{i}{2} \partial_\omega^\DD \, \partial_t^\Pi - \frac{i}{2} \partial_\omega^\Pi \, \partial_t^\DD\right] \left[\DD^R_{n_1, n_1 - n}(\omega; t)\, i\Pi^{K}_{\ell; n - n_1, n_1}(\omega; t) + i\DD^K_{n_1, n_1 - n}(\omega, t)\, \Pi^{A}_{\ell; n - n_1, n_1}(\omega; t)\right]+ \mathcal{O}(\partial_t^2).
\end{align}
\end{subequations}
\end{widetext}
Even though Eqs.~\eqref{eq:FBDR}-\eqref{eq:FBDKadj} have a more complex presentation compared to the original KB equations, they are significantly simpler to work with in practice: convolution integrals have been reduced to discrete Floquet index summations, and the two-time structure has been reduced to the COM time and the relative frequency $\omega$ which does not appear in a convolution. Finally, we remark that the electron-hole bubble $\Pi_\ell$ acts as a dissipation source (``bath'') for phonons, and must be determined by solving the evolution equations for $\hGG$ in a fully self-consistent treatment. We will argue later that it can be approximately calculated using bare electron propagators at the initial temperature as long as heating does not bring up the energy density of electrons to phonon energy scales. Approximate expressions for $\Pi_\ell$ have been provided in Sec.~\ref{sec:landau_damping} (see Eq.~\ref{eq:equilibrium_bath}). We show $\Pi_\ell$ acts as an quantum Ohmic bath for phonons, and gives rise to a Lamb shift of the Einstein oscillator. 

\subsection{Floquet-Boltzmann quantum kinetic equations for electrons \label{sec:QFB_electron}}
We can obtain quantum kinetic equations for the nonequilibrium evolution of electrons in a similar to phonons. Before embarking on deriving these equations, we take a short detour to derive explicit expressions for the Migdal-Eliashberg and employ the approximations discussed in Sec.~\ref{sec:minmodel}.

\subsubsection{Migdal-Eliashberg self-energy: the general case}
Our goal in this section is derive explicit expressions for various real-time components of the Migdal-Eliashberg self-energy, starting from Eq.~\eqref{eq:ME0}. To keep the notation uncluttered, we will work out the results for a general Wigner-transformed propagators as a first step. We employ the approximations discussed in Sec.~\ref{sec:minmodel} step by step. The simplified self-energy expressions for Floquet-Wigner-transformed propagators are readily found in the end as an special case.\\

As a starting point, we transform Eq.~\eqref{eq:ME0} to the Wigner representation and employ the decomposition of propagators defined in Eqs.~\eqref{eq:G_decomp}-\eqref{eq:D_decomp} in terms of their Keldysh and spectral components. The calculation is elementary and the final result is:
\begin{widetext}
\begin{align}
\hat{\Sigma}^R_\kk(\omega,t) &= \frac{1}{2N} \sum_{\kk'} |\low{g}_{\kk,\kk'}|^2 \int \frac{\dd\omega'}{2\pi} \, \frac{\dd \nu}{2\pi} \, \frac{1}{\omega - \omega' - \nu + i0^+} \left\{i\DD^K_{\kk-\kk'}(\nu,t)\,\check{\AAA}_{\kk'}(\omega', t) + \rho_{\kk-\kk'}(\nu,t)\,i\check{\GG}^K_{\kk'}(\omega', t)\right\},\nonumber\\
i\hat{\Sigma}^K_\kk(\omega,t) &= \frac{1}{2N} \sum_{\kk'} |\low{g}_{\kk,\kk'}|^2 \int \frac{\dd\omega'}{2\pi} \, \frac{\dd \nu}{2\pi} \, (2\pi)\delta(\omega - \omega' - \nu) \left\{i\DD^K_{\kk-\kk'}(\nu,t)\,i\check{\GG}^K_{\kk'}(\omega', t) + \rho_{\kk-\kk'}(\nu,t)\,\check{\AAA}_{\kk'}(\omega', t)\right\}.
\end{align}
\end{widetext}
where the capped Nambu propagators are defined according to Eq.~\eqref{eq:cap}.  We define the Eliashberg function~\footnote{In the Migdal-Eliashberg theory literature, it is customary to refer to az $F^\rho$ as $\alpha^2 F$. We do not employ this cluttered notation in our treatment.} $F^\rho_{\xi,\xi'}(\nu)$, as well as a 
Keldysh Eliashberg function $iF^K_{\xi,\xi'}(\nu)$ as:
\begin{subequations}
\begin{align}
\label{eq:Frho}
F^\rho_{\xi,\xi'}(\nu,t) \equiv&\, \frac{\nu(0)}{\nu(\xi)\,\nu(\xi')}\frac{1}{N^2}\sum_{\kk,\kk'} \frac{|\low{g}_{\kk,\kk'}|^2}{2\pi} \,\rho_{\kk-\kk'}(\nu,t)\nonumber\\
&\times \delta(\xi_\kk - \xi) \, \delta(\xi_{\kk'} - \xi'),\\
\label{eq:FK}
iF^K_{\xi,\xi'}(\nu,t) \equiv&\, \frac{\nu(0)}{\nu(\xi)\,\nu(\xi')}\frac{1}{N^2}\sum_{\kk,\kk'} \frac{|\low{g}_{\kk,\kk'}|^2}{2\pi} \,i\DD^K_{\kk-\kk'}(\nu,t)\nonumber\\
&\times \delta(\xi_\kk - \xi) \, \delta(\xi_{\kk'} - \xi').
\end{align}
\end{subequations}
In the special case of Einstein phonons where $i\DD^K_\qq$ and $\rho_\qq$ have no $\qq$-dependence, we find:
\begin{subequations}
\begin{align}
\label{eq:Frho1}
F^\rho_{\xi,\xi'}(\nu,t) &\equiv \frac{\alpha^2(\xi,\xi')}{2\pi} \,\rho(\nu,t),\\
\label{eq:FK1}
iF^K_{\xi,\xi'}(\nu,t) &\equiv \frac{\alpha^2(\xi,\xi')}{2\pi} \,iD^K(\nu,t),
\end{align}
\end{subequations}
where the energy-resolved dimensionless coupling constant $\alpha^2(\xi,\xi')$ is defined as:
\begin{equation}
\alpha^2(\xi,\xi') \equiv \frac{\nu(0)}{\nu(\xi)\,\nu(\xi')}\frac{1}{N^2}\sum_{\kk,\kk'} |\low{g}_{\kk,\kk'}|^2 \, \delta(\xi_\kk - \xi) \, \delta(\xi_{\kk'} - \xi'),
\end{equation}
which in turn in the limit of flat EDOS and constant $\low{g}_{\kk,\kk'}$ evaluates to $\alpha^2(\xi,\xi') \rightarrow \nu(0)\, g^2 \sim \mathrm{const}$. 
The Fermi-surface averaged (FSA) self-energy is readily found as:
\begin{subequations}
\begin{widetext}
\begin{align}
\label{eq:sig_R_FSA}
\hat{\Sigma}^R(\omega,t) &\equiv \llangle \hat{\Sigma}^R_\kk(\omega,t) \rrangle_\mathrm{FS} = \int_{-W/2}^{+W/2}\dd \xi' \, \frac{\nu(\xi')}{\nu(0)} \, \int_{-\infty}^{+\infty}\frac{\dd\omega'}{2\pi}\int_{-\infty}^{+\infty} \frac{\dd\nu}{\omega - \omega' - \nu + i0^+}\nonumber\\
&\hspace{200pt}\times \frac{1}{2}\left\{iF^K_{0,\xi'}(\nu,t) \, \check{\AAA}_{\xi'}(\omega',t)+ F^\rho_{0,\xi'}(\nu,t) \, i\check{\GG}^K_{\xi'}(\omega',t)\right\},\\
\label{eq:sig_K_FSA}
i\hat{\Sigma}^K(\omega,t) &\equiv \llangle i\hat{\Sigma}^K_\kk(\omega,t) \rrangle_\mathrm{FS} = \int_{-W/2}^{+W/2}\dd \xi' \, \frac{\nu(\xi')}{\nu(0)} \, \int_{-\infty}^{+\infty}\frac{\dd\omega'}{2\pi}\int_{-\infty}^{+\infty} \dd\nu \, (2\pi)\delta(\omega - \omega' - \nu)\nonumber\\
&\hspace{200pt}\times \frac{1}{2}\left\{iF^K_{0,\xi'}(\nu,t) \, i\check{\GG}^K_{\xi'}(\omega',t) + F^\rho_{0,\xi'}(\nu,t) \, \check{\AAA}_{\xi'}(\omega',t)\right\}.
\end{align}
\end{widetext}
\end{subequations}
Since the FSA self-energy has no momentum dependence, the $\kk$-dependence of the resulting electron propagators is induced from the bare electron dispersion $\xi_\kk$. Hence, we have legitimately replaced the $\kk'$ momentum labels with $\xi'$, and $\kk'$ momentum summations with EDOS-weighted $\xi'$ integrals over the bandwidth $[-W_\mathrm{el}/2, +W_\mathrm{el}/2]$. At this point, we employ the remaining approximations discussed in Sec.~\ref{sec:minmodel}, i.e. flat EDOS, infinitely large electronic bandwidth $W_\mathrm{el}$ compared to the phonon scale, constant $\low{g}_{\kk,\kk'}$, and dispersionless phonons. In particular, the last two imply that the Eliashberg functions do not depend on $\xi'$. The final result reads:
\begin{subequations}
\begin{align}
\label{eq:senRfin}
\hat{\Sigma}^R(\omega,t) &= \frac{1}{2}\int_{-\infty}^{+\infty}\frac{\dd\omega'}{2\pi}\int_{-\infty}^{+\infty} \frac{\dd\nu}{\omega - \omega' - \nu + i0^+}\nonumber\\
&\hspace{-20pt}\times \Big[iF^K(\nu,t)\,\check{\AAA}(\omega',t) + F^\rho(\nu,t)\,i\check{\GG}^K(\omega',t)\Big],\\
\label{eq:senKfin}
i\hat{\Sigma}^K(\omega,t) &= \frac{1}{2}\int_{-\infty}^{+\infty}\frac{\dd\omega'}{2\pi}\int_{-\infty}^{+\infty} \dd\nu \, (2\pi)\delta(\omega - \omega' - \nu)\nonumber\\
&\hspace{-20pt}\times \Big[iF^K(\nu,t)\,i\check{\GG}^K(\omega',t) + F^\rho(\nu,t)\,\check{\AAA}(\omega',t)\Big].
\end{align}
\end{subequations}
where:
\begin{subequations}
\begin{align}
\label{eq:A_xi_integ}
\check{\AAA}(\omega,t) &\equiv \int_{-\infty}^{+\infty} \dd\xi \, \check{\AAA}_{\xi}(\omega,t),\\
\label{eq:GK_xi_integ}
i\check{\GG}^K(\omega,t) &\equiv \int_{-\infty}^{+\infty} \dd\xi \, i\check{\GG}^K_\xi(\omega,t),
\end{align}
\end{subequations}
are {\em local} electronic spectral and Keldysh functions as obtained by summing over all momentum states:
\begin{subequations}
\begin{align}
\label{eq:Frho_simplified}
F^{\rho}(\nu,t) &\equiv F^{\rho}_{0,0}(\nu,t) = \nu(0) \, g^2 \, \rho(\nu,t),\\
\label{eq:FK_simplified}
F^{K}(\nu,t) &\equiv F^{K}_{0,0}(\nu,t) = \nu(0) \, g^2 \, \DD^K(\nu,t).
\end{align}
\end{subequations}
Finally, we find it useful to parametrize the strength of the electron-phonon coupling in terms of the dimensionless {\em mass enhancement factor} of an ideal Einstein oscillator at equilibrium~\cite{bennemann2008superconductivity}:
\begin{equation}\label{eq:lambda_def}
\lambda \equiv 2\int_0^\infty \dd \nu \, \frac{F^\rho_\mathrm{ideal}(\nu)}{\nu} = \frac{2 g^2 \nu(0)}{\omega_0}.
\end{equation}

\subsubsection{Migdal-Eliashberg self-energy: the normal state}
The results of the previous section were worked out for a general Nambu electron propagator. We specialize the result to the normal non-paired state in this section. A Nambu functions are diagonal in the normal state and the Nambu structure of the self-energy and the ensuing KB equations can be simplified. Starting with the general ansatz,
\begin{equation}
\hGG^{R}_\xi = \left(
  \begin{array}{cc}
    \GG^R_\xi & 0\\
    0 & \bar{\GG}^R_\xi
  \end{array}
\right), \qquad
\hSIG^{R/K} = \left(
  \begin{array}{cc}
    \Sigma^{R/K} & 0\\
    0 & \bar{\Sigma}^{R/K}
  \end{array}
\right),
\end{equation}
and using Lemma 1(b) from Appendix~\ref{sec:props}, we find:
\begin{equation}
\check{\AAA}_\xi(\omega,t) = \hat{\AAA}_\xi(\omega,t) = \left(
  \begin{array}{cc}
    \AAA_\xi(\omega,t) & 0\\
    0 & \AAA_\xi(-\omega,t)
  \end{array}
\right),
\end{equation}
where we have defined $\AAA_\xi(\omega,t) = -2\,\mathrm{Im} \, \GG_\xi^R(\omega,t)$ as the normal spectral function. The kinetic energy variable $\xi$ appears as a convenient scalar surrogate for $\kk$ after using FSA self-energies (see the discussion after Eq.~\ref{eq:sig_K_FSA}). We have also used Lemma 1(d) to relate the time-reversed spectral function $\bar{\AAA}_\xi(\omega,t) \equiv \mathrm{Im} \, \bar{\GG}_\xi^{R}(\omega,t)$ to $\AAA_\xi(\omega,t)$, i.e. $\bar{\AAA}_\xi(\omega,t) = \AAA_\xi(-\omega,t)$. We further define an unrestricted ansatz for $i\GG_\xi^K$ in compliance with Lemma 1(c) and 1(e):
\begin{equation}\label{eq:psidef}
i\hGG^{K}_\xi(\omega,t) = \left(
  \begin{array}{cc}
    \psi_\xi(\omega,t)\,\AAA_\xi(\omega,t) & 0\\
    0 & -\psi_\xi(\omega,t)\, \AAA_\xi(-\omega,t)
  \end{array}
\right),
\end{equation}
where $\psi(\omega,t)$ is an {\em odd real function of $\omega$} that encodes the statistics of electrons in the normal state. For example, in thermal equilibrium, the Kubo-Martin-Schwinger (KMS) boundary condition implies:
\begin{equation}
\psi_\xi^\mathrm{eq}(\omega,t) \rightarrow 1 - 2 n_\mathrm{FD}(\omega) = \tanh(\beta\omega/2).
\end{equation}
We further define the {\em local} electron statistics as:
\begin{equation}\label{eq:psi_loc_def}
\psi(\omega, t) = \frac{1}{2\pi}\int_{-\infty}^{+\infty} \dd\xi \, i\GG^K_\xi(\omega, t).
\end{equation}
In theory, the $\Sigma^{R/K}$ can be expressed as functionals of $\psi$, $i\DD^K$, and $\rho$. An explicit formula for $\Sigma^{R/K}[\psi, i\DD^k, \rho]$ can be found using Eqs.~\eqref{eq:senRfin}-\eqref{eq:senKfin}, Eqs.~\eqref{eq:Frho_simplified}-\eqref{eq:FK_simplified}, Eq.~\eqref{eq:psidef}, and the following crucial lemma:\\

\remark{Lemma ($\GG^R$ momentum summation formula)} {\em Assuming (1) infinite electronic bandwidth limit, and (2) a momentum-independent retarded self-energy as in Fermi surface averaging approximation and local approximation (DMFT), the following identity holds:}
\begin{equation}
\label{eq:GR_sum_formula}
\int_{-\infty}^{+\infty} \dd \xi \, \GG^R_\xi(\omega,t) = -i\pi.
\end{equation}
The proof is given in Appendix~\ref{sec:GR_sum_lemma_proof}. An immediate corollary is:
\begin{equation}
\int_{-\infty}^{+\infty} \dd \xi \, \AAA_\xi(\omega,t) = 2\pi.
\end{equation}

Combining Eqs.~\eqref{eq:senRfin}-\eqref{eq:senKfin}, Eqs.~\eqref{eq:Frho_simplified}-\eqref{eq:FK_simplified}, Eq.~\eqref{eq:psidef}, and Eq.~\eqref{eq:GR_sum_formula}, we find the sought after explicit self-energy functionals:
\begin{subequations}
\begin{align}
\label{eq:senRN}
\Sigma^R(\omega,t) =&\, \int_{0}^{+\infty} \dd\omega'\int_{-\infty}^{+\infty} \dd\nu \bigg[\frac{N^+(\omega',\nu;t)}{\omega - \omega' - \nu + i0^+}\nonumber\\
&+ \frac{N^-(\omega',\nu;t)}{\omega + \omega' - \nu + i0^+}\bigg],\\
\label{eq:senKN}
i\Sigma^K(\omega,t) &= \pi \int_{-\infty}^{+\infty} \dd\nu \,\psi(\omega-\nu,t) \, iF^K(\nu,t),
\end{align}
\end{subequations}
where:
\begin{equation}
N^\pm(\omega,\nu;t) \equiv \frac{1}{2}\left[iF^K(\nu,t) \pm \psi(\omega,t)\,F^\rho(\nu)\right].
\end{equation}
It is noteworthy that $\Sigma^{R/K}$ only depends on the {\em local} electron statistics $\psi(\omega,t)$ and not the $\xi$-response $\psi_\xi(\omega, t)$. Finally, Lemma 3(a) and 3(b) from Appendix~\ref{sec:props} and the above result imply $\bar{\Sigma}^{R/K}(\omega,t) = \Sigma^{R/K}(\omega,t)$, i.e. $\hat{\Sigma}$ is proportional to the identity matrix in the normal state. This result is strictly a consequence of the ideal Migdal limit. One can show $\Sigma^R - \bar{\Sigma}^R \propto 1/W_\mathrm{el}$ in a finite bandwidth model. This completes our discussion of the Migdal-Eliashberg self-energy in the normal state.

\subsubsection{Floquet-Boltzmann kinetic equation for electrons}\label{sec:QFB_el}
An explicit Floquet-Boltzmann kinetic equation can be derived for electrons in the normal state using the result of the previous section and the KB equations. To this end, we write the Keldysh component of Eqs.~\eqref{eq:KBG0}-\eqref{eq:KBG0adj} using FSA self-energy:
\begin{align}
\label{eq:KBGK}
\left[+i\partial_{t_1} - \xi\right] \GG^K_\xi &= \Sigma^R \star \GG^K_\xi + \Sigma^K \star \GG^A_\xi,\\
\label{eq:KBGKadj}
\left[-i\partial_{t_2} - \xi\right] \GG^K_\xi &= \GG_\xi^R \star \Sigma^K + \GG_\xi^K \star \Sigma^A,
\end{align}
Subtracting the sides of these equations from one another and performing a Wigner transformation, we find:
\begin{equation}
\partial_t \, i\GG^K_\xi = \Sigma^R \star \GG^K_\xi + \Sigma^K \star \GG^A_\xi - \GG^R_\xi \star \Sigma^K - \GG^K_\xi \star \Sigma^A.
\end{equation}
Integrating both sides over $\xi$, using Eq.~\eqref{eq:GR_sum_formula}, and the definition of the local electron statistics (see Eq.~\ref{eq:psi_loc_def}), we find a simple evolution equation for $\psi(\omega,t)$:
\begin{equation}
\partial_t \psi = i\Sigma^K - i\left(\Sigma^R \star \psi - \psi \star \Sigma^A\right).
\end{equation}
As an intermediate consistency check, at equilibrium where $\psi=\tanh(\beta\omega/2)$ and the GM product reduces to an algebraic product, the right hand side evaluates to $i\Sigma^K(\omega) + 2i\mathrm{Im}[\Sigma^R]\tanh(\beta\omega/2)$ which vanishes in light of the KMS boundary condition. Thus, the thermal state remains stationary as expected.

For a periodically driven system with a slowly varying drive envelope, expanding the convolution integrals using first-order FGM product formula yields:
\begin{widetext}
\begin{align}\label{eq:FBE}
\partial_t \psi_{n,m} =&\,\, in\Omega \, \psi_{n,m} + i\Sigma^K_{n,m}[\psi] - i\sum_{n'}\Big( \Sigma^R_{n',-n+n'+m}[\psi]\,\psi_{n-n',n'+m} - \psi_{n',-n+n'+m}\,\Sigma^A_{n-n',n'+m}[\psi]\nonumber\\
&+ \frac{i}{2}\,\partial_\omega\Sigma^R_{n',-n+n'+m}[\psi]\,\partial_t \psi_{n-n',n'+m} - \frac{i}{2}\,\partial_t\Sigma^R_{n',-n+n'+m}[\psi]\,\partial_\omega \psi_{n-n',n'+m}\nonumber\\
& - \frac{i}{2}\,\partial_\omega\psi_{n',-n+n'+m}\,\partial_t\Sigma^A_{n-n',n'+m}[\psi] + \frac{i}{2}\,\partial_t\psi_{n',-n+n'+m}\,\partial_\omega\Sigma^A_{n-n',n'+m}[\psi]\Big) + \mathcal{O}(\partial_t^2).
\end{align}
\end{widetext}
The Floquet components of self-energy functionals are worked out easily from Eqs.~\eqref{eq:senRN}-\eqref{eq:senKN}:
\begin{subequations}
\begin{align}
\label{eq:senRN_floquet}
\Sigma^R_n(\omega;t) =&\, \int_{0}^{+\infty} \dd\omega'\int_{-\infty}^{+\infty} \dd\nu \bigg[\frac{N^+_n(\omega',\nu; t)}{\omega - \omega' - \nu + i0^+}\nonumber\\
&+ \frac{N^-_n(\omega',\nu; t)}{\omega + \omega' - \nu + i0^+}\bigg],\\
\label{eq:senKN_floquet}
i\Sigma^K_n(\omega; t) &= \pi \sum_{n'}\int_{-\infty}^{+\infty} \dd\nu \,\psi_{n-n'}(\omega-\nu; t) \, iF^K_{n'}(\nu; t),
\end{align}
\end{subequations}
where:
\begin{equation}\label{eq:N_def}
N^\pm_n(\omega',\nu; t) = \frac{1}{2}\left[iF^K_n(\nu; t) \pm \sum_{n'}\psi_{n-n'}(\omega; t)\,F^\rho_{n'}(\nu; t)\right].
\end{equation}
The advanced self-energy is readily obtained using the symmetry relation $\Sigma^A_{n,m}(\omega; t) = \Sigma^A_{n}(\omega - m\Omega/2; t) = [\Sigma^R_{-n}(\omega - m\Omega/2; t)]^* = [\Sigma^R_{-n,m}(\omega; t)]^*$. Self-energies with finite Floquet quasi-momentum, e.g. $\Sigma^R_{n,m}(\omega;t)$ are found by shifting $\Sigma^R_{n,m}(\omega;t) \equiv \Sigma^R_{n}(\omega - m\Omega/2;t)$ (see Eq.~\ref{eq:nm_def}).

In the fully self-consistent scheme, one must integrate Eq.~\eqref{eq:FBE} together with the previously derived kinetic equations for the lattice displacement and phonon propagator self-consistently. We note that Eq.~\eqref{eq:FBE} is an {\em implicit} integro-differential equation for $\partial_t \psi_{n,m}$ in disguise due to the presence of derivative terms $\partial_t \Sigma^{R/A}[\psi]$. We will discuss the numerical approach for solving this equation in Sec.~\ref{app:numerics}.

\section{Migdal-Eliashberg theory of Floquet superconducting instability}\label{sec:FME_instab}
We derived a set of tractable evolution equations for the driven system in the normal state. In this section, we derive a criterion to identify the instability of the normal state toward forming a Floquet superconducting state. This criterion follows from a careful linear response analysis as follows: we introduce a small fictitious {\em pairing potential} (i.e. an off-diagonal self-energy term) to the time-dependent self-energy obtained in the normal state: $\hat{\Sigma}^R(\omega,t) \rightarrow \Sigma^R(\omega,t) \mathbb{I} + \hat{\phi}(\omega,t)$. Here, $\hat{\phi}(\omega,t)$ is off-diagonal in the Nambu space. The off-diagonal self-energy, in turn, induces an anomalous (off-diagonal) propagator $\delta \hat{\FF}[\hat{\phi}]$ which in turn generates the pairing potential. The introduced pair potential may only persist if and only if $\hat{\Sigma}^R[\hat{\GG} + \delta \hat{\FF}[\hat{\phi}]]  - \hat{\Sigma}^R[\hat{\GG}] \equiv \hat{\phi}$. This procedure is shown diagrammatically as:
\begin{equation}\label{eq:diag_pairing}
\eqfigscl{0.8}{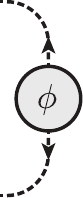} \,\, = \,\, \eqfigscl{0.8}{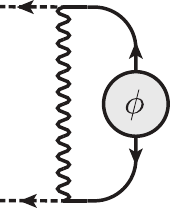}  
\end{equation}
Since the self-energy is a linear functional of $\hat{\phii}$, satisfiability of the above equation for a non-trivial $\hat{\phi}$ requires the linear operator $\mathbb{I} - \DD \GG \GG$ to have a non-trivial null space. This operator is precisely the inverse vertex operator that appears in the two-particle propagator $\sim \GG\GG(\mathbb{I} - \DD \GG \GG)^{-1}$. This, the pairing condition is formally equivalent to requiring a zero-energy pole in the two-particle propagator, the well-known Thouless criterion for spontaneous symmetry breaking~\cite{thouless1960perturbation}. The precise condition for a driven nonequilibrium system is complicated due to nonequilibrium propagators and Floquet bands, and requires a careful implementation of the outlined steps, which is the goal of the next sections. As a first step, we will work out the pairing instability criterion for an arbitrary normal state. The results will be used to find the pairing condition for quasi-steady Floquet states.

\subsection{Pairing instability criterion for arbitrary nonequilibrium states}
We start the analysis by revisiting the KB equations for the retarded Nambu propagator using the FSA self-energy and in the Wigner representation:
\begin{subequations}
\begin{multline}
\label{eq:dyswigNambu}
\big[+(i/2)\,\partial_t + \omega]\,\hGG^R_\xi(\omega,t) - \xi\,\hsz\,\hGG^R_\xi(\omega,t) = \mathbb{I} \\+ \left[\hat{\Sigma}^R \star \hat{\GG}^R_\xi\right](\omega,t),
\end{multline}
\vspace{-15pt}
\begin{multline}
\label{eq:dyswigadjNambu}
\big[-(i/2)\,\partial_t + \omega]\,\hGG^R_\xi(\omega,t) - \xi \,\hGG^R_\xi(\omega,t)\,\hsz = \mathbb{I} \\+ \left[\hat{\GG}^R_\xi \star \hat{\Sigma}^R\right](\omega,t).
\end{multline}
\end{subequations}
As discussed in the earlier remarks, we assume the following ansatz at the onset of pairing:
\begin{align} 
\hat{\Sigma}^R[\phi](\omega,t) &= \Sigma^R(\omega,t) \, \mathbb{I} + \hat{\phi}(\omega,t),\nonumber\\
\hGG_\xi^R[\phi](\omega,t) &= \hGG_{\xi}^R(\omega,t) + \delta \hat{\FF}^R_{\xi}[\phi](\omega,t) + \mathcal{O}(\phi^2),\nonumber\\
\hGG_\xi^K[\phi](\omega,t) &= \hGG_{\xi}^K(\omega,t) + \psi_\xi(\omega,t)\, \delta \hat{\AAA}_\xi[\phi](\omega,t) + \mathcal{O}(\phi^2),
\end{align}
where $\hat{\phi}$ is the infinitesimal pairing potential, $\hGG^{R/K}_\xi$ and $\Sigma^R \mathbb{I}$ denote the unperturbed Nambu propagators and self-energy in the normal state, respectively, $\psi_\xi$ is the electron statistics in the unperturbed normal state (see Eq.~\ref{eq:psidef}, and $\delta \hFF^R_{\xi}[\phi]$ and $\delta \hat{\AAA}_\xi[\phi]$ denote the off-diagonal linear response of the retarded propagator and the spectral function, respectively. Using Lemma 1(b) of Appendix~\ref{sec:props}, we have:
\begin{equation}
\label{eq:dA}
\delta \check{\AAA}_\xi[\phi](\omega, t) = i\hsx \left[\delta \hFF^R_\xi[\phi](\omega,t) - \delta \hFF^R_\xi[\phi](\omega,t)^*\right] \hsx.
\end{equation}
Here, we have assumed that the pairs formed at the onset of transition have the same statistics $\psi_\xi$ as the normal electrons. Inserting the ansatz in Eqs.~\eqref{eq:dyswigNambu}-\eqref{eq:dyswigadjNambu}, summing the sides and keeping terms to linear order in $\hat{\phi}$, we find:
\begin{multline}\label{eq:pair0}
2\omega \, \delta \hFF^R_\xi[\phi] - \xi\,\big\{\hsz, \delta \hFF^R_\xi[\phi]\big\} = \hat{\phi} \star \hGG^R_\xi + \hGG^R_\xi \star \hat{\phi}\\
+ \Sigma^R \star \delta \hFF^R_\xi[\phi] + \delta \hFF^R_\xi[\phi] \star \Sigma^R.
\end{multline}
Since $\delta \hFF^R_{\xi}$ is fully off-diagonal, $\big\{\hsz, \delta \hFF^R_\xi\big\}=0$. Integrating both sides of Eq.~\eqref{eq:pair0} over $\xi$, using Eq.~\eqref{eq:GR_sum_formula} to replace the $\xi$-summed normal retarded propagators with the universal value of $-i\pi$, we find:
\begin{equation}\label{eq:generaldG}
2\omega \, \delta \FF^R[\phi] = -2\pi i \, \phi + \Sigma^R \star \delta \FF^R[\phi] + \delta \FF^R[\phi] \star \Sigma^R.
\end{equation}
Without the loss of generality, we have assumed $\hat{\phi} = \phi \, \hsx$ and $\delta \hFF^R[\phi] = \delta \FF^R[\phi] \, \hsx$ to distill the Nambu matrix structure on the last equation.

Solving the equation above for arbitrary $\Sigma(\omega,t)$ and $\phi(\omega,t)$ is a formidable task due to the intricate differential structure of the GM product formula. However, we will show later that it can be reduced to a simpler algebraic structure using the properties of Floquet-Boltzmann states. Here, we proceed with the general observation that $\delta \FF^R[\phi]$, and subsequently $\delta \check{\AAA}[\phi]$ as given by Eq.~\eqref{eq:dA}, are computable linear functionals of $\phi$. Projecting out the off-diagonal component of the retarded Migdal-Eliashberg self-energy (Eq.~\ref{eq:senRfin}), we find:
\begin{widetext}
\begin{equation}\label{eq:phieq0}
\phi(\omega, t) = \frac{1}{4}\int_{-\infty}^{+\infty}\frac{\dd\omega'}{2\pi}\int_{-\infty}^{+\infty} \frac{\dd\nu}{\omega - \omega' - \nu + i0^+} \,\Big\{iF^K(\nu,t)\,\mathrm{Tr}\left[\hsx \, \delta\check{\AAA}[\phi](\omega',t)\right]
+ F^\rho(\nu,t)\,\mathrm{Tr}\left[\hsx \, i\delta\check{\GG}^K(\omega',t)\right]\Big\}.
\end{equation}
\end{widetext}
This is the  sought after self-consistency relation between the pairing potential and the induced anomalous response. Eq.~\eqref{eq:dA} implies $\mathrm{Tr}\left[\hsx \, \delta\check{\AAA}_\xi[\phi](\omega,t)\right] = +4 \, \mathrm{Im} \, \delta \FF^R_\xi[\phi](\omega,t)$. Furthermore, Lemma 1(d) of Appendix~\ref{sec:props} implies that this quantity is a real odd function of $\omega$. These considerations allow us to simplify the pairing self-consistency condition, Eq.~\eqref{eq:phieq0}, to:
\begin{equation}\label{eq:phieq1}
\phi(\omega, t) = \int_{0}^{+\infty} \frac{\dd\omega'}{\omega'} \, K(\omega,\omega'; t) \, \Delta[\phi](\omega',t),
\end{equation}
where:
\begin{subequations}
\begin{align}
\label{eq:gapdef}
\Delta[\phi](\omega,t) & \equiv -\frac{\omega}{\pi} \, \mathrm{Im}\, \delta \FF^R[\phi](\omega,t),\\
K(\omega,\omega';t) & \equiv \int_{-\infty}^{+\infty} \dd\nu \bigg[\frac{N^-(\omega',\nu;T)}{\omega + \omega' - \nu + i0^+}\nonumber\\
&\qquad\qquad\qquad- \frac{N^+(\omega',\nu;t)}{\omega - \omega' - \nu + i0^+}\bigg].
\end{align}
\end{subequations}
Eq.~\eqref{eq:phieq1} is a functional eigenvalue equation for $\phi(\omega,t)$. As mentioned earlier, the paired state is stable if and only if Eq.~\eqref{eq:phieq1} admits a non-trivial solution for $\phi(\omega,t)$.

\subsection{Pairing instability criterion for quasi-steady Floquet states}
As mentioned earlier, solving Eq.~\eqref{eq:generaldG} for arbitrary nonequilibrium states is a challenging task and requires resorting to numerical methods in general. This task is significantly simpler in special cases such as stationary states where all time derivatives vanish, or quasi-steady Floquet states where the Fourier amplitudes of all involved quantities are approximately stationary. In both cases, Eq.~\eqref{eq:generaldG} can be cast into an algebraic equation and be solved either numerically or by perturbation. To study the case of quasi-steady Floquet states, we take a Fourier transform of the sides of Eq.~\eqref{eq:generaldG} in $t$ and neglect the time derivatives of Fourier amplitudes in convolutions. Physically, the latter is justified if the pair formation rate is faster than the macroscopic time scale over which the quasi-stationary Floquet-Boltzmann state evolves. Replacing the GM products appearing in the right hand side of Eq.~\eqref{eq:generaldG} with the leading order FGM product formula, we find:
\begin{subequations}
\begin{multline}\label{eq:dFFloq2}
(2\omega - m\Omega) \, \delta \FF^R_{n,m}(\omega;t) = -2\pi i \, \phi_{n,m}(\omega;t)\\
+ \sum_{n'}\Sigma^R_{n', m - n + n'}(\omega;t) \, \delta \FF^R_{n-n', m + n'}(\omega;t)\\
+ \sum_{n'}\Sigma^R_{n', m + n - n'}(\omega;t) \, \delta \FF^R_{n-n', m-n'}(\omega;t).
\end{multline}
The FGM product formula mixes different Floquet quasi-momentum states of $\Sigma^R$ and $\delta \FF^R$. We have further introduced an arbitrary Floquet quasi-momentum label by shifting $\omega \rightarrow \omega - m\Omega/2$ on both sides toward a more uniform notation. The above equation can be thought of as an infinite dimensional linear system for $\delta \FF^R_{n,m}$. In practice, one truncates Floquet bands and quasi-momenta to obtain a proper finite linear system (e.g. see Ref.~\cite{tsuji2008correlated}). The finite system is then solved numerically or by perturbation to find an explicit linear relation between the Floquet components of $\delta \FF$ and $\phi$:
\begin{equation}\label{eq:Q_def}
\delta \FF^R_{n,m}(\omega; t) = \sum_{n',m'}\QQ^{n,m}_{n',m'}(\omega; t)\,\phi_{n',m'}(\omega; t).
\end{equation}
\end{subequations}
We note that $\QQ^{n,m}_{n',m'}(\omega; t)$ only depends on the unperturbed retarded self-energy in the normal state. Diagrammatically, this step is equivalent to attaching the pair propagator to $\phi$ an in the right hand side of Eq.~\eqref{eq:diag_pairing}. This expression, together with Eq.~\eqref{eq:gapdef}, yield the sought after explicit relation between the gap and the pairing potential:
\begin{multline}\label{eq:gap2}
\Delta_n(\omega; t) = \frac{i\omega}{2\pi}\sum_{n',m'}\Big\{\QQ^{n,0}_{n',m'}(\omega; t)\,\phi_{n',m'}(\omega; t)\\
- \left[\QQ^{-n,0}_{n',m'}(\omega; t)\right]^*\,\phi^*_{n',m'}(\omega; t)\Big\}.
\end{multline}
Taking a Fourier transform of the sides of Eq.~\eqref{eq:phieq1} in $t$ and neglecting time derivatives as before, we find:
\begin{multline}\label{eq:phieq2}
\phi_{n',m'}(\omega; t) = \sum_{n''}\int_{0}^{+\infty} \frac{\dd\omega'}{\omega'} \, K_{n'',m'}(\omega,\omega';t)\\
\times \Delta_{n'-n''}(\omega';t),
\end{multline}
where:
\begin{multline}\label{eq:K_gap_def}
K_{n,m}(\omega,\omega';t) = \int_{-\infty}^{+\infty} \dd\nu \bigg[\frac{N^-_n(\omega',\nu;t)}{\omega - m\Omega/2 + \omega' - \nu + i0^+}\\
- \frac{N^+_n(\omega',\nu;t)}{\omega -m \Omega/2 - \omega' - \nu + i0^+}\bigg].
\end{multline}
Plugging $\phi_{n',m'}$ from Eq.~\eqref{eq:phieq2} into Eq.~\eqref{eq:gap2} yields the final functional eigenvalue equation for the Floquet gap at the onset of pairing:
\begin{widetext}
\begin{multline}\label{eq:scfin}
\Delta_n(\omega;t) = \frac{i\omega}{2\pi}\sum_{n',n'',m'}\bigg\{\QQ^{n,0}_{n',m'}(\omega;t)\,\int_{0}^{+\infty} \frac{\dd\omega'}{\omega'} \, K_{n'',m'}(\omega,\omega';t) \, \Delta_{n'-n''}(\omega';t) \\
- \left[\QQ^{-n,0}_{n',m'}(\omega;t)\right]^*\,\int_{0}^{+\infty} \frac{\dd\omega'}{\omega'} \, K^*_{n'',m'}(\omega,\omega';t) \, \Delta^*_{n'-n''}(\omega';t)\bigg\}.
\end{multline}
\end{widetext}
It is worthwhile to take a moment and study this equation in some detail. We recall that $\QQ^{n,m}_{n',m'}$ is derived from the normal-state self-energy and relates the anomalous response to the pairing potential (see Eq.~\ref{eq:Q_def}). Thus, this quantity brings in the physics of quasiparticle propagation in the normal state such as lifetime and residue. On the other hand, the phonon propagator enters through $K_{n,m}$ and therefore, it brings in the retarded phonon-mediated attraction between the quasiparticles (see Eqs.~\ref{eq:K_gap_def} and~\ref{eq:N_def}). 

Finally, let us consider the static non-driven limit of Eq.~\eqref{eq:scfin} as a consistency check where all Floquet indices and summations can be dropped. In this limit, Eq.~\eqref{eq:dFFloq2} admits a simple algebraic solution:
\begin{equation}
\delta\FF(\omega) = -\frac{i\pi}{\omega - \Sigma^R(\omega)}\,\phi(\omega),
\end{equation}
implying $\QQ^{n,0}_{n',m'}(\omega) = -i\pi \, \delta_{n,0}\,\delta_{n',0} \, \delta_{m',0} \left[\omega - \Sigma^R(\omega)\right]^{-1}$. Plugging this into Eq.~\eqref{eq:scfin}, we find:
\begin{equation}
\Delta(\omega) = \mathrm{Re}\left[\frac{1}{Z(\omega)}\int_{0}^{+\infty} \frac{\dd\omega'}{\omega'} \, K(\omega,\omega') \, \Delta(\omega')\right],
\end{equation}
where $Z(\omega) = [\omega - \Sigma^R(\omega)]/\omega$ as it is usually defined in the context of Migdal-Eliashberg theory. This coincides with the result obtained earlier by by Scalapino, Schrieffer, and Wilkins~\cite{scalapino1966strong} for equilibrium systems. Our result is a proper generalization of the static Migdal-Eliashberg pairing criterion to arbitrary nonequilibrium states [Eq.~\eqref{eq:phieq1}], and particularly to Floquet states [Eq.~\eqref{eq:scfin}].

In practice, we monitor the eigenvalue spectrum of the linear functional posed by Eq.~\eqref{eq:scfin} as the system evolves in the normal state. The normal state is deemed unstable as soon as the lowest eigenvalue crosses zero. The same procedure applies to equilibrium states, where one calculates the normal-state self-energy of electrons at different temperatures and decreases the temperature until the lowest eigenvalue crosses zero. 

\section{Revisiting the problem: results from the Floquet-Migdal-Eliashberg theory}\label{sec:FME_results}

The machinery of Floquet-Migdal-Eliashberg (FME) quantum kinetics allows us to fill out the missing details in the preliminary analysis given in Sec.~\ref{sec:param}. In particular, we can study the role of competing factors such as parametric phonon generation and the heating of electrons in order to assess whether the mechanism laid out in Sec.~\ref{sec:param} persists in transient dynamics.

At this point, we have developed all the necessary tools to solve the problem using the full FME formalism, Fig.~\ref{fig:flowchart}. As outlined above, the electron-phonon system is initially prepared in an equilibrium normal state with temperature $T_i > T_c$, where $T_c$ is the critical superconducting transition temperature for the given system parameters. The drive is smoothly ramped up according to:
\begin{equation}\label{eq:ramp_def}
|F(t)|^2 = \frac{I_0}{2}\left[1 + \tanh(t/\tau_\mathrm{drv})\right]\,\cos^2(\Odrv t),
\end{equation}
where $I_0$ denotes the intensity of the drive. We restrict our numerical analysis to weak and intermediate couplings where the phononic and electronic quantities can be calculated iteratively as described below.

As a first step, the electrons are assumed to remain in the equilibrium state, effectively providing a fixed-temperature Ohmic quantum bath $\Pi^{(0)}_\ell(\omega)$ for the phonons. Explicit expressions for $\Pi^{(0)}_\ell$ are given in Appendix~\ref{sec:landau_damping}). The Floquet-Boltzmann equations for the lattice displacement $\{\phii_n(t)\}$ and phonon propagators $\{i\DD^K_n(\omega;t), \rho_n(\omega;t)\}$ are then numerically integrated forward in time as described in Appendix~\ref{sec:QFB_phonon_numerics}. Subsequently, the Floquet-Boltzmann equations for the energy distribution of electrons $\{\psi_n(\omega;t)\}$ are numerically solved as described in Appendix~\ref{sec:QFB_electron_numerics}. If deemed necessary, this two-pass iterative calculation is looped until a self-consistent nonequilibrium solution is obtained for both phonons and electrons. For our choice of parameters, we found additional iterations to be unnecessary by the virtue of the large separation of energy scales between phonons and electrons and weak coupling. Finally, we study the FME pairing instability condition throughout the evolution in order to assess whether the normal state exhibits the pairing instability at some time.

We remark that the system is assumed to evolve in the normal state throughout the simulation. The present formalism detects the instability toward Cooper pairing. Studying the full Floquet superconducting gap formation and its dynamics is a more challenging problem and is better suited to be studied via a Floquet extension of the time-dependent Landau-Ginzburg (TDGL) formalism. 

We present the results in two stages. As a first step, in order to gain insight into parameter regimes of maximally enhanced superconductivity, we hold the electrons in thermal states with different temperatures (e.g. by coupling them to a large and infinitely efficient heat bath). We proceed by letting the driven phonons settle to a stationary Floquet driven-dissipative state and calculate the $T_c$ of electrons on its backdrop. This procedure is similar to calculating $T_c$ in equilibrium by studying the eigenvalue spectrum of Eq.~\eqref{eq:scfin}
, however, using the driven phonon propagators.

Finally, we study the full evolution of the coupled system by allowing the electrons evolve on par with phonons. This allows us to investigate the transient nature of the superconducting instability. We find that the heating of electrons destroys the instability at late times and stabilizes the normal state as seen in the experiments~\cite{mitrano2016possible}.\\

\noindent{\em Choice of physical parameters---} The trimmed-down model is fully specified by a few physical parameters: electron-phonon coupling $g$, electronic density of states at the Fermi level $\nu(0)$, optical phonon frequency $\omega_0$, the phenomenological damping of the coherent lattice displacement $\gamma_0$, and the drive coupling strength $\Lambda$. We trade $g$ and $\nu(0)$ with the mass enhancement factor at equilibrium $\lambda_\mathrm{eq}$ and local phonon damping $\gamma_\ell$. These dimensionless quantities are defined in Eqs.~\eqref{eq:lambda_def} and~\eqref{eq:equilibrium_bath} and we quote them again here for reference:
\begin{equation}
\lambda_\mathrm{eq} \equiv \frac{2 g^2 \nu(0)}{\omega_0}, \qquad \gamma_\ell \equiv 4\pi g^2\nu(0)^2.
\end{equation}
We set $\lambda = 0.5$ and $\gamma_\ell = 0.2$ in the numerics, which correspond to typical values for fulleride superconductors~\cite{varma1991superconductivity}. We study cubic and quartic nonlinearities separately. The majority of the results are presented for a cubic nonlinearity. As we argued earlier in Sec.~\ref{sec:higher_order_nonlins}, both types of nonlinearity lead to qualitatively similar phenomena. We set $\kappa_3 = 0.1\,\omega_0$, $\kappa_4 = 0$ for ``cubic results'', and  $\kappa_3 = 0$, $\kappa_4 = 0.1\,\omega_0$ for ``quartic results''. These values are expected to reflect the typical intrinsic lattice nonlinearities.\\

\noindent{\em Normalization constants---} Quantities with the dimension of energy are presented in the units of $\Omega_0$, the renormalized phonon frequency at equilibrium defined in Eq.~\eqref{eq:renorm_phonon_freq}, an the time axes are scaled with respect to $\tau_\mathrm{ph}$, the renormalized period of phonons at equilibrium:
\begin{equation}
\Omega_0 \equiv \sqrt{\omega_0^2 + 2\omega_0 \bar{\omega}_{L} + 2\omega_0 U_0}, \qquad \tau_\mathrm{ph} \equiv \frac{2\pi}{\Omega_0}.
\end{equation}
Here, $\bar{\omega}_L$ is the effective Lamb shift of phonons as a matter of coupling to electrons which can be neglected in the weak coupling regime. $U_0$ is determined by self-consistently solving the set of equations given in Eq.~\eqref{eq:phonon_equilibrium_sc_eqs} and represents the phonon frequency correction due to lattice nonlinearities. Finally, we parametrize the fully ramped-up drive amplitude by the dimensionless quantity $\mathcal{A}$ defined as:
\begin{equation} 
\mathcal{A} \equiv \lim_{t\rightarrow \infty} \frac{\Lambda}{2\omega_0} I_0 \, F_\mathrm{env}^2(t) = \frac{\Lambda\, I_0}{2\omega_0},
\end{equation}
and set $\tau_\mathrm{drv} = 5\,\tau_\mathrm{ph}$ in Eq.~\eqref{eq:ramp_def}.\\

\noindent{\em Notation---} We often discuss period-averaged quantities along with their temporal variances, respectively defined as:
\begin{align}
\langle O(t) \rangle &\equiv \frac{\Odrv}{\pi} \int_{-\pi/(2\Odrv)}^{+\pi/(2\Odrv)}\dd \tau \, O(t + \tau),\nonumber\\
\mathrm{Var}[O(t)] &\equiv \frac{\Odrv}{\pi} \int_{-\pi/(2\Odrv)}^{+\pi/(2\Odrv)}\dd \tau \, \left[ O(t + \tau) - \langle O(t) \rangle \right]^2
\end{align}
where $O(t)$ is an arbitrary observable. Note that the effective drive period is $\pi/\Odrv$ since the principal harmonic of all observables is $2\Odrv$ (see the discussion after Eq.~\ref{eq:FW_inverse}). If $O(t)$ is given as a Fourier series with slowly-varying amplitudes, i.e. $O(t) = \sum_{n = -\infty}^{\infty} O_n(t)\,e^{2ni\Odrv t}$, then $\langle O(t) \rangle = O_0(t)$ and $\mathrm{Var}[O(t)] \equiv \sum_{n=1}^\infty |O_n(t)|^2$.

\begin{figure*}
\includegraphics[width=\linewidth]{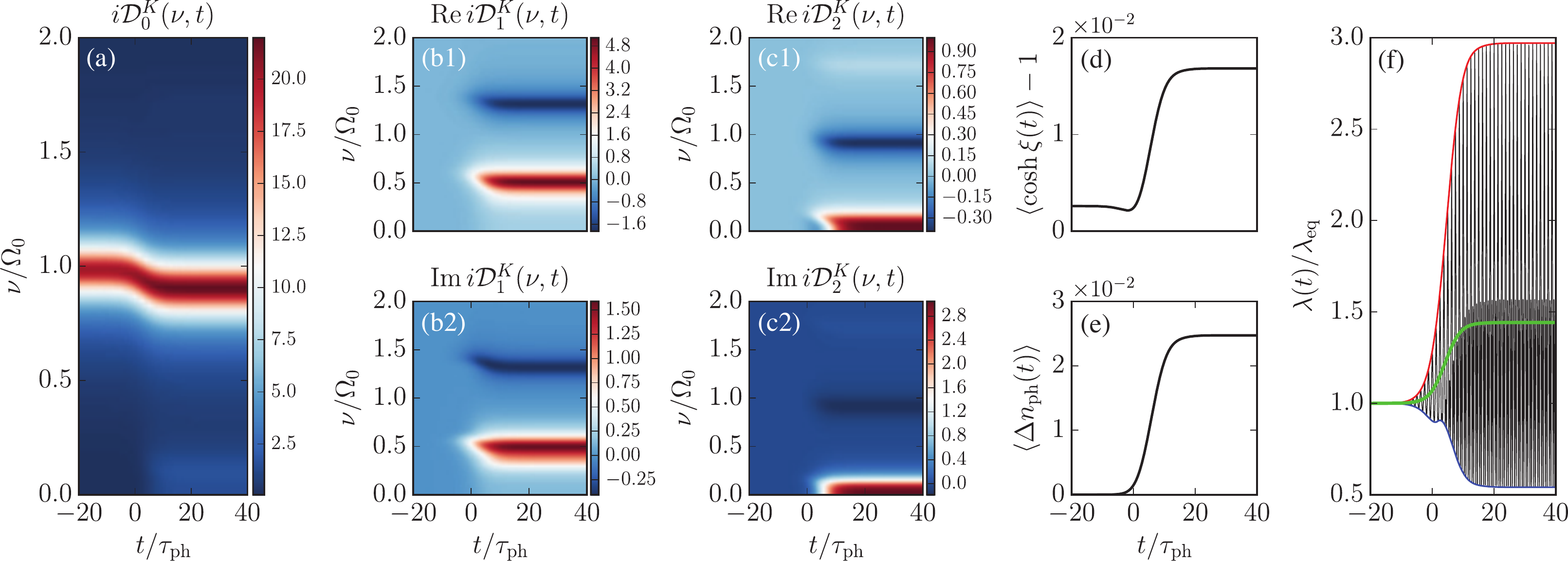}
\caption{{\bf Evolution of phonon propagators in response to a ramped-up external drive for drive frequency $\Odrv = 0.4\,\Omega_0$ and maximum drive amplitude $\mathcal{A} = 0.75$.} The physical parameters are set to $\kappa_3 = 0.1$, $\kappa_4 = 0$, and $\gamma_\ell = 0.2\,\Omega_0$. The leftmost panel shows $n=0$ (periods-averaged) Keldysh phonon propagator. The redshift of the phonon peak is clearly noticeable. The next two columns show the real and imaginary parts of $n=1,2$ propagators. Notice the absence of a single quasiparticle peak. Panels (d) and (e) show the period-averaged squeezing correlations and the density of phonon excitations, respectively. Both quantities increase as the external field is ramped up. Finally, panel (f) shows the time-dependent mass enhancement factor as defined in the text, along with its time average (green solid line) and the lower and upper envelopes (blue and red lines, respectively). Notice the significant increase in the mass enhancement factor, as well as its high amplitude oscillations.}
\label{fig:phonon_dyn}
\end{figure*}

\begin{figure}[h!]
\includegraphics[width=\linewidth]{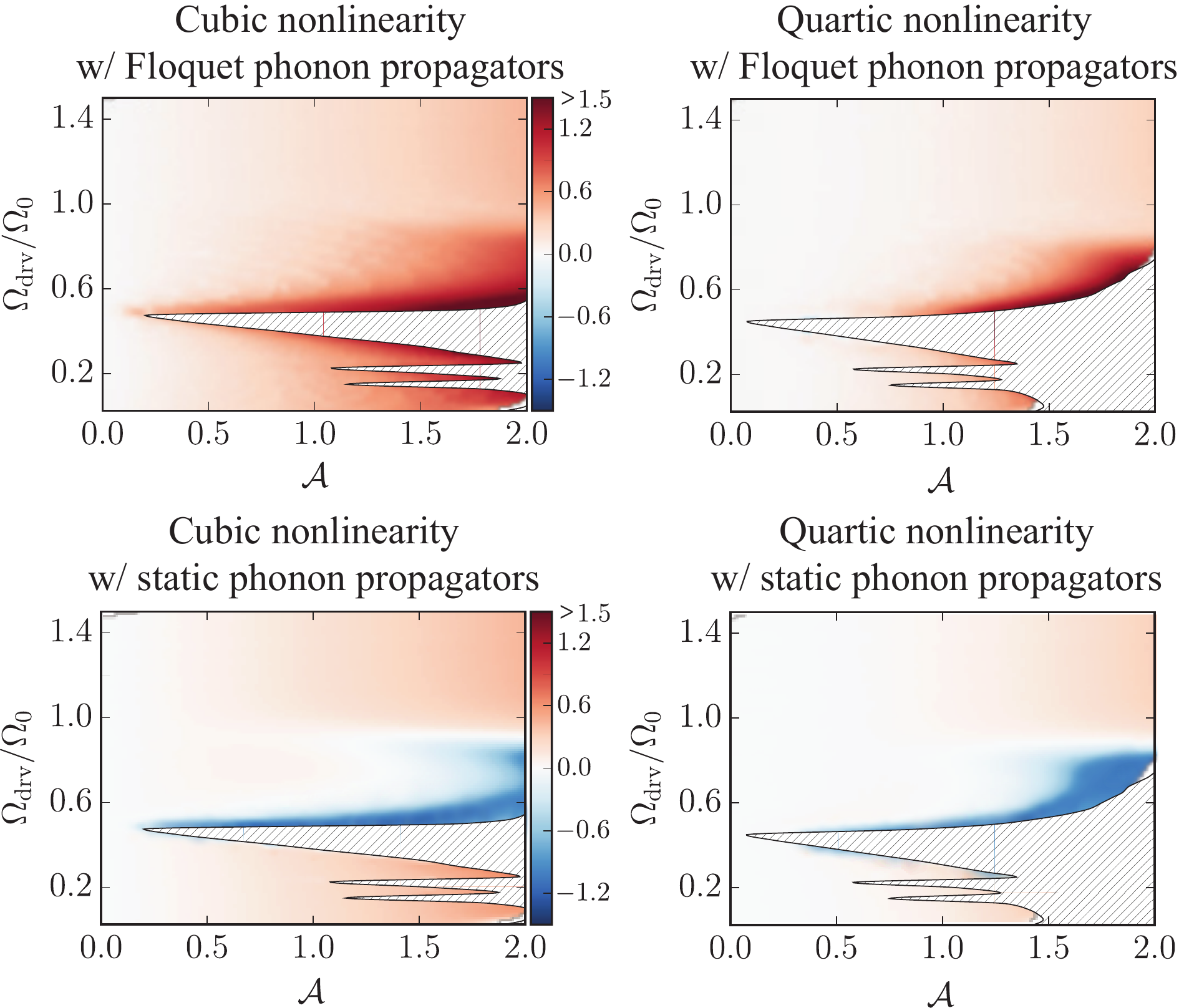}
\caption{{\bf The relative change of Floquet superconducting transition temperature with respect to equilibrium, $T_c^\mathrm{Floq}/T_c^{\mathrm{eq}} - 1$, in the equilibrium electron approximation.} The left and right columns show the results for cubic and quartic nonlinearities, $\kappa_3 = 0.1\,\Omega_0, \kappa_4 = 0$ and $\kappa_3 = 0, \kappa_4 = 0.1\,\Omega_0$, respectively. The top row is obtained using full Floquet phonon propagators whereas $n>0$ Floquet components (dynamical effects) are neglected in the bottom row.}
\label{fig:T_c}
\end{figure}

\begin{figure*}
\includegraphics[width=\linewidth]{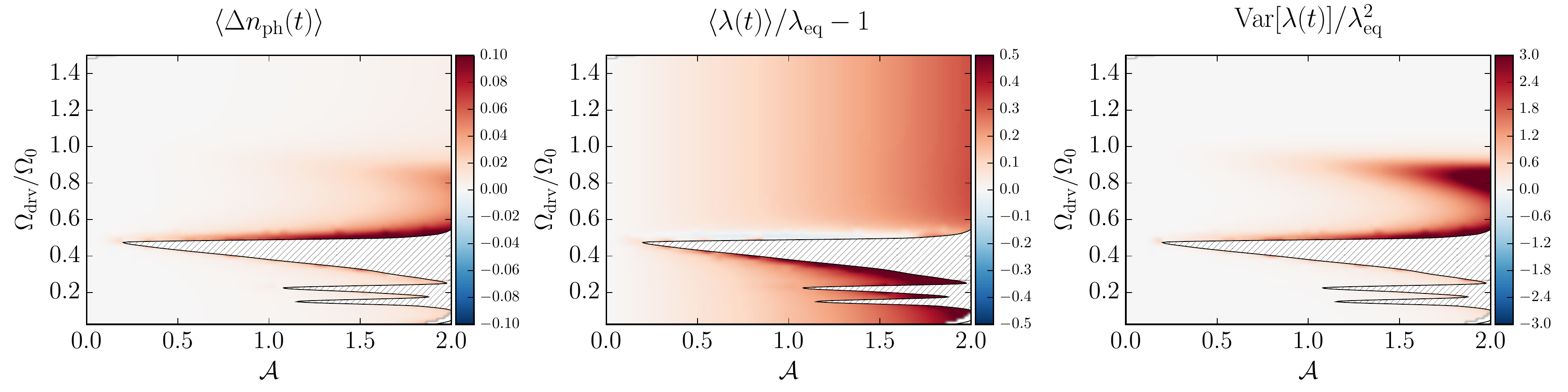}
\caption{{\bf Density of phonon excitations (left), period-average (middle) and time-variance (right) of the mass enhancement factor.} The electrons are kept in a thermal state at temperature $T = 0.04\,\Omega_0$. These quantities are calculated in the stationary driven-dissipative state of phonons.}
\label{fig:T_c_suppl}
\end{figure*}

\begin{figure*}
\includegraphics[width=\linewidth]{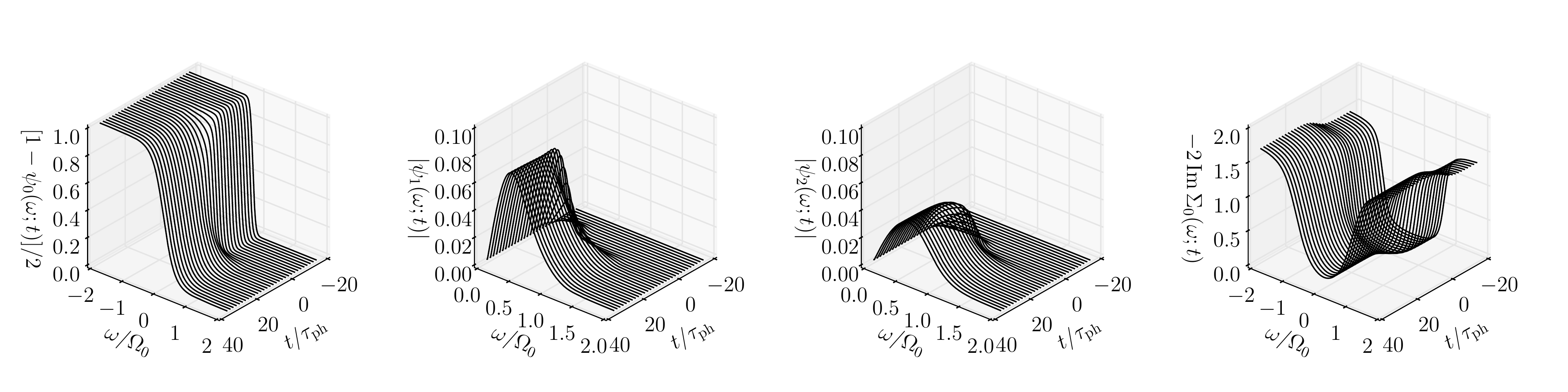}
\caption{{\bf Time evolution of the energy statistics of electrons for $\Odrv=0.4\,\Omega_0$, $\mathcal{A} = 0.75$, and $\tau_\mathrm{drv} = 5\,\tau_\mathrm{ph}$.} From left to right, the plots show the period-averaged $(n=0)$ energy statistics, its first and second Floquet components, and the period-averaged scattering rate $\langle \Gamma(\omega,t) \rangle \equiv -2i\,\mathrm{Im}[\Sigma^R_{n=0}(\omega;t)]$ of electrons. The heating of electrons is noticeable in panel (a) as the drive is ramped up, as well as the increase in the scattering rate in panel (d).}
\label{fig:electron_stats}
\end{figure*}

\begin{figure*}
\includegraphics[width=\linewidth]{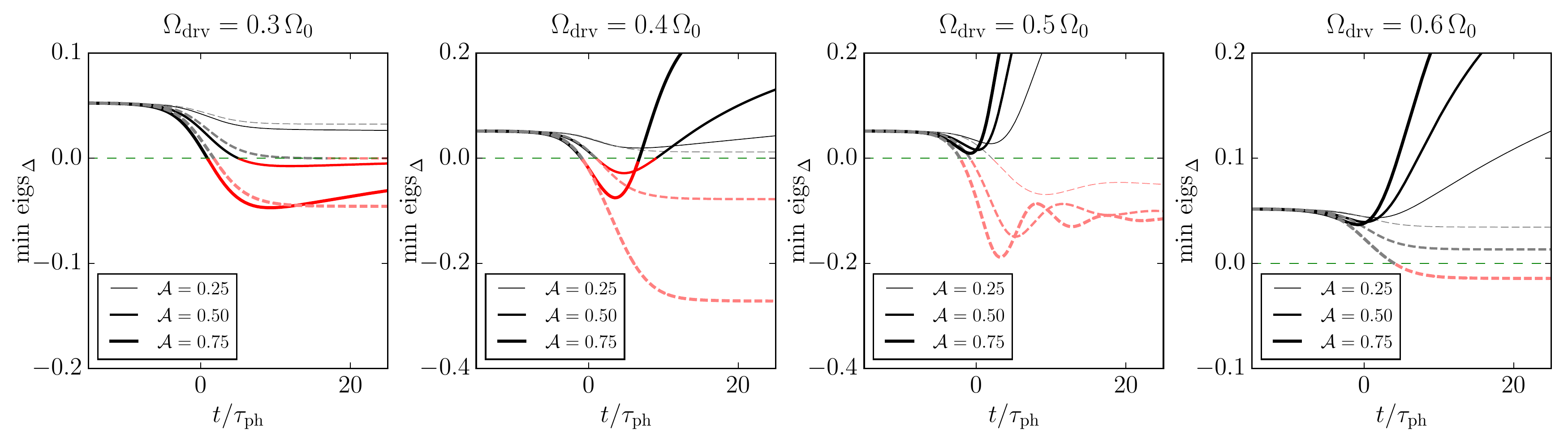}
\caption{{\bf Assessment of the Floquet-Migdal-Eliashberg (FME) pairing condition for a ramped up external drive with different frequencies and amplitudes.} The red segments indicate regions where the lowest eigenvalue of the FME gap functional is negative, signaling the pairing instability. The dashed lines show the hypothetical case if the electrons were to remain in their initial thermal state (no heating). The nonlinearity is cubic, the initial temperature is set to $T_i = 0.04\,\Omega_0 \simeq 1.2\,T_c^\mathrm{eq}$, and the physical parameters are chosen as described in Sec.~\ref{sec:FME_results}.}
\label{fig:spect}
\end{figure*}

\begin{figure*}
\includegraphics[width=\linewidth]{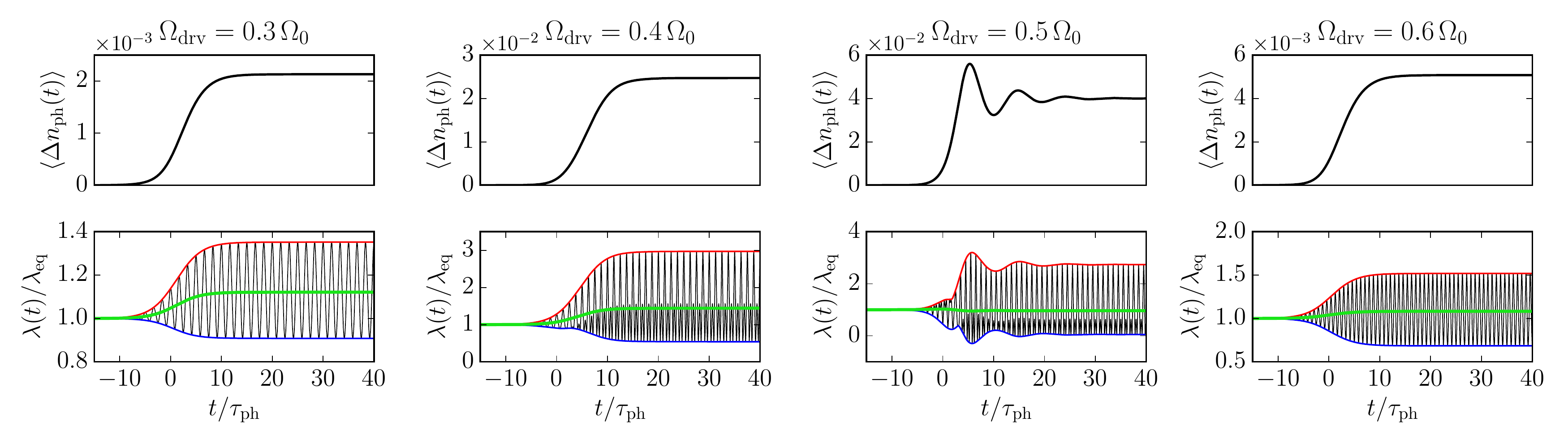}
\caption{{\bf Evolution of the density of phonon excitations $\langle \Delta n_\mathrm{ph}(t) \rangle$ and instantaneous mass enhancement factor $\lambda(t)$ for different drive frequencies and amplitudes.} In the bottom row, the green, blue, and red lines corresponds to the period-averaged, lower, and upper envelopes of $\lambda(t)$. The nonlinearity is cubic, the initial temperature is $T_i = 0.04\,\Omega_0 \simeq 1.2\,T_c$. The choice physical parameters is as given in Sec.~\ref{sec:FME_results}.}
\label{fig:suppl_phonon_stats}
\end{figure*}

\begin{figure*}
\includegraphics[width=\linewidth]{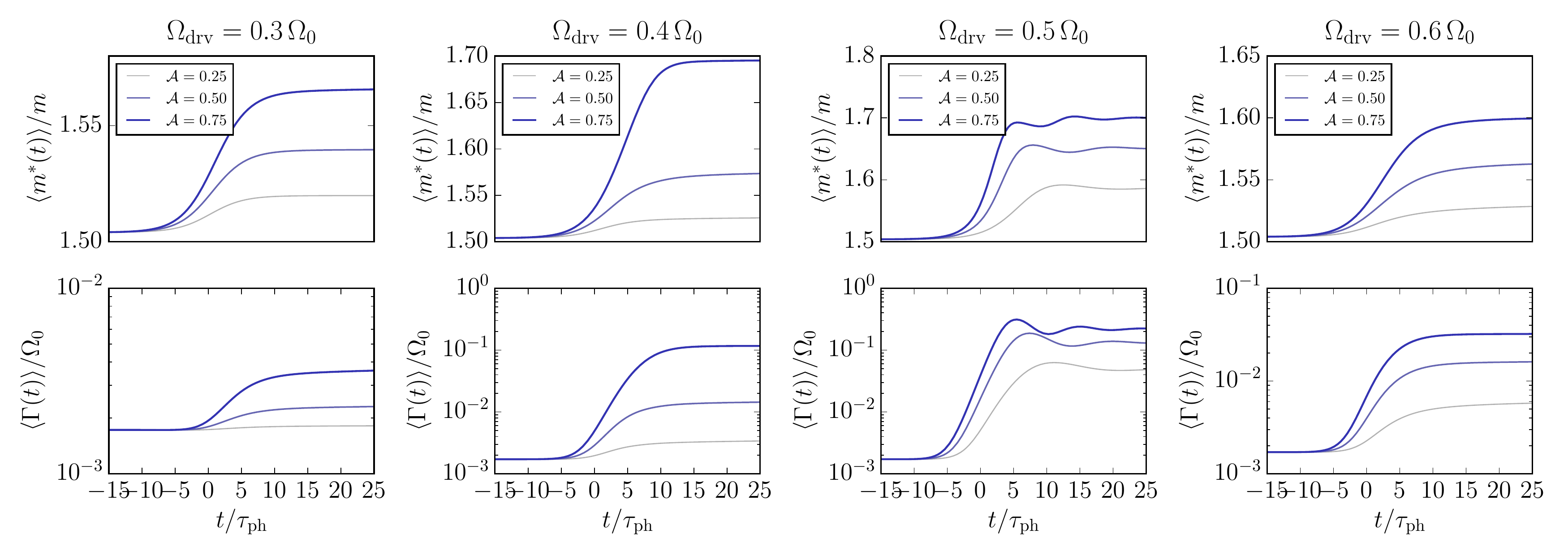}
\caption{{\bf Evolution of the electronic effective mass $\langle m^*(t) \rangle$ and scattering rate $\langle \Gamma(t) \rangle$ for different drive frequencies and amplitudes.} The nonlinearity is cubic, the initial temperature is $T_i = 0.04\,\Omega_0 \simeq 1.2\,T_c$. The choice physical parameters is as given in Sec.~\ref{sec:FME_results}.}
\label{fig:suppl_electron_stats}
\end{figure*}

\subsection{Stage I: Driven phonons, thermal electrons}
Fig.~\ref{fig:phonon_dyn} shows a typical example of phonon propagators subject to nearly-resonant drive in the presence of lattice nonlinearities. Panel (a) shows a heat map of the period-averaged Keldysh phonon propagator $iD^K_{n=0}(\nu;t)$  at a function of relative frequency $\nu$ and COM time $t$. The most prominent feature is the red-shift of the phonon peak frequency. The red-shift is a direct consequence of the lattice nonlinearity: with a cubic nonlinearity $\sim \kappa_3\,\hat{\phii}^3$, the drive shifts the equilibrium position of the lattice on average, producing a frequency renormalization $\Delta \omega^2_\mathrm{ph} \sim \kappa_3 \langle \hat{\phii} \rangle$. For a nearly-resonant drive, nonlinear effects dominate the value of $\langle \hat{\phii} \rangle$ such that $\mathrm{sign}[\langle \hat{\phii} \rangle] = - \mathrm{sign}(\kappa_3)$. As a result, $\Delta \omega^2_\mathrm{ph} < 0$ regardless of the sign of $\kappa_3$ for a strong nearly-resonant drive. In other words, the cubic nonlinearity always softens the lattice. This phenomenon resembles the usual physics of thermal expansion where the drive plays the role of heating.

Panels (b1-2) and (c1-2) show the higher Floquet components of the Keldysh phonon propagator, both of which show emergent features as the drive is ramped up. It is noticed that $|i\DD^K_2| < |i\DD^K_1|$ suggesting that the role of higher Floquet bands become increasingly smaller. Most strikingly, it is noticed that $n>0$ Floquet phonons do not admit a single coherent peak in contrast to the $n=0$ case. As a consequence, Eqs.~\eqref{eq:FBDR}-\eqref{eq:FBDKadj} do not admit a reliable Boltzmann ``quasiparticle'' limit, justifying our usage of the more cumbersome {\em quantum} kinetic formalism.

One consequence of the drive is parametric generation of phonons~\cite{berges2003parametric} and squeezing of lattice momentum fluctuation~\cite{knap2015dynamical}. Panel (d) and (e) show the evolution of these quantities as a function of COM time $t$. Appendix~\ref{sec:squeezing_keldysh} shows how these quantities can be calculated from $i\DD^K(\omega, t)$. As expected, both of these observables significantly increase as the drive is ramped up. The emergence of squeezed states is a well-known hallmark of parametrically driven harmonic oscillators. Finally, panel (f) shows the instantaneous mass enhancement factor $\lambda(t)$ defined as:
\begin{equation}
\lambda(t) =  \omega_0 \lambda_\mathrm{eq} \sum_n\int_0^\infty \frac{\dd \nu}{\nu} \, \rho_n(\nu;t)\, e^{2in\Odrv t}.
\end{equation}
This quantity plays a similar role in the Migdal-Eliashberg theory as $\nu(0) U(t)$ in the BCS theory (the latter was defined in Sec.~\ref{sec:param}). For instance, $T_c \approx \omega_0 \, e^{-1/\lambda_\mathrm{eq}}$ in the equilibrium Migdal-Eliashberg theory~\cite{bennemann2008superconductivity}. It is noticed that $\lambda(t)$ has a strong oscillatory component. The period-average of $\lambda(t)$ during one drive period and the lower and upper envelopes are shown as green, blue, and red solid lines.

The Floquet superconducting transition temperature $T_c^\mathrm{Floq}$ can be determined for each choice of $\Odrv$ and $\mathcal{A}$ by calculating the eigenvalue spectrum of the FME gap functional (Eq.~\ref{eq:scfin}, $\mathrm{eigs}_\Delta$, for different electronic temperatures and locating the first zero-crossing of the lowest eigenvalue $\mathrm{min}~\mathrm{eigs}_\Delta$.

Fig.~\ref{fig:T_c} shows the results separately for cubic and quartic nonlinearities. The top row corresponds to a calculation using full Floquet phonon propagators. The bottom row is obtained using only $n=0$ (period-averaged) phonon propagators. Strong driving near parametric resonances lead to the instability of the lattice due to nonlinearities. This stems of the our choice of $\mathcal{V}^\mathrm{ph}(\phii)$ (see Eq.~\ref{eq:Vnonlin}) which is only valid for low-amplitude deformations and becomes unphysical for large deformations. These unstable regions are hatched in the heat map plots and indeed coincide with the three first parametric resonances at $\Odrv/\Omega_0 \approx 1/2, 1/4, 1/8$.

The full Floquet result (top row) shows a dramatic enhancement of $T_c$, reaching beyond three times the equilibrium value near the resonances. Neglecting the ac components, only a moderate enhancement of $T_c$ is found, and only away from the resonances. In particular, $T_c$ is suppressed above the main resonance in the static approximation in contrast to the full Floquet result. This finding is strikingly similar to the analytic Floquet BCS analysis of Sec.~\ref{sec:floquet_BCS}; see Fig.~\ref{fig:T_c_analytic}.

To shed light on this finding, we have plotted the density of excited phonons $\langle \Delta n_\mathrm{ph}(t) \rangle$ as well as the mean and variance of $\lambda(t)$ during a drive period as a function of $\Odrv$ and $\AM$ in Fig.~\ref{fig:T_c_suppl}. It is noticed that (1) $\langle \lambda(t) \rangle$ is enhanced and suppressed below and above the main resonance, respectively, similar to the analysis of Sec.~\ref{sec:param} and as summarized in Fig.~\ref{fig:semiclassical_stats}; (2) both $\langle \Delta n_\mathrm{ph} \rangle$ and $\mathrm{Var}[\lambda(t)]$ are increase significantly above the main resonance. Neglecting $n>0$ components of the phonon propagator and neglecting $\mathrm{Var}[\lambda(t)]$ go hand in hand. The high density of phonon excitations and the suppression of $\langle \lambda(t) \rangle$ above the resonance imply decreased quasiparticle lifetime and electron-electron attraction, respectively, both of which are unfavorable for Cooper pairing. This explains suppression of $T_c$ above the main resonance in the static approximation. Away from the resonances, the moderate enhancement of $\langle \lambda(t) \rangle$, which has its roots in the phonon frequency red-shift and is present in the static approximation, explains the moderate enhancement of $T_c$. Finally, we remark that including $n>0$ components brings in large $\mathrm{Var}[\lambda(t)]$. In the example shown in Fig.~\ref{fig:phonon_dyn}(f), the upper envelope of $\lambda(t)$ is nearly three times as large as $\lambda_\mathrm{eq}$. As we argued earlier in Sec.~\ref{sec:floquet_BCS}, $T_c$ is a convex functional of the interaction parameter in the weak coupling limit such that temporal variation of interaction can increase $T_c$ even if the period-average remains fixed or even decrease.

\subsection{Stage II: Driven phonons, evolving electrons}\label{sec:full_picture}
The notion of superconducting transition temperature is only well-defined in thermal states. Once the electrons are allowed to evolve as a matter of coupling to phonons, a different diagnostic will be needed to assess the enhancement or suppression of the superconducting transition. Here, we attempt to model a realistic experimental scenario: we prepare the electron-phonon system in $T_i > T_\mathrm{eq}$, ramp up the drive, and calculate the ensuing nonequilibrium dynamics of phonons and electrons. Since $T_i>T_c$ in the beginning, the lowest eigenvalue of the FME gap functional begins as a positive value. Whether it remains positive throughout the evolution or crosses zero at some point is our diagnostic. This allows us to study the instability of the normal state toward Cooper pair formation but does not describe the physics of gap formation. The latter can be addressed with an extended formalism based on the the present developments.

Fig.~\ref{fig:electron_stats} shows an example of the evolution of the Floquet components of the $\xi$-summed (local) energy statistics of electrons $\{\psi_n(\omega;t)\}$ defined in Eq.~\eqref{eq:psi_loc_def}, along with their period-averaged scattering rate $\langle \Gamma(\omega,t) \rangle \equiv -2i\,\mathrm{Im}[\Sigma^R_{n=0}(\omega;t)]$. The prominent features are (1) the heating of electrons, manifested as the decreased slope of $\psi_0(\omega;t)$ at $\omega=0$ as the drive is ramped up, (2) emergence of electrons in $n>0$ Floquet bands, and (3) increased (decreased) spectral broadening (lifetime) of quasiparticles.

Fig.~\ref{fig:spect} shows the evolution of the lowest eigenvalue of the FME gap functional, $\mathrm{min\,eigs}_\Delta$, for $\Odrv/\Omega_0 = 0.3, 0.4, 0.5, 0.6$ and $\mathcal{A} = 0.25, 0.50, 0.75$. The nonlinearity is cubic and the choice of physical parameters is as described in the introductory remarks of this section, implying $T^\mathrm{eq}_c \simeq 0.034\,\Omega_0$. The initial temperature chosen as $T_i = 0.04\,\Omega_0 \simeq 1.2\,T_c^\mathrm{eq}$. The dashed lines show the hypothetical case where the electrons are kept at $T_i$ (no heating). Red segments indicate where $\mathrm{min\,eigs}_\Delta < 0$. The most favorable outcome occurs for lower frequency driving, e.g. $\Odrv = 0.3\,\Omega_0$, where the pairing instability persists for a long time. In all cases, heating of electrons tends to stabilize the normal state with long enough driving. This is most easily noticeable for $\Odrv = 0.4\,\Omega_0$ where the instability is confined to a short interval. For $\Odrv = 0.5\,\Omega_0, 0.6\,\Omega_0$, we find $\mathrm{min\,eigs}_\Delta > 0$ at all times for all three drive strengths. The strong heating of electrons prohibits pairing even though in the absence of heating (dashed lines), pairing would have ensued.

The desirability of lower frequency driving for enhancing the pairing instability can be understood by appealing to the different nature the two competing effects, parametric amplification of the retarded response on one hand, and parametric phonon generation on the other hand. As we argued early on, the former is the main mechanism for enhancing $T_c$ and the latter is the main suppressant. Heating of electrons and the decreased coherence of quasiparticles are both consequences of the interaction with the generated high-energy phonons.

Parametric generation of phonons is an on-shell process. For low frequency driving, phonons may only be generated through accumulation of multiple energy quanta from the drive. These higher order processes, however, become increasingly less probable. In contrast, (1) the retarded response does not need to satisfy an on-shell energetic condition, and (2) the cascade of low-frequency parametric resonances at $\Omega_0/(2n)$ extend the range of parametric amplification to very low frequencies. For a fixed driving strength and finite damping, the infinite cascade of parametric resonance ``tongues'' will be truncated at a certain lower frequency. This is exemplified in Fig.~\ref{fig:semiclassical_U} in which only two resonances are present, or in Fig.~\ref{fig:T_c} where only three lattice instability tongues are found. Nevertheless, the presence of even a few higher order resonances enables the amplification of the retarded response for reasonably low frequency drives.

To substantiate these arguments with numerical results, we have plotted the evolution of $\langle \Delta n_\mathrm{ph}(t) \rangle$ and $\lambda(t)$ in Fig.~\ref{fig:suppl_phonon_stats}, and electronic effective mass $\langle m^*(t) \rangle$ and damping $\langle \Gamma(t) \rangle$ in Fig.~\ref{fig:suppl_electron_stats}. As before, time-averaging is performed during one drive period. We notice that for $\Odrv = 0.3\,\Omega_0$ both $\langle \Delta n_\mathrm{ph}(t) \rangle$ and $\langle \Gamma(t) \rangle$ remain nearly two orders of magnitude smaller than the on-resonance drive $\Odrv = 0.5\,\Omega_0$. In contrast, $\langle m^*(t) \rangle$ and $\lambda(t)$ are at most a factor of four smaller. Thus, we indeed expect a negligible undesirable suppression while still benefiting from parametric amplification of $\lambda(t)$.

\begin{figure}
\includegraphics[width=\linewidth]{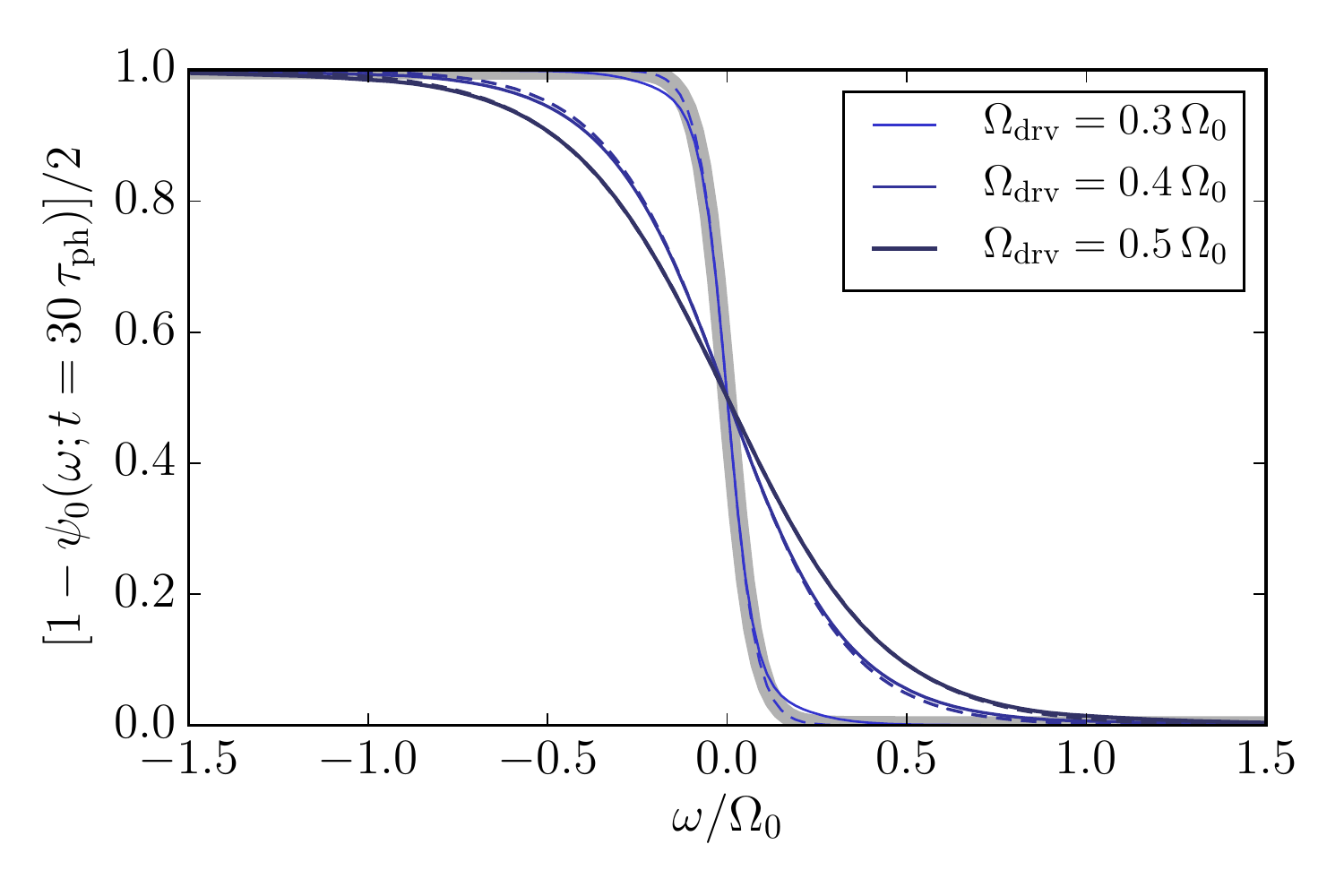}
\caption{{\bf The energy distribution of electrons after the drive is ramped up.} The physical parameters are the same as Fig.~\ref{fig:spect}. The dashed lines show the Fermi-Dirac fits. The thick gray line is the Fermi-Dirac distribution at the initial temperature $T_i = 0.04\,\Omega_0$. The thin solid lines correspond to $\Odrv/\Omega_0 = 0.3, 0.4, 0.5$ with decreasing slope, respectively. The measurement time is $t = 30\,\tau_\mathrm{ph}$. The final effective temperatures are $T^\mathrm{eff}_f/\Omega_0 \approx 0.04, 0.17, 0.22$ for $\Odrv/\Omega_0 = 0.3, 0.4, 0.5$, respectively.}
\label{fig:final_electron_dist}
\end{figure}

The energy distribution of the electrons at $t = 30\,\tau_\mathrm{ph}$ is shown in Fig.~\ref{fig:final_electron_dist}.  It is noticed that the distribution barely changes from the initial thermal state for $\Odrv = 0.3\,\Omega_0$, in agreement with the aforementioned arguments regarding suppressed parametric phonon generation below the resonance. In all cases, a decent Fermi-Dirac fit can be obtained. For a on-resonant driving frequency $\Odrv = 0.5\,\Omega_0$, the effective temperature reaches $T_f^\mathrm{eff} \approx 0.22\,\Odrv$ (see the figure caption). Even in such cases, the energy density of electrons remains low enough, obviating performing a self-consistency feedback loop to phonons (see Fig.~\ref{fig:flowchart}).

Finally, we note that the light-induces superconducting state is expected to persist beyond the predictions of the present analysis in the experiments. On the one hand, the formation of a superconducting gap leads to increased coherence of electrons and suppress scatterings. On the other hand, the bulk electrons and phonons that remain unaffected by the pump pulse act as a low temperature heat bath for the driven subsystem and keep it from excessive heating. Exploring these aspects of the problem is left for future works and is further discussed in Sec.~\ref{sec:conclusions}.

\subsection{Predictions for time-resolved angle-resolved photo-emission spectroscopy experiments (tr-ARPES)}\label{sec:arpes}
\begin{figure*}
\includegraphics[width=\linewidth]{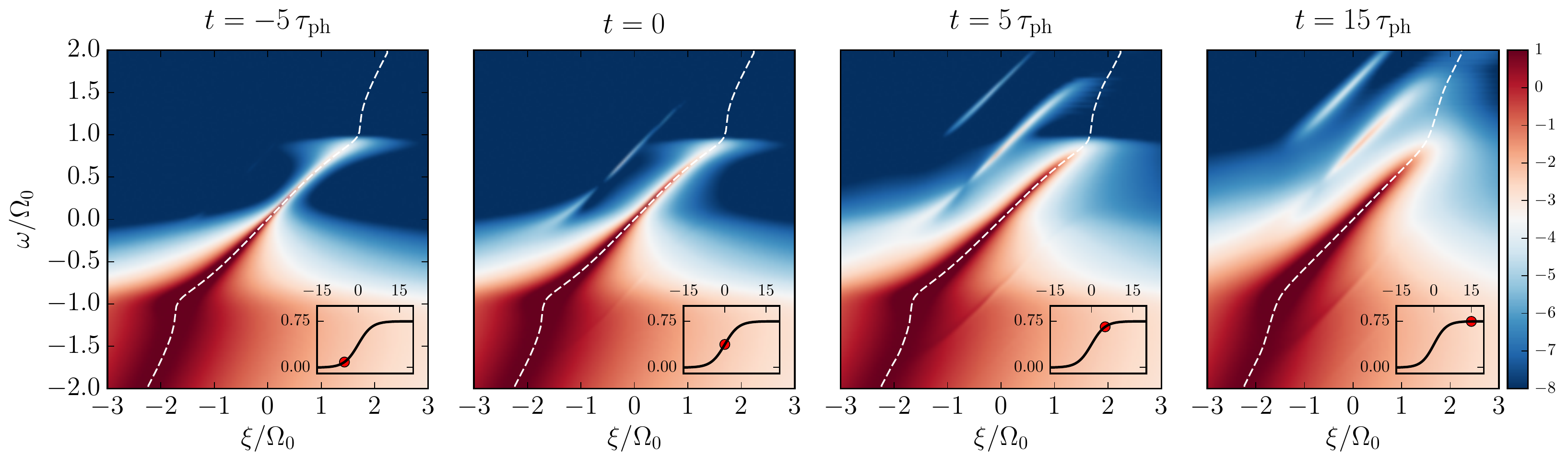}
\caption{{\bf Probing the formation of electronic Floquet bands via tr-ARPES experiments.} The heat map plots show the intensity of the signal at different times in the logarithmic scale. The inset plots show the instantaneous amplitude of the drive during ramp-up. The white dashed lines indicate the dispersion of the main quasiparticle peak. The drive parameters are chosen as $\Odrv = 0.4\,\Omega_0$ and $\AM = 0.75$. The nonlinearity is cubic, and the physical parameters are chosen as described in Sec.~\ref{sec:FME_results}. Note the progressive formation of Floquet quasiparticle bands and the softening of the polaronic kink in the main quasiparticle dispersion as the system heats up. The time-averaged effective mass at the Fermi surface is inversely proportional to the slope of the main quasiparticle dispersion at $\xi=0$ and is shown separately in Fig.~\ref{fig:suppl_electron_stats} for better visibility.}
\label{fig:arpes}
\end{figure*}
Up to this point, we have developed the theory of parametric amplification of electron-phonon coupling within a model that is general and material-agnostic. Whether or not, and how, the mechanism is realized in a specific material depends on a number of factors such as the phonon spectra, strength of nonlinearities, and the selection rules that dictate and presence or absence of the required nonlinear phonon couplings. These questions can be investigated either by performing {\em ab initio} calculation or through further experimental scrutiny. The interaction of electrons with periodically driven phonons will necessarily induce a certain degree of periodicity in electronic observables. This can be see, for example, in Fig.~\ref{fig:electron_stats} where the energy statistics develop Floquet components as the drive is ramped up. Here, we will show that the signal measured in tr-ARPES experiments will help reveal the formation of Floquet bands.

The tr-ARPES signal can be theoretically calculated from the {\em lesser} electron Green's function~\cite{sentef2013examining} as follows:
\begin{align}
I(\kk, \omega, t) &\propto \mathrm{Im} \frac{1}{2\pi \sigma_\mathrm{pr}^2}\int \dd t_1 \int \dd t_2\, \GG^<_{\xi_\kk}(t,t')\, e^{-(t_1 - t)^2/2\sigma_\mathrm{pr}^2}\nonumber\\
&\qquad\times e^{-(t_2 - t)^2/2\sigma_\mathrm{pr}^2}\,e^{i\omega(t-t')},
\end{align}
where $\sigma_\mathrm{pr}$ is the temporal resolution of the probe field which generically satisfies $\sigma_\mathrm{pr} \gg \Odrv^{-1}$. The Gaussian window functions thus simply serve as picking up the ``period-averaged'' lesser Green's function, which coincides with $n=0$ Floquet component of $\GG$. Thus,
\begin{equation}\label{eq:arpes}
I(\kk, \omega, t) \propto \mathrm{Im}\,\GG^<_{\xi_\kk; n=0,m=0}(\omega;t).
\end{equation}
Up to transitory effects, the lesser Green's function can be calculated as $\GG^< = \GG^R \star \Sigma^< \star \GG^A$~\cite{danielewicz1984quantum}. To leading order, we may approximate convolutions with zeroth-order FGM product formula. Each of the required ingredients for calculating $\GG^<$ can be obtained using the results already available to us. As a first step, we solve the Dyson's equation for $\GG^R_{n;\xi}(\omega;t)$:
\begin{multline}
(\omega - m\Omega/2 - \xi) + \, \GG^R_{\xi;n,m}(\omega;t) = \delta_{n,0}\\
+ \sum_{n'}\Sigma^R_{n', n' - n + m}(\omega;t)\,\GG^R_{\xi; n-n',n' + m}(\omega;t) + \mathcal{O}(\partial_t).
\end{multline}
Having calculated $\Sigma^R_{n,m}(\omega;t)$ previously from solving the Floquet-Boltzmann for electrons, $\GG^R_{\xi;n,m}(\omega;t)$ is obtained by truncating the above equations in Floquet bands and solving it as a proper linear system. The advanced Green's function is found immediately using the identity $\GG^A_{\xi;n,m}(\omega;t) = \GG^{R*}_{\xi;-n,m}(\omega;t)$. By definition, the lesser electron self-energy $\Sigma^<$ is related to $\Sigma^{R/A/K}$ as $\Sigma^< = (\Sigma^K - \Sigma^R + \Sigma^A)/2$. Taking a FW transform, we find:
\begin{equation}
\Sigma^<_{n,m}(\omega;t) = \frac{1}{2}\left[\Sigma^K_{n,m}(\omega;t) - \Sigma^R_{n,m}(\omega;t) + \Sigma^{R*}_{-n,m}(\omega;t)\right],
\end{equation}
where we have used the identity $\Sigma^A_{n,m}(\omega;t) = \Sigma^{R*}_{-n,m}(\omega;t)$. With the knowledge of $\Sigma^<_{n,m}$, $\GG_{\xi;n,m}^R$, and $\GG_{\xi;n,m}^A$, $\GG^<_{\xi;n=0,m=0}$ is calculating by employing the zeroth-order FGM formula twice. The final result is:
\begin{multline}
\GG^<_{\xi;0,0} = \sum_{n_R, n_L, n_A} \delta(n_R + n_L + n_A)\\
\times \GG^R_{\xi; n_R, -n_L-n_A}\,\Sigma^<_{n_L, -n_A + n_R}\,\GG^A_{\xi;n_A, n_R + n_L},
\end{multline}
where we have dropped the common $(\omega;t)$ arguments for brevity. This procedure is formed for a range of $\xi$, and Eq.~\eqref{eq:arpes} is used to find the intensity of the tr-ARPES signal.

Fig.~\ref{fig:arpes} shows an example of the predicted tr-ARPES signal for $\Odrv = 0.4\,\Omega_0$ and $\AM = 0.75$ as a function of $\omega$ and $\xi$. We have traded the momentum transfer with $\xi \equiv \epss_\kk - \epss_F$, the kinetic energy measures from the Fermi surface. The period-averaged quasiparticle dispersion $E_\xi$ is obtained by locating the main coherent peak of the period-averaged retarded propagator by solving $\mathrm{Re} \, [\GG^R_{\xi;n=0}(E_\xi;t)]^{-1} = 0$ and is shown as white dashed lines. Before the drive is ramped up, the signal matches what is expected from a coupled electron / optical phonon system at equilibrium~\cite{bennemann2008superconductivity}: filled states for $\omega<0$, decreased quasiparticle coherence at $\omega = \pm \Omega_0$, and a larger effective mass for $|\omega| < \Omega_0$. The effective mass is obtained as $m/\langle m^*(t)\rangle = \dd E_\xi(\omega=0)/\dd\xi|_{\xi=0}$ and is shown in the second column of Fig.~\ref{fig:suppl_electron_stats} for the same drive parameters as Fig.~\ref{fig:arpes}.

An intriguing feature of Fig.~\ref{fig:arpes} is the progressive formation of electronic Floquet bands as the drive is ramped up. The dynamical formation of Floquet bands in driven systems has been observed before experimentally in tr-ARPES spectroscopy of topological insulators~\cite{wang2013observation}. The frequency spacing between the emerging Floquet bands is set by $2\Odrv$. Therefore, the ARPES experiment along with the available spectroscopic measurements of the phonon spectra will inform about the origin of the persistent drive after the pump pulse is gone. We remark that the higher Floquet bands might be challenging to observe from noisy measurements due to the small weight of these extra features. For example, at $t=15\,\tau_\mathrm{ph}$ and for the strong drive parameters used in Fig.~\ref{fig:arpes}, the intensity of the first Floquet band is nearly four orders of magnitude smaller than the main quasiparticle peak.

\section{Conclusions and Outlook}\label{sec:conclusions}
In this paper, we studied the parametric resonances of driven nonlinear lattices and discussed its role in enhancing the effective phonon-mediated electron-electron attraction. We presented the analysis in two stages. First, we gave a qualitative and intuitive account using perturbation theory, classical dynamics, and the Floquet BCS theory in Sec.~\ref{sec:param} in order to elucidate the mechanism of parametric amplification of phonon-mediated Cooper pairing. Next, we developed a quantum kinetic formalism based on an extension of the Migdal-Eliashberg theory to driven systems and nonlinear lattices in Sec.~\ref{sec:qfb} and revisited the problem one more time and in full detail in Sec.~\ref{sec:FME_results}. The numerically tractable quantum kinetic formalism allowed us to study both the intricate transient and long-time dynamics of the system following the pump pulse. In particular, we investigated the role of parametric phonon generation and subsequent heating of electrons in destroying the transient superconducting instability. Finally, we predicted the transient formation of electronic Floquet bands as an experimentally observable consequence of parametrically driven phonons in Sec.~\ref{sec:arpes}. This prediction can be tested in time-resolved ARPES experiments and can be used to better understand material-specific mechanisms of parametric drive generation.

This work can be extended in several directions. So far, we have studied the evolution of the system in the normal-conducting state and treated Cooper pairing as an instability. An important extension of this work is to take into account dynamical symmetry breaking and the formation of the Floquet superconducting gap. This can be done most naturally by generalizing the Floquet-Boltzmann kinetic equation of electrons to symmetry broken states and deriving a time-dependent Ginzburg-Landau theory for the slowly-varying Floquet components of the gap $\{\Delta_n(\mathbf{x}, t)\}$. This extension allows us to address a broad range of largely unexplored theoretical questions, such as the scaling behavior of the coherently-driven system in the critical regime, and the nature of Kibble-Zurek defects~\cite{kibble1980some,zurek1996cosmological} formed as a result of non-adiabatic preparation of the ordered state. Furthermore, extension to gapped states allows us to calculate the nonequilibrium optical conductivity and make a more direct connection to pump-probe experiments~\cite{mitrano2016possible}. A related problem is the question of the lifetime of the transient superconducting state. We find superconductivity as a transient phenomenon as shown in Fig.~\ref{fig:spect}. It arises when electron-phonon interaction is already enhanced and before electrons have been heated too much. 
To give a more detailed analysis of the duration of the transient regime, we need to allow for opening of the quasiparticle gap which we expect to make transient superconductivity last longer. 

The role of light-induced changes in the screened Coulomb interaction has been recently highlighted in the phenomenology of the light-induced superconductivity in $K_3 C_{60}$ in Refs.~\cite{kim2016enhancing,mitrano2016possible}. Furthermore, the shortcomings of Migdal-Eliashberg theory for providing an accurate description of fullerene superconductors and necessity of beyond-Migdal vertex corrections have been indicated in Ref.~\cite{cappelluti2001superconductivity}. Therefore, it is desirable to extend the present formalism to include both Coulomb interaction and beyond-Migdal vertex corrections~\cite{grimaldi1995nonadiabatic} and to study their role to the extent relevant to the mechanism discussed in this paper. In equilibrium, the effects of retarded Coulomb interaction can be incorporated in the Migdal-Eliashberg theory using the Morel-Anderson (MA) pseudo-potential~\cite{morel1962calculation,schrieffer1983theory,bauer2012theory}. A nonequilibrium extension of this result is lacking and must be worked out. A naive application of the equilibrium result suggests that the MA pseudo-potential $\mu^* = \mu_c/[1 + \mu_c \log(\epss_F/\omega_\mathrm{ph})]$ directly decreases the mass enhancement factor, i.e. $\lambda(t) \rightarrow \lambda(t) - \mu^*$. Here, $\mu_c = \nu(0) U_c$ and $U_c$ is the typical screened Coulomb interaction between conduction electrons. In this paper, we showed that parametric driving enhances Cooper pairing by increasing $\langle \lambda(t) \rangle$ and its temporal variations. Since the Coulomb interaction does not directly play a role in the parametric resonance of the lattice, we expect our conclusions to remain valid. Moreover, Ref.~\cite{kim2016enhancing} suggests that $\mu_c$ effectively decreases in the pumped system, in which case, the parametric amplification of $\lambda(t)$ and decreased $\mu^*$ both work toward enhancing Cooper pair formation in $K_3 C_{60}$. The role of dynamical vertex corrections and the status of Migdal's theorem, in particular in the presence of the external drive, is less clear and must be carefully reassessed via real-time techniques in the spirit of the analysis provided for dynamical electron-mediated nonlinearities in Appendix~\ref{sec:nonlin}.

We note that photo-induced enhancement of superconductivity has also been observed in high-Tc cuprates \cite{mankowsky2014nonlinear} along with several theoretical proposals for explaining these experiments~\cite{raines2015enhancement,okamoto2016theory,hoppner2015redistribution,coulthard2016enhancement,sentef2016theory,patel2016light}. Cuprate superconductors are considerably more complicated than conventional electron-phonon superconductors that we considered in this paper. Superconductivity in these materials is likely to be of non-phononic origin and there are several competing orders. However, we expect that the ideas explored here may be relevant for light-enhanced superconductivity in these materials as well. For example, periodic lattice modulation changes the strength of magnetic exchange interactions and may lead to parametrically amplified electron-paramagnon coupling. Paramagnons are expected to play the role of phonons in unconventional superconductors.

Last but not least, another intriguing future research direction which is also of much technological interest, is to extend the present analysis to open driven-dissipative systems along with accurate material-specific {\em ab initio} calculations. The transient light-induced superconducting state can be enhanced further or even stabilized by continuous pumping of the lattice and simultaneous cooling. Such a hybrid ``pumped-and-cooled'' device may operate more efficiently compared to the usual refrigerated superconductor depending on the highest achievable effective critical temperature and the pump absorption power of the material.

\section{Acknowledgments}
We thank A. Cavalleri, A. Georges, V. Galitski, C. Kollath, A. Millis, and B. Halperin for useful discussions. MB and GR acknowledge support from the Institute for Quantum Information and Matter (IQIM), an NSF Physics Frontiers Center with support of the Gordon and Betty Moore Foundation. MK acknowledges support from the Technical University of Munich - Institute for Advanced Study, funded by the German Excellence Initiative and the European Union FP7 under grant agreement 291763, and from the DFG grant No. KN 1254/1-1. IM acknowledges support from the Materials Sciences and Engineering Division, Basic Energy Sciences, Office of Science, US Dept. of Energy. ED acknowledges support from Harvard-MIT CUA, NSF Grant No. DMR-1308435, AFOSR Quantum Simulation MURI, and AFOSR MURI Photonic Quantum Matter.

\appendix

\section{The parametrically driven harmonic oscillator}\label{sec:mathieu}
We studied the problem of parametrically driven harmonic oscillator formally in Sec.~\ref{sec:param} in the context of the resonant amplification of phonon response. Some of the technical details were left out and we present them here.

We presented the solution of the Heisenberg equation in terms of four special functions $\mathfrak{M}_{\alpha\beta}(t,t')$, $\alpha,\beta = P, Q$ (see Eq.~\ref{eq:QP_dyn}). These functions can be expressed in terms of even and odd Mathieu functions and their derivatives as follows:
\begin{align}\label{eq:mathieu}
\mathfrak{M}_{QQ}(t,t') &= \frac{\mathrm{s}(\Odrv t)\,\mathrm{c}'(\Odrv t') - \mathrm{c}(\Odrv t)\,\mathrm{s}'(\Odrv t')}{\mathrm{s}(\Odrv t')\,\mathrm{c'}(\Odrv t') - \mathrm{c}(\Odrv t')\,\mathrm{s}'(\Odrv t')},\nonumber\\
\mathfrak{M}_{QP}(t,t') &= \frac{\mathrm{s}(\Odrv t)\,\mathrm{c}(\Odrv t') - \mathrm{c}(\Odrv t)\,\mathrm{s}(\Odrv t')}{\mathrm{s}(\Odrv t')\,\mathrm{c'}(\Odrv t') - \mathrm{c}(\Odrv t')\,\mathrm{s}'(\Odrv t')},\nonumber\\
\mathfrak{M}_{PQ}(t,t') &=\frac{\mathrm{s}'(\Odrv t)\,\mathrm{c}'(\Odrv t') - \mathrm{c}'(\Odrv t)\,\mathrm{s}'(\Odrv t')}{\mathrm{s}(\Odrv t')\,\mathrm{c'}(\Odrv t') - \mathrm{c}(\Odrv t')\,\mathrm{s}'(\Odrv t')},\nonumber\\
\mathfrak{M}_{PP}(t,t') &=\frac{\mathrm{s}(\Odrv t)\,\mathrm{c}'(\Odrv t') - \mathrm{s}'(\Odrv t)\,\mathrm{c}(\Odrv t')}{\mathrm{s}(\Odrv t')\,\mathrm{c'}(\Odrv t') - \mathrm{c}(\Odrv t')\,\mathrm{s}'(\Odrv t')}.
\end{align}
where $\mathrm{s}(z) \equiv \mathrm{Se}(\omega_\qq^2/\Odrv^2, -\alpha \omega_\qq^2/\Odrv^2, z)$ and $\mathrm{c}(z) \equiv \mathrm{Ce}(\omega_\qq^2/\Odrv^2, -\alpha \omega_\qq^2/\Odrv^2 ,z)$ denote the odd and even Mathieu functions with characteristic value $\omega_\qq^2/\Odrv^2$ and parameter $-\alpha \omega_\qq^2/\Odrv^2$, respectively, and the prime sign denotes derivatives with respect to $z$. We showed that $\mathfrak{M}_{QP}(t,t')$ is of particular interest and determines the retarded phonon response $\DD^R_{\QQ}(t,t')$ (see Eq.~\ref{eq:D_ret_mathieu}). Here, we present a series expansion of this function in terms of the parameter $\alpha$ (see Eq.~\ref{eq:alpha_def}), i.e. $\mathfrak{M}_{QQ}(t,t') = \sum_{n=0}^\infty a^n \, \mathfrak{M}^{(n)}_{QQ}(t,t')$. The first two terms in the series are given as:
\begin{align}\label{eq:traj_pert}
\mathfrak{M}^{(0)}_{QQ}(t,t') &= \frac{\Odrv}{\omega_\qq}\sin[\omega_\qq(t-t')],\nonumber\\
\mathfrak{M}^{(1)}_{QQ}(t,t') &= -\frac{\omega_\qq\cos[\Odrv(t+t')]}{2(\omega_\qq^2 -\Odrv^2)}\nonumber\\
&\times\Big\{(\omega_\qq + \Odrv)\sin[(\omega_\qq-\Odrv)(t-t')]\nonumber\\
&\qquad- (\omega_\qq - \Odrv)\sin[(\omega_\qq + \Odrv)(t-t')]\Big\}.
\end{align}
The higher order terms are increasingly more complex but can be easily worked out using a computer algebra system. 

\section{Definition and properties of the CTP Green's functions}\label{sec:props}
In this appendix, we briefly review the definition of CTP Green's functions, their various real-time components, and their symmetries.

The CTP Nambu electron propagator is defined as:
\begin{equation}
\hGG_\kk(t_1, t_2) = -i \left\langle T_\CC\left[\PSI_\kk(t_1) \PSID_\kk(t_2)\right]\right\rangle,
\end{equation}
where $\PSI_{\kk} = (c_{\kk\up}, c_{-\kk\dn}^\dagger)^T$, $\PSID_{\kk} = (c_{\kk\up}^\dagger, c_{-\kk\dn})$, and $\hGG_\kk(t_1, t_2)$ is a $2 \times 2$ matrix in the Nambu space. Here, $\CC = \CC^+ \cup \CC^-$ denotes the round-trip Keldysh contour where $\CC^+ = [t_0, +\infty)$ and $\CC^- = (+\infty, t_0]$, and $T_\CC$ is the fermionic (anti-symmetric) time ordering operator on $\CC$. Similarly, the real phonon propagator is defined as:
\begin{equation}\label{eq:D_prof_def}
\DD_\qq(t_1, t_2) = -i \left\langle T_\CC\left[\phii_{\qq}(t_1) \, \phii_{-\qq}(t_2)\right]\right\rangle,
\end{equation}
where $\phii_{\qq} = b_\qq^\dagger + b_{-\qq}^{\phantom{\dagger}}$ is the Fourier transform of the lattice displacement operator, and $T_\CC$ is the bosonic (symmetric) time ordering operator in $\CC$. The {\em lesser} ($<$) and {\em greater} ($>$) real-time Green's functions are defined as specific orderings of the two contour times where $t_1 <_\CC t_2$ and $t_1 >_\CC t_2$, respectively:
\begin{subequations}
\begin{align}
\hGG^<_\kk(t_1, t_2) &= +i \left\langle \PSID_\kk(t_2) \, \PSI_\kk(t_1)\right\rangle,\\
\hGG^>_\kk(t_1, t_2) &= -i \left\langle \PSI_\kk(t_1) \, \PSID_\kk(t_2) \right\rangle,\\
\DD^<_\qq(t_1, t_2) &= -i \left\langle \phii_{-\qq}(t_2) \, \phii_{\qq}(t_1)\right\rangle,\\
\DD^>_\qq(t_1, t_2) &= -i \left\langle \phii_{\qq}(t_1) \, \phii_{-\qq}(t_2)\right\rangle.
\end{align}
\end{subequations}
The retarded ($R$), advanced ($A$), and Keldysh ($K$) propagators are defined as $\AM^R(t_1, t_2) = \theta(t_1 - t_2)[\AM^>(t_1, t_2) - \AM^<(t_1, t_2)]$, $\AM^A(t_1, t_2) = -\theta(t_2 - t_1)[\AM^>(t_1, t_2) - \AM^<(t_1, t_2)]$, and $\AM^K(t_1, t_2) = \AM^>(t_1, t_2) + \AM^<(t_1, t_2)$, respectively, where $\AM$ is either $\hGG$ or $\DD$. We define spectral/statical decomposition of lesser/greater electron and phonon Green's functions as follows:
\begin{subequations}
\begin{align}
\label{eq:G_decomp}
i\hGG^{\gtrless}_\kk(t_1, t_2) &= \frac{1}{2}\left[i\hGG^K_\kk(t_1, t_2) \pm \hat{\AAA}_\kk(t_1, t_2)\right],\\
\label{eq:D_decomp}
i\DD^\gtrless_\qq(t_1, t_2) &= \frac{1}{2}\left[i\DD_\qq^K(t_1, t_2) \pm \rho_\qq(t_1, t_2)\right],
\end{align}
\end{subequations}
These definitions can be thought of as definitions of electron and phonon spectral functions:
\begin{subequations}
\begin{align}
\hat{\AAA}_\kk(t_1, t_2) &\equiv i\left[\hGG^>_\kk(t_1, t_2) - \hGG^<_\kk(t_1, t_2)\right],\\
\rho_\qq(t_1, t_2) &\equiv i\left[\DD^>_\qq(t_1, t_2) - \DD^<_\qq(t_1, t_2)\right].
\end{align}
\end{subequations}
Similar definitions apply to Green's functions in (Floquet-)Wigner representation, and for momentum-summed Green's functions. For all Nambu matrix quantities such as $\hat{\AAA}$, $i\hGG^K$, etc., we define {\em capped} Nambu matrices as:
\begin{equation}
\label{eq:cap}
\check{\AAA} \equiv \hat{\sigma}_z \, \hat{\AAA} \, \hat{\sigma}_z.
\end{equation}
We finish this appendix by listing a number of useful symmetry relations in the Wigner representation.\\

\remark{Lemma 1 (symmetries of Nambu functions)} {\em We define time reversal symmetric (TRS) states as being invariant under operation $(\kk, \up) \leftrightarrow (-\kk,\dn)$. The following identities hold for a TRS state:}
\begin{subequations}
\begin{align}
\label{eq:lemma_1_a}&\hat{\AAA}_\kk(\omega,t)^\dagger = \hat{\AAA}_\kk(\omega,t)\\
\label{eq:lemma_1_b}&\hat{\AAA}_\kk(\omega,t) = i\big[\hGG_\kk^R(\omega,t)-\hGG_\kk^R(\omega,t)^\dagger\big]\\
\label{eq:lemma_1_c}&\big[i\hat{\GG}_\kk^K(\omega,t)\big]^\dagger = i\check{\GG}_\kk^K(\omega,t)\\
\label{eq:lemma_1_d}&\hat{\AAA}_\kk(-\omega,t) = \hat{\sigma}_x \, \check{\AAA}_\kk(\omega, t)^* \, \hat{\sigma}_x\\
\label{eq:lemma_1_e}&i\hat{\GG}^K_\kk(-\omega,t) = -\hat{\sigma}_x \, \big[i\check{\GG}^K_\kk(\omega,t)\big]^* \, \hat{\sigma}_x
\end{align}
\end{subequations}
\noindent (proof) The proofs are elementary and readily follow from the definitions. The last two identities are less trivial and require a careful examination of the matrix elements of $\hGG_\kk(\omega, t)$.\\

\remark{Lemma 2 (symmetries of Eliashberg functions)} {\em We define an inversion symmetric (IS) state as being invariant under operation $\qq \leftrightarrow -\qq$. The following identities hold exactly for inversion symmetric states:}
\begin{subequations}
\begin{align}
\label{eq:lemma_2_a}&F_{\xi,\xi'}^\rho(\nu,t) = \big[F_{\xi,\xi'}^\rho(\nu,t)]^* = -F_{\xi,\xi'}^\rho(-\nu,t)\\
\label{eq:lemma_1_b}&iF_{\xi,\xi'}^K(\nu,t) = \big[iF_{\xi,\xi'}^K(\nu,t)\big]^* = iF_{\xi,\xi'}^K(-\nu,t)
\end{align}
\end{subequations}
\noindent (proof) The proofs are elementary and follow from the definition of Eliashberg functions (Eqs.~\ref{eq:Frho}-\ref{eq:FK}) and phonon propagators.\\

\remark{Lemma 3 (symmetries of the Nambu self-energy)} {\em The following identities holds for TRS and IS states:}
\begin{subequations}
\begin{align}
\label{eq:lemma_3_a}&\hat{\Sigma}^R(-\omega,T) = -\hat{\sigma}_x \, \big[\check{\Sigma}^R(\omega,T)\big]^* \, \hat{\sigma}_x\\
\label{eq:lemma_3_b}&i\hat{\Sigma}^K(-\omega,T) = -\hat{\sigma}_x \, \big[i\check{\Sigma}^K(\omega,T)\big]^* \, \hat{\sigma}_x
\end{align}
\end{subequations}
\noindent (proof) Both identities are easily established by calculating $\hat{\Sigma}^{R/K}(-\omega,T)$ using Eqs.~\eqref{eq:senRfin}-\eqref{eq:senKfin}, changing integration variables $\omega', \nu \rightarrow -\omega', -\nu$ and using Lemma 1 and 2 identities to change the sign of the frequencies that appear in the electron and phonon propagators.\\

\section{Proof of $\GG^R$ momentum summation formula}\label{sec:GR_sum_lemma_proof}
In this section, we give a proof for $\GG^R$ momentum summation formula (Eq.~\ref{eq:GR_sum_formula}) using perturbation theory. One of the assumptions of the lemma is the independence of $\Sigma^R$ from the momentum variable $\kk$. As a result, $\GG^R$ depends on $\kk$ only via the electronic dispersion $\xi_\kk$. Therefore, we may trade the momentum variable in $\GG^R$ with $\xi$ without loss of generality. The Dyson series for $\GG^R$ is:
\begin{multline}\label{eq:dys_wig}
\GG^R_\xi = \GG^R_{0,\xi} + \GG^R_{0,\xi} \star \Sigma^R \star \GG^R_{0,\xi}\\
+ \GG^R_{0,\xi} \star \Sigma^R \star \GG^R_{0,\xi} \star \Sigma^R \star \GG^R_{0,\xi} + \ldots,
\end{multline}
where:
\begin{equation}
\GG^R_{0,\xi} = \frac{1}{\omega - \xi - i0^+},
\end{equation}
is the non-interacting retarded Green's function. Let us consider the second term in the series:
\begin{multline}
\GG^R_{0,\xi} \star \Sigma^R \star \GG^R_{0,\xi} = \GG^R_{0,\xi}\,\exp\left[\frac{i}{2}\,\vec \partial_t \, \cev \partial_\omega - \frac{i}{2} \, \cev \partial_t \, \vec \partial_\omega\right]\\
\times \left(\Sigma^R \exp\left[\frac{i}{2}\,\vec \partial_t \, \cev \partial_\omega - \frac{i}{2} \, \cev \partial_t \, \vec \partial_\omega\right] \GG^R_{\xi,0}\right).
\end{multline}
Since $\partial_t \GG^R_{\xi,0} = 0$, if in addition we had $\partial_t \Sigma^R(\omega,t) = 0$, we would simply get $[\GG^R_{0,\xi}]^2 \, \Sigma_R$. Expanding the differential operators in the exponents, it is easily noticed that every $t$-derivative of $\Sigma^R$ is accompanied either by $\partial_\omega \GG^R_{0,\xi} = - [\GG^R_{0,\xi}]^2$, or by $\GG^R_{0,\xi} \partial_\omega \Sigma^R$. Therefore, derivative corrections due to $t$-dependence of $\Sigma^R$ are accompanied by {\em at least} one extra power of $\GG^R_{0,\xi}$. Thus,
\begin{equation}\label{eq:dys_wig_2}
\GG^R_{0,\xi} \star \Sigma^R \star \GG^R_{0,\xi} = [\GG^R_{0,\xi}]^2\,\Sigma^R + [\GG^R_{0,\xi}]^3 \times \mathcal{O}(\partial_t \Sigma^R) + \ldots
\end{equation}
This result is easily generalized to the $n$th term in the Dyson series:
\begin{multline}\label{eq:dys_wig_n}
\GG^R_{0,\xi} \star \Sigma^R \star \GG^R_{0,\xi} \star \ldots \star \GG^R_{0,\xi} = [\GG^R_{0,\xi}]^{n}\,[\Sigma^R]^{n-1}\\
+ [\GG^R_{0,\xi}]^{n+1} \times \mathcal{O}(\partial_t \Sigma^R) + \ldots
\end{multline}
With this observation, let us integrate the sides of Eq.~\eqref{eq:dys_wig} over $\xi$, considering only the first $n$ terms in the series. The integral over the first term is trivial:
\begin{equation}
\int_{-\infty}^{+\infty} \mathrm{d}\xi\,\GG^R_{0,\xi} = \int_{-\infty}^{+\infty}\frac{\dd \xi}{\xi - \omega + i0^+} = -i\pi.
\end{equation}
Using Eq.~\eqref{eq:dys_wig_2}, it is easily shown that the integral over the second term vanishes:
\begin{align}
&\int_{-\infty}^{+\infty}\dd\xi \, \GG^R_{0,\xi} \star \Sigma^R \star \GG^R_{0,\xi} =\nonumber\\
&\qquad\qquad\Sigma^R \int_{-\infty}^{+\infty}\frac{\dd\xi}{(\omega - \xi + i0^+)^2}\nonumber\\
&\qquad\qquad+ \mathcal{O}(\partial_t \Sigma^R) \int_{-\infty}^{+\infty}\frac{\dd\xi}{(\omega - \xi + i0^+)^3} + [\ldots] = 0.
\end{align}
This result is due to the fact that every term in the expansion of the Groenewold-Moyal series has at least a second order pole. This the same result holds for all higher order terms in the Dyson series. Assuming that the order of limit and integrations can be interchanged, and that the Dyson series converges, we find that the only non-trivial contribution stems from the non-interacting Green's function. This proves the sought after result:
\begin{equation}
\int_{-\infty}^{+\infty}\dd\xi\,\GG^R_{\xi} = \int_{-\infty}^{+\infty}\dd\xi\,\GG^R_{0,\xi} = -i\pi.
\end{equation}

\section{Calculating phonon excitation density and phonon squeezing from $i\DD^K$}\label{sec:squeezing_keldysh}
We formulated the problem in Sec.~\ref{sec:FME_main} in terms of the real phonon propagator $\DD_\qq(t_1, t_2) \equiv -i \left\langle T_\CC\left[\phii_{\qq}(t_1) \, \phii_{-\qq}(t_2)\right]\right\rangle$. While this formulation is convenient and compact, it does not immediately yield useful physical observables such a phonon number $n_\qq(t) \equiv \langle b_\qq^\dagger(t) b^{\phantom{\dagger}}_\qq(t) \rangle$ or the anomalous correlations $\kappa_\qq(t) \equiv \mathrm{Re}[\langle b_\qq^{\phantom{\dagger}}(t) b^{\phantom{\dagger}}_{-\qq}(t) \rangle]$. Here, we show that both quantities can be readily calculated from the Keldysh phonon correlator in the Wigner representation, $\DD^K_\qq(\omega, t)$, by performing appropriate frequency integrations. This is enabled by the observation that the interaction and drive terms in the Hamiltonian both commute with $\hat{\phii}_\qq$. The only non-commuting term is the lattice kinetic energy. Thus, the Heisenberg equation for $\hat{\phii}_\qq$ takes the following simple form:
\begin{equation}\label{eq:pi_heisenberg}
\partial_t \hat{\phii}_\qq(t) = 2\omega_\qq\hat{\pi}_\qq(t). 
\end{equation}
We assume the $\qq \leftrightarrow -\qq$ symmetry in this section and set $\hbar=1$. The last equation allows us obtain $\pi\pi$ correlators by calculating appropriate time derivatives of $\DD^K$. Defining $\mathcal{P}_\qq(t_1, t_2) \equiv -i \left\langle T_\CC\left[\pi_{\qq}(t_1) \, \pi_{-\qq}(t_2)\right]\right\rangle$, Eq.~\ref{eq:pi_heisenberg} immediately implies:
\begin{equation}\label{eq:P_D_rel}
\mathcal{P}_\qq(t_1, t_2) = \frac{1}{4\omega_\qq^2}\,\partial_{t_1}\partial_{t_2}\,\DD_{\qq}(t_1, t_2).
\end{equation}
At equal times, the Keldysh $\phii\phii$ and $\pi\pi$ correlators evaluate to a combination of our sought-after observables, $n_\qq$ and $\kappa_\qq$: 
\begin{subequations}
\begin{align}
(i/2)\DD^K_\qq(t, t) &\equiv 1 + 2n_\qq(t) + 2\kappa_\qq(t),\\
(2i)\mathcal{P}^K_\qq(t, t) &\equiv 1 + 2n_\qq(t)  - 2\kappa_\qq(t),
\end{align}
\end{subequations}
which together with Eq.~\eqref{eq:P_D_rel} yields:
\begin{subequations}
\begin{align}
n_\qq(t) &= \frac{1}{4}\,i\DD^K_\qq(t,t) + \frac{1}{4}\,\partial_{t_1}\partial_{t_2}\,i\DD^K_\qq(t_1,t_2)|_{t_1 = t_2 = t} - \frac{1}{2},\\
\kappa_\qq(t) &= \frac{1}{4}\,i\DD^K_\qq(t,t) - \frac{1}{4}\,\partial_{t_1}\partial_{t_2}\,i\DD^K_\qq(t_1,t_2)|_{t_1 = t_2 = t}.
\end{align}
\end{subequations}
In the Wigner representation, $\partial_{t_1}\partial_{t_2} \rightarrow \nu^2$ and we find:
\begin{subequations}
\begin{align}
n_\qq(t) &= \frac{1}{4}\int_{-\infty}^{+\infty} \frac{\mathrm{\dd\nu}}{2\pi}\left(1 + \frac{\nu^2}{\omega_\qq^2}\right) i\DD^K_\qq(\nu, t) - \frac{1}{2},\\
\kappa_\qq(t) &= \frac{1}{4}\int_{-\infty}^{+\infty} \frac{\mathrm{\dd\nu}}{2\pi}\left(1 - \frac{\nu^2}{\omega_\qq^2}\right) i\DD^K_\qq(\nu, t).
\end{align}
\end{subequations}

The anomalous phonon density can be related to phonon squeezing with additional considerations. First, we observe $\mathrm{Im}[\langle b_\qq^{\phantom{\dagger}}(t) b^{\phantom{\dagger}}_{-\qq}(t) \rangle] = 0$ in our problem since nonlinearities produce modulation of $\phii_\qq\,\phii_{-\qq}$ (as opposed to $\pi_\qq\,\pi_{-\qq}$; see Sec.~\ref{sec:param}). Assuming that a low temperature state is maintained at all times and weak electron-phonon coupling, the phonon state can be approximated as product of two-modes squeezed states of $\pm \qq$ on the top of a coherent state for $\qq=0$, i.e. $|\Phi(t)\rangle \sim \{\prod_\qq \exp[\xi_\qq(t) (b^\dagger_\qq\,b^{\dagger}_{-\qq} - b^{\phantom{\dagger}}_{-\qq}\,b^{\phantom{\dagger}}_{\qq})]\}\,\exp[\sqrt{N}\phii_0(t)(b^\dagger_{0} - b^{\phantom{\dagger}}_{0})/2]\,|0\rangle$ throughout the evolution. Here, $\xi_\qq(t)$ is the momentum-squeezing strength and $\phii_0(t)$ is the coherent displacement. For $\qq \neq 0$, this ansatz provides the following relation between the squeezing parameter $\xi_\qq(t)$ and the anomalous phonon density:
\begin{equation}
\kappa_\qq(t) = \frac{1}{2}\,\sinh[2\xi_\qq(t)] 
\end{equation}
For weak nonlinearities, the squeezing is also weak $|\xi_\qq(t)| \ll 1$ and we find $\xi_\qq(t) = \kappa_\qq(t) + \mathcal{O}(\kappa_\qq^3)$. Thus, the anomalous phonon density directly yields the squeezing parameter. We have shown the period-averaged $\langle \cosh[\xi_\qq(t)] \rangle - 1 \approx 2 \langle \kappa_\qq^2(t) \rangle$ in Fig.~\ref{fig:phonon_dyn}(d). It is noticed that squeezing significantly increases as the drive is ramped-up, consistent with the physics of the parametrically driven harmonic oscillator.

\section{Electron-mediated phonon dissipation and nonlinearities}\label{sec:nonlin}
The evolution equation for phonon propagators was derived in Sec.~\ref{sec:2PI_EOM} as well as their counterparts in the Floquet-Boltzmann kinetic approximation in Sec.~\ref{sec:QFB_lattice}. So long as the evolution of phonons is concerned, electrons play the role of a {\em quantum bath} through memory convolution integrals $\Pi_\qq \star \DD_\qq$ and $\DD_\qq \star \Pi_\qq$ appearing in Eqs.~\eqref{eq:KBD0} and~\eqref{eq:KBD0adj}, respectively. 

In this section, we derive approximate expressions for $\Pi_\qq$ assuming that the electrons remain in the initial low-temperature degenerate regime. Meanwhile, we also study the contribution of electrons to phonon nonlinearities. Both objectives can be achieved by integrating out the electrons from the Lagrangian $\mathcal{L}[\phii,\Psi]$ and obtaining a phonon-only effective action $S_\mathrm{eff}[\phii]$. Expanding the effective action in the electron-phonon coupling, we obtain the bath term at the second order. Higher order terms give the electronic contribution to lattice nonlinearities. Since these corrections have a strong dynamical nature, it is conceivable that they could become large when certain resonance conditions are met; indeed we find this to be case. In other words, even though the {\em intrinsic} lattice nonlinearities might be small, coupling to conduction electrons effectively produces large nonlinearities in the presence of a near-resonant drive.

We start our discussion with the electron-phonon Lagrangian:
\begin{multline}\label{eq:lag1}
\mathcal{L}[\phii,\Psi] = \mathcal{L}_0[\phii] + \sum_{\kk} \Psi^\dagger_{\kk}\left(i\partial_t\mathbb{I} - \xi_\kk\hat{\sigma}_3\right)\Psi^{\phantom{\dagger}}_{\kk}\\
- \frac{1}{\sqrt{N}}\sum_{\kk,\kk'} \low{g}_{\kk,\kk'} \, \low{\phii}_{\kk-\kk'}\,\PSID_{\kk'}\hat{\sigma}_3\PSI_{\kk},
\end{multline}
where $\mathcal{L}_0[\phii] = - \sum_\qq (2\omega_\qq)^{-1} \, \low{\phii}_\qq \left(\partial_t^2 + \omega_\qq^2\right)\low{\phii}_{-\qq}/2 + (\Lambda/2)\,|F(t)|^2\,\sqrt{N} \, \low{\phii}_{\qq=0}$ is the quadratic part, including the external drive. It is most convenient to perform the calculations in the real time formalism in order to avoid tedious analytical continuation procedure required in the Matsubara formalism. Integrating out the electrons, we find:
\begin{multline}
S_\mathrm{eff}[\phii] = \int_\CC \dd t \, \mathcal{L}_0[\phii] - i \mathrm{Tr \, ln}\bigg[\hGG_{0,\kk}^{-1}(t,t') \, \delta_{\kk,\kk'}\\
- \frac{1}{\sqrt{N}}\,\low{g}_{\kk,\kk'}\, \hat{\sigma}_3 \, \phii_{\kk-\kk'}(t) \, \delta_\CC(t,t') \bigg.
\end{multline}
Here, $\hGG_{0,\kk}^{-1}(t,t') = (i\partial_t - \xi_\kk\,\hat{\sigma}_3) \delta_\CC(t,t')$ and the trace implies momentum summation, contour time integration, and Nambu space summation. Expanding the second term in powers of $g$, we find:
\begin{align}
S_\mathrm{eff}[\phii] &= S_0[\phii] + \sum_{n=1}^\infty S_n[\phii],\nonumber\\
S_n[\phii] &= \frac{i}{n N^{n/2}}\sum_{\kk_i}\int_\CC \dd t_1 \ldots \dd t_n \, \mathrm{Tr}\Big[\hGG_{0,\kk_1}(t_1, t_2)\, \hsz\nonumber\\
&\times \hGG_{0,\kk_2}(t_2, t_3) \, \hsz \, \ldots \hGG_{0,\kk_n}(t_n, t_1) \, \hsz \Big]\nonumber\\
&\times \low{g}_{\kk_1,\kk_2} \, \low{g}_{\kk_2,\kk_3} \, \ldots \, \low{g}_{\kk_n,\kk_1}\nonumber\\
&\times \low{\phii}_{\kk_1 - \kk_2}(t_2) \, \low{\phii}_{\kk_2 - \kk_3}(t_3) \, \ldots \, \low{\phii}_{\kk_n - \kk_1}(t_1).
\end{align}
The first order correction $S_1[\phii]$ vanishes for Holstein-type screened electron-phonon couplings (see the discussion after Eq.~\ref{eq:g_scr}). The sum of higher order vertices can be diagrammatically represented as:
\begin{equation}\label{eq:S_eff_diag}
\sum_{n=2}^\infty S_n[\phii] = \eqfigscl{0.75}{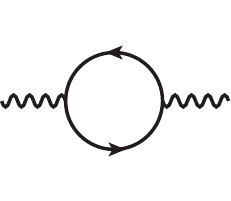} \,\, + \,\, \eqfigscl{0.75}{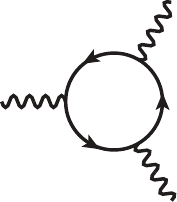} \,\, + \,\, \eqfigscl{0.75}{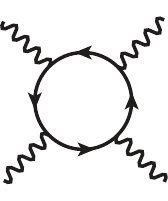} \,\, + \ldots
\end{equation}
It is convenient to make the forward/backward contour time indices explicit and perform a Keldysh rotation of $\phii^\pm$ fields into symmetric (``classical'') and anti-symmetric (``quantum'') components, $\uphii^\alpha = U_{\alpha\beta}\,\phii_\beta$:
\begin{equation}
\left(\begin{array}{c} \uphii^c \\ \uphii^q\end{array}\right) = \frac{1}{\sqrt{2}}\left(\begin{array}{cc} 1 & 1 \\ 1 & -1\end{array}\right) \left(\begin{array}{c} \phii^+ \\ \phii^-\end{array}\right).
\end{equation}
The bare action in Keldysh representation reads:
\begin{multline}
S_0[\phii] = -\frac{1}{2}\sum_\qq \frac{1}{2\omega_\qq}\int_{-\infty}^\infty \dd t \big[ \uphii_\qq^c \, (\partial_t^2 + \omega_\qq^2) \, \uphii_{-\qq}^q\\
+ \uphii_\qq^q \, (\partial_t^2 + \omega_\qq^2) \, \uphii_{-\qq}^c \big]\\
+ \frac{\Lambda}{2} \int_{-\infty}^{+\infty} \dd t \,|F(t)|^2 \, \sqrt{2N}\, \bar{\phii}^q_{\qq=0}(t).
\end{multline}
Likewise, the higher order terms in the Keldysh representation read:
\begin{align}
S_n[\phii] &= \frac{1}{n! \, N^{n/2-1}}\sum_{\qq_i}\int_{-\infty}^{+\infty} \dd t_1 \, \ldots \, \dd t_n \nonumber\\
&V^{\alpha_1 \, \ldots \, \alpha_n}_{\qq_1 \, \ldots \, \qq_n}(t_1, \ldots, t_n)\, \uphii_{\qq_1}^{\alpha_1}(t_1) \ldots \uphii_{\qq_n}^{\alpha_n}(t_n) \, \delta\left(\sum_i \qq_i\right),
\end{align}
where:
\begin{multline}\label{eq:Vdef}
V^{\alpha_1 \, \ldots \, \alpha_n}_{\qq_1 \, \ldots \, \qq_n}(t_1, \ldots, t_n) = \frac{i(n-1)!}{N}\sum_\kk \mathrm{Tr}\Big[\uhGG_{0,\kk + \qq_1}^{\mu_1\nu_2}(t_1, t_2)\\
\times \hsz \, \uhGG_{0,\kk + \qq_1 + \qq_2}^{\mu_2 \nu_3}(t_2, t_3) \, \hsz \, \ldots \uhGG^{\mu_n\nu_1}_{0,\kk}(t_n, t_1) \, \hsz \Big]\\
\times \low{g}_{\kk,\kk+\qq_1} \, \low{g}_{\kk+\qq_1,\kk+\qq_1+\qq_2} \, \ldots \, \low{g}_{\kk + \qq_1 + \ldots \qq_{n-1},\kk}\\
\times \Gamma_{\alpha_1 \mu_1 \nu_1} \, \ldots \, \Gamma_{\alpha_n \mu_n \nu_n},
\end{multline}
and $\Gamma_{\alpha\mu\nu} = \sum_\beta U_{\alpha\beta} U_{\mu\beta} U_{\nu\beta} \sigma^z_{\beta\beta}$ is a vertex in the Keldysh space. Finally, $\uhGG_0$ is the bare propagator in the Keldysh space:
\begin{equation}
\uhGG = \left(\begin{array}{cc} \hGG^K & \hGG^R \\ \hGG^A & 0\end{array}\right).
\end{equation}
Note that each matrix element additionally carries a $2 \times 2$ Nambu structure. We restrict our analysis to the normal state hereafter, in which case, the Nambu structure is immaterial. The Nambu space traces reduce to a multiplicative factor of $2$ (= total spin degeneracy) for each electron loop. In the following sections, we briefly study the first few vertices in succession.

\subsection{The second order correction: Landau damping}\label{sec:landau_damping}
A direct calculation using Eq.~\eqref{eq:Vdef} gives the matrix elements of $\Pi_\qq^{\alpha_1\alpha_2}(t_1, t_2) \equiv -V_{\qq,-\qq}^{\alpha_1\alpha_2}(t_1,t_2)$:
\begin{equation}\label{eq:Keldysh_self_energy}
\Pi_{\qq,-\qq}^{\alpha_1\alpha_2}(t_1,t_2) = \left(
\begin{array}{cc}
  0 & \Pi_{\qq}^{A}(t_1,t_2)\\
  \Pi_{\qq}^{R}(t_1,t_2) & \Pi_{\qq}^{K}(t_1,t_2)
\end{array}
\right),
\end{equation}
where:
\begin{align}\label{eq:Pi_def_0}
\Pi_\qq^A(t_1, t_2) =&\, -\frac{i}{N}\sum_\kk|\low{g}_{\kk,\kk+\qq}|^2 \big[\GG^A_{0,\kk+\qq}(t_1, t_2) \, \GG^K_{0,\kk}(t_2, t_1)\nonumber\\
&\,\,+ \GG^K_{0,\kk+\qq}(t_1, t_2) \, \GG^R_{0,\kk}(t_2, t_1)\big],\nonumber\\
\Pi_\qq^R(t_1, t_2) =&\, -\frac{i}{N}\sum_\kk|\low{g}_{\kk,\kk+\qq}|^2 \big[\GG^K_{0,\kk+\qq}(t_1, t_2) \, \GG^A_{0,\kk}(t_2, t_1)\nonumber\\
&\,\,+ \GG^R_{0,\kk+\qq}(t_1, t_2) \, \GG^K_{0,\kk}(t_2, t_1)\big],\nonumber\\
\Pi_\qq^K(t_1, t_2) =&\, -\frac{i}{N}\sum_\kk|\low{g}_{\kk,\kk+\qq}|^2 \big[\GG^R_{0,\kk+\qq}(t_1, t_2) \, \GG^A_{0,\kk}(t_2, t_1)\nonumber\\
&\hspace{-50pt}+ \GG^A_{0,\kk+\qq}(t_1, t_2) \, \GG^R_{0,\kk}(t_2, t_1) + \GG^K_{0,\kk+\qq}(t_1, t_2) \, \GG^K_{0,\kk}(t_2, t_1)\big].
\end{align}
The bare propagators in equilibrium are functions of $t_1 - t_2$ and admit the following standard Fourier representation:
\begin{align}
\GG^{R/A}_\kk(\omega) &= \frac{1}{\omega - \xi_\kk \pm i0^+},\nonumber\\
\GG^{K}_\kk(\omega) &= -2\pi i \, \delta(\omega - \xi_\kk)[1-2n_F(\xi_\kk)].
\end{align}
Calculating $\Pi^{R/A/K}_\qq$ in equilibrium is standard and yields the well-known Lindhard function~\cite{mahan2013many}:
\begin{subequations}
\begin{align}
\label{eq:PiRA}
\Pi_\qq^{R/A}(\omega) &= \frac{2}{N}\sum_\kk |\low{g}_{\kk,\kk+\qq}|^2 \frac{n_F(\xi_\kk) - n_F(\xi_{\kk+\qq})}{\omega - \xi_{\kk+\qq} + \xi_{\kk} \pm i0^+},\\
\label{eq:PiK}
i\Pi_\qq^{K}(\omega) &= -2\mathrm{Im}[\Pi_\qq^R(\omega)]\,\coth(\beta\omega/2).
\end{align}
\end{subequations}
Combining $S_2[\phii]$ with the bare action $S_0[\phii]$ yields the full quadratic part of the effective phonon-only action $S_\mathrm{quad}[\phii]$:
\begin{multline}
S_{\mathrm{quad}}[\phii] = -\frac{1}{2}\sum_\qq \frac{1}{2\omega_\qq}\int_{-\infty}^\infty \dd t_1 \, \dd t_2 \, \uphii_\qq^{\alpha_1}(t_1)\\
\times \big[(\partial_{t_1}^2 + \omega_\qq^2)\, \hsx^{\alpha_1\alpha_2} \, \delta(t_1 - t_2) + 2\omega_\qq \Pi^{\alpha_1\beta_1}(t_1, t_2) \big]\,\uphii_{-\qq}^{\alpha_2}(t_2)\\
+ (\Lambda/2) \int_{-\infty}^{+\infty} \dd t \, |F(t)|^2 \, \sqrt{2N}\, \uphii^q_{\qq=0}(t)
\end{multline}
The quantity $\Gamma^{(\mathrm{ph})}_\qq(\omega) \equiv -2\mathrm{Im}[\Pi_\qq^R(\omega)]$ is of particular interest and represents the spectrum of the dissipative bath the electronic degrees of freedom provide for phonons. We observe that $\Gamma_{\qq=0}(\omega) = 0$ for finite $\omega$, which is an expected consequence of momentum conservation. Thus, the uniform lattice displacement $\langle \uphii^c(t) \rangle$ experiences no friction from electrons. Employing the simplifications introduced in Sec.~\ref{sec:minmodel}, i.e. flat EDOS, local approximation for $\Pi$, and $\kk,\qq$-independent e-p coupling, we can calculate the local bath spectrum $\Gamma_\ell^{(\mathrm{ph})}$analytically at $T=0$:
\begin{align}
\Gamma_\ell^{(\mathrm{ph})}(\omega; T=0) &\equiv \frac{1}{N}\sum_\qq \Gamma_\qq(\omega; T=0)\nonumber\\
&\approx 4\pi |g|^2 \int_{-\omega}^0 \dd \xi \, \nu(\xi) \, \nu(\xi + \omega)\nonumber\\
&\approx \left(4\pi \, |g|^2 \nu(0)^2\right)\omega. 
\end{align}
Given that the energy density of electrons is much lower than Einstein phonon energy scale in the present context, it is an excellent approximation to use the last equation even for nonequilibrium electronic states. The last result is akin to the well-known Allen's formula~\cite{allen1987theory} which is often used to infer electron-phonon coupling from the phonon linewidth broadening. Finally, one can calculate $\mathrm{Re} \, \Pi^{(\ell)}(\omega; T=0)$ within the same approximations to find:
\begin{multline}
\mathrm{Re} \, \Pi^R_\ell(\omega) \approx\\
2|g|^2\nu(0)^2 \int_{0}^{W_\mathrm{el}/2} \int_{0}^{W_\mathrm{el}/2}  \frac{(\xi_1 + \xi_2) \, \dd \xi_1 \, \dd \xi_2}{(\omega + i0^+)^2 - (\xi_1 + \xi_2)^2}.
\end{multline}
In the Migdal limit, $\omega \ll W_\mathrm{el}$ for all $\omega$ of interest. To leading order in $\omega/W_\mathrm{el}$, we find $\mathrm{Re} \, \Pi^R_\ell(\omega) \approx -4|g|^2 \nu(0)^2 \ln(2)\,W_\mathrm{el}$ which corresponds to a constant Lamb shift.

In summary, within the validity limits of the simplified model of Sec.~\ref{sec:minmodel}, the dissipative effect of electrons on the dynamics of optical phonons can be modeled as a local quantum Ohmic bath:
\begin{align}\label{eq:equilibrium_bath}
\Pi^{R}_\ell(\omega) &\simeq \omega_L - i \gamma_\ell \omega/2,\nonumber\\
\Pi^{K}_\ell(\omega) &\simeq \gamma_\ell \, \omega \, \coth(\beta \omega/2),
\end{align}
where $\omega_L \approx -4 |g|^2 \nu(0)^2\,\ln(2)\,W$ is the Lamb shift and $\gamma_\ell = 2\pi \, \omega_0\, \nu(0) \, \lambda$ is the dimensionless friction constant expressed in terms of the mass enhancement factor $\lambda$ (see Eq.~\ref{eq:lambda_def}). Furthermore, the above expressions remain valid as long as the electrons approximately remain in a quantum degenerate state.

\subsection{The third-order correction: electron-mediated cubic nonlinearity}
The cubic vertex $V^{\alpha_1, \alpha_2, \alpha_3}_{\qq_1, \qq_2, \qq_3}$ has a complicated spatial and temporal structure due the nonlocality of electrons. Here, we rather focus on calculating the retarded phonon self-energy correction that arises from this cubic vertex rather than a general analysis. Recalling that the lattice has a large coherent uniform displacement in our problem, we find that the leading self-energy correction is obtained by contracting one of the legs (the third leg without the loss of generality) with the classical displacement $\phii_0$. The resulting self-energy correction, $\Pi_{\Delta,\qq}(t_1, t_2)$, has the following diagrammatic representation:
\begin{equation*}
\includegraphics[scale=0.75]{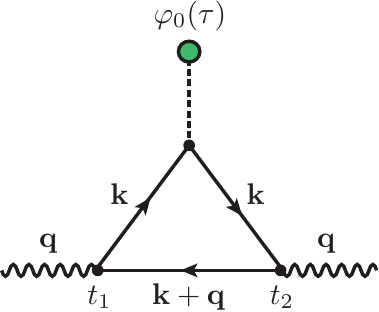}
\end{equation*}
Integration over $\tau$, the time argument of $\phii_0(\tau)$, is implied. Contracting the third leg with $\phii_0$ sets $\alpha_3$ to 1, i.e. to the ``classical'' Keldysh index. The retarded phonon self-energy is obtained by further choosing $\alpha_1 = 2$ and $\alpha_2 = 1$ (e.g. see Eq.~\ref{eq:Keldysh_self_energy}). Performing the intermediate Keldysh space traces in Eq.~\eqref{eq:Vdef}, we find: 
\begin{multline}
V^{2,1,1}_{\qq,-\qq, \mathbf{0}}(t_1, t_2, \tau) = \frac{i\sqrt{2}}{N}\sum_\kk |\low{g}_{\kk,\kk+\qq}|^2 \low{g}_{\kk,\kk}\\
\times\big[\GG_{\kk+\qq}^K(t_1,t_2) \, \GG_\kk^A(t_2,\tau) \, \GG_\kk^A(\tau,t_1)\\
+ \GG_{\kk+\qq}^R(t_1,t_2) \, \GG_\kk^K(t_2,\tau) \, \GG_\kk^A(\tau,t_1)\\
+ \GG_{\kk+\qq}^R(t_1,t_2) \, \GG_\kk^R(t_2,\tau) \, \GG_\kk^K(\tau,t_1)\big].
\end{multline}
Notice that the electron lines, starting form the $\kk+\qq$ line and traversing counter clockwise, assume the following Keldysh space labels: KAA, RKA, RRK. It can be shown that the same structure applies to higher order single-electron-loop vertices: with $N$ fermion propagators in a loop and $N-2$ classical field contractions, the retarded self-energy comprises $N$ terms, and the electron propagators in each term have Keldysh space labels $[R \ldots R] K [A \ldots A]$ in a counter clockwise fashion. The index subsets $[R \ldots R]$ and $[A \ldots A]$ comprise $N-1$ indices, and either subset can be empty (for example, see the next section for the quartic vertex).

The overall symmetry factor can be worked out as follows: $1/3!$ from the definition of $S_3[\phii]$, 3 choices for the classical leg, 2 choices for attaching one of the two remaining legs to the left external point, and a factor of $i^2$ from the two phonon propagators, amounting to $2 \times 3 \times i^2/3! = -1$. Thus, we obtain:
\begin{equation}
\Pi_{\Delta,\qq}^R(t_1, t_2) = -\int_{-\infty}^{+\infty} \dd\tau \, \frac{2\sqrt{N}\phii_0(\tau)}{\sqrt{2}} \, V^{2,1,1}_{\qq,-\qq, \mathbf{0}}(t_1, t_2, \tau)
\end{equation}
Note that $\low{\phii}_{\qq=0}(\tau) = (1/\sqrt{N})\sum_j \phii_j(\tau) = \sqrt{N} \phii_0(\tau)$, where $\phii_j(\tau) = \phii_0(\tau)$ is the uniform ionic displacement at site $j$. Also, the factor $2/\sqrt{2}$ arises from the definition of the ``classical'' component, i.e. $\uphii^c = (\phii^+_0 + \phii^-_0)\sqrt{2} = 2\phii_0/\sqrt{2}$. To proceed, we assume $\phii_0(\tau) = A\,e^{i\Omega\tau}$ and a thermal state for electrons. Taking a Wigner transform $(t_1, t_2) \rightarrow (\omega, t)$, we find:
\begin{widetext}
\begin{multline}
\Pi_{\Delta,\qq}^R(\omega,t) = (-2i) \, A \, \frac{1}{N}\sum_\kk |\low{g}_{\kk,\kk+\qq}|^2 \, \low{g}_{\kk,\kk} \int_{-\infty}^{+\infty} \dd \tau \, e^{i\Omega \tau} \int_{-\infty}^{+\infty} \dd t \, e^{i\omega s} \int \frac{\dd \omega_1}{2\pi} \, \frac{\dd \omega_2}{2\pi} \, \frac{\dd \omega_3}{2\pi} e^{-i\omega_1 s} \, e^{-i\omega_2(t - s/2 - \tau)} \\
\times e^{-i\omega_3(\tau - t - s/2)} \big[\GG_{\kk+\qq}^K(\omega_1) \, \GG_\kk^A(\omega_2) \, \GG_\kk^A(\omega_3) + \GG_{\kk+\qq}^R(\omega_1) \, \GG_\kk^K(\omega_2) \, \GG_\kk^A(\omega_3)
+ \GG_{\kk+\qq}^R(\omega_1) \, \GG_\kk^R(\omega_2) \, \GG_\kk^K(\omega_3)\big].
\end{multline}
\end{widetext}
Performing the time integrals and subsequently the frequency integrals over $\omega_1$ and $\omega_2$ is a lengthy calculation and we quote the final result:
\begin{multline}\label{eq:Pi_3_fin}
\Pi_{\Delta,\qq}^R(\omega,t) = -4 A \, e^{i\Omega t} \, \frac{1}{N}\sum_\kk |\low{g}_{\kk,\kk+\qq}|^2 \low{g}_{\kk,\kk}\\
\times \frac{n_F(\xi_\kk)-n_F(\xi_{\kk+\qq})}{(\omega - \xi_{\kk+\qq} + \xi_\kk + i0^+)^2 - \Omega^2/4}.
\end{multline}
We recall that the uniform displacement $\phii_0(\tau)$ approximately takes the form $\phii_0(\tau) \approx \phii_0 + \phii_1\,\cos(2\Odrv t) = \phii_0 + (\phii_1/2)\,e^{2i\Odrv t} + (\phii_1/2)\,e^{-2i\Odrv t}$ for a slowly ramped-up drive (see Eq.~\ref{eq:phi0_approx_sol}). Accordingly, the complete cubic self-energy correction is the sum of three terms obtained from replacing $(A,\Omega) \rightarrow (\phii_0, 0)$, $(\phii_1/2, +2\Odrv)$ and $(\phii_1/2, -2\Odrv)$ in Eq.~\eqref{eq:Pi_3_fin}.

Since $\Pi_{\Delta,\qq}^R(\omega,t) \propto g_{\kk,\kk}$, it vanishes for screened Holstein-type electron-phonon couplings. Had $g_{\kk,\kk}$ been finite, however, $\Pi_{\Delta,\qq}^R(\omega,t)$ would show a divergent behavior for $\qq \approx 0$ and $\omega \simeq \Odrv \simeq \omega_\qq$. In any event, $\Pi_{\Delta,\qq}^R(\omega,t)$ remains $\mathcal{O}(g^3)$ and non-divergent for $\Odrv \sim \omega_\qq/2$. Therefore, the electronic contribution to cubic lattice nonlinearity and its corresponding phonon self-energy corrections are negligible. We show in the next section that the situation is very different for the forth order correction.

\subsection{The forth-order correction: electron-mediated quartic nonlinearity}
We can similarly calculate the contribution of the quartic vertex (the last diagram in Eq.~\ref{eq:S_eff_diag}) to the phonon self-energy. In this case, two diagrams with different topologies comprise the leading order contribution to the quartic self-energy correction:
\begin{equation*}
\begin{array}{cc}
(a)\,\,\parbox[innerpos=c]{100pt}{\includegraphics[scale=0.75]{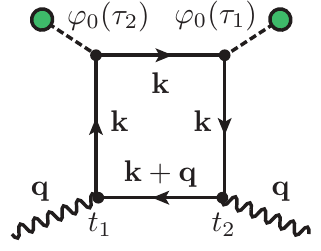}} & (b)\,\,\parbox[innerpos=c]{100pt}{\includegraphics[scale=0.75]{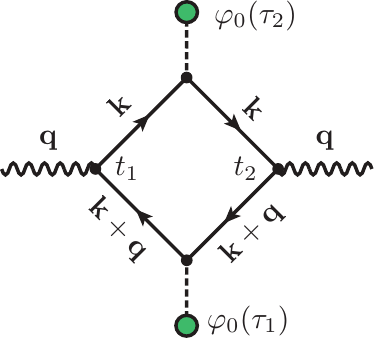}}
\end{array}
\end{equation*}
We attach $A_1 e^{i\Omega_1\tau_1}$ and $A_2 e^{i\Omega_1\tau_2}$ to two of the external legs. This can be done in $3 \times 2 = 6$ and $2 \times 2 = 4$ different ways for (a) and (b) topologies, respectively, this:
\begin{multline}
\Pi^{R}_{\square, \qq}(\omega,t) = -\frac{1}{4!\,N}\int \dd s\, e^{i\omega s} \, \int \dd \tau_1 \sqrt{2N}\, A_1 \, e^{i\Omega_1 \tau_1}\\
\int \dd \tau_2 \sqrt{2N}\, A_2 \, e^{i\Omega_2 \tau_2} \Big[6\,V_{\qq,-\qq,\mathbf{0},\mathbf{0}}^{2,1,1,1}(t + s/2, t - s/2, \tau_1, \tau_2)\\
+ 4\,V_{\qq,\mathbf{0},-\qq,\mathbf{0}}^{2,1,1,1}(t + s/2, \tau_1, t - s/2, \tau_2)\Big].
\end{multline}
Performing the intermediate Keldysh space summations in Eq.~\eqref{eq:Vdef}, we find:
\begin{align}
&V^{2,1,1,1}_{\qq,-\qq,\mathbf{0},\mathbf{0}}(t_1,t_2,\tau_1,\tau_2) = \frac{3i}{N}\sum_{\kk}\sum_{(a_1\ldots a_4) \in \mathcal{I}_4}|\low{g}_{\kk,\kk+\qq}|^2 |\low{g}_{\kk,\kk}|^2\nonumber\\
&\qquad \times \GG^{a_1}_{\kk+\qq}(t_1,t_2)\,\GG^{a_2}_{\kk}(t_2,\tau_1)\,\GG^{a_3}_{\kk}(\tau_1,\tau_2)\,\GG^{a_4}_{\kk}(\tau_2,t_2),\nonumber\\
&V^{2,1,1,1}_{\qq,\mathbf{0},-\qq,\mathbf{0}}(t_1,\tau_1,t_2,\tau_2) = \frac{3i}{N}\sum_{\kk}\sum_{(a_1\ldots a_4) \in \mathcal{I}_4}|\low{g}_{\kk,\kk+\qq}|^2 |\low{g}_{\kk,\kk}|^2\nonumber\\
&\qquad \times \GG^{a_1}_{\kk+\qq}(t_1,\tau_1)\,\GG^{a_2}_{\kk+\qq}(\tau_1,t_2)\,\GG^{a_3}_{\kk}(t_2,\tau_2)\,\GG^{a_4}_{\kk}(\tau_2,t_1),
\end{align}
where $\mathcal{I}_4 = \{KAAA, RKAA, RRKA, RRRK\}$ denotes the set of Keldysh space labels of the four electron propagators. After a lengthy but straightforward calculation, we find the contribution of the first diagram to be:
\begin{multline}\label{eq:Pi_4_a}
\Pi^{(a),R}_{\square, \qq}(\omega,t) = -\frac{3i}{2} A_1 A_2 \, e^{i(\Omega_1+\Omega_2)t} \, \frac{1}{N}\sum_{\kk}|\low{g}_{\kk,\kk+\qq}|^2\\
\times |\low{g}_{\kk,\kk}|^2  \sum_{(a_1\ldots a_4) \in \mathcal{I}_4}\int\frac{\dd\omega_4}{2\pi}\,\GG^{a_1}_{\kk+\qq}(\omega+\omega_4-\Omega_1/2-\Omega_2/2)\\
\times  \GG^{a_2}_{\kk}(\omega_4 -\Omega_1-\Omega_2) \, \GG^{a_3}_{\kk}(\omega_4-\Omega_2)\,\GG^{a_4}_{\kk}(\omega_4).
\end{multline}
Similarly, the contribution of the second diagram is found as:
\begin{multline}\label{eq:Pi_4_b}
\Pi^{(b),R}_{\square, \qq}(\omega,t) = -2i \, A_1 A_2 \,e^{i(\Omega_1+\Omega_2)t}\,\frac{1}{N}\sum_{\kk}|\low{g}_{\kk,\kk+\qq}|^2 \\
\times |\low{g}_{\kk,\kk}|^2 \sum_{(a_1\ldots a_4) \in \mathcal{I}_4}\int\frac{\dd\omega_4}{2\pi}\,\GG^{a_1}_{\kk+\qq}(\omega+\omega_4-\Omega_1/2-\Omega_2/2)\\
\times \GG^{a_2}_{\kk+\qq}(\omega+\omega_4 + \Omega_1/2-\Omega_2/2) \, \GG^{a_3}_{\kk}(\omega_4-\Omega_2)\,\GG^{a_4}_{\kk}(\omega_4).
\end{multline}
The $\omega_4$-integration is easily performed since for each choice of Keldysh space labels $(a_1\ldots a_4) \in \mathcal{I}_4$, one of the electrons is on shell (Keldysh) and fixes the value of $\omega_4$. Assuming a coherent displacement like $\phii_0(\tau) = \phii_0 + \phii_1\,\cos(2\Odrv \tau)$, the resulting self-energy contributions will have 3 contributions: a constant (dc) contribution, a contribution $\propto \cos(2\Odrv t)$, and a contribution $\propto \cos(4\Odrv t)$. Each contribution can be found by making appropriate choices for $(A_1, \Omega_1)$ and $(A_2, \Omega_2)$. The most interesting contribution is the one $\propto \cos(4\Odrv t)$ which is found by substituting $(A_1, A_2; \Omega_1, \Omega_2) \rightarrow (\phii_1/2, \phii_1/2; \pm 2\Odrv, \pm \Odrv)$ in Eqs.~\eqref{eq:Pi_4_a}-\eqref{eq:Pi_4_b} and summing up the four contributions. We quote the final result from this lengthy calculation:
\begin{multline}\label{eq:Pi_4_ac}
\Pi^{R,\mathrm{ac}}_{\square,\qq}(\omega,t) = \frac{3}{4}\,\phii_1^2\,\cos(4\Odrv t) \, \frac{1}{N}\sum_\kk |\low{g}_{\kk,\kk+\qq}|^2 |\low{g}_{\kk,\kk}|^2\\
\times \left\{\left[1-2n_F(\xi_{\kk})\right] \frac{A_{\kk,\qq}}{B_{\kk,\qq}} + \left[1-2n_F(\xi_{\kk+\qq})\right]\frac{C_{\kk,\qq}}{D_{\kk,\qq}}\right\},
\end{multline}
where:
\begin{align}
A_{\kk,\qq} &= \xi_\kk^2 \, (\xi_{\kk+\qq}-\omega ) (2 \xi_\kk+3 \xi_{\kk+\qq}-3 \omega )\nonumber\\
&\qquad +4 \xi_\kk \Odrv ^2 (\xi_\kk-2 \xi_{\kk+\qq}+2 \omega )-64 \Odrv^4,\nonumber\\
B_{\kk,\qq} &= \left(\xi_\kk^4-20 \xi_\kk^2 \Odrv^2+64 \Odrv^4\right) (\xi_{\kk+\qq}-\omega )\nonumber\\
&\qquad \times \left[(\xi_{\kk+\qq}-\omega )^2-4\Odrv^2\right],\nonumber\\
C_{\kk,\qq} &= \xi_{\kk+\qq} (2 \xi_{\kk}+\xi_{\kk+\qq}+2 \omega)+ 4\Odrv^2,\nonumber\\
D_{\kk,\qq} &= (\xi_{\kk}+\omega) (\xi_{\kk+\qq}^4 - 4\Odrv^2) [(\xi_{\kk+\qq}+\omega)^2 -4\Odrv^2].
\end{align}
As in the cubic vertex case, this contribution also vanishes for a perfectly screened Holstein-type electron-phonon coupling since $\Pi^{R,\mathrm{ac}}_{\square,\qq}(\omega,t) \propto |g_{\kk,\kk}|^2$. In a more realistic model, $g_{\kk,\kk}$ is generically non-vanishing. 

The $\kk$-integral in Eq.~\eqref{eq:Pi_4_ac} can be calculated in the limit $\qq \approx 0$ and assuming a constant electronic density of states and zero temperature. In the vicinity of the parametric resonance $\Odrv \sim \omega_\qq/2$, we find:
\begin{multline}
\Pi^{R,\mathrm{ac}}_{\square,\qq \approx 0}(\omega_\qq, t) \simeq \cos(2\omega_\qq t)\,\frac{3\nu(0)\,|g_{\kk,\kk}|^4\, \phii_1^2}{2\omega_\qq^2}\\
\times \ln\left[\frac{2\Odrv - \omega_\qq}{4\omega_\qq}\right] + \mathcal{O}(1).
\end{multline}
The logarithmic divergence could be anticipated from $D_{\kk,\qq} \propto (\omega_\qq^2 - 4\Odrv^2)$ in the limit $\omega = \omega_\qq$ and $\qq \approx 0$.

The above finding has a consequential implication: the electronic contribution to the lattice nonlinearity, even though is $\sim \mathcal{O}(g^2)$ and small in general, in the presence of coherent lattice oscillations leads to a self-energy correction that diverges logarithmically in the vicinity of $\Odrv \sim \omega_\qq/2$. Thus, even if purely ionic contributions to the lattice nonlinearity is small, large nonlinearities will be {\em dynamically generated} as a matter of coupling to electrons. Also, note that $\Pi^{R,\mathrm{ac}}_{\square,\qq \approx 0}(\omega_\qq, t) \propto \cos(2\omega_\qq t)$ which is precisely the COM time-dependence required for giving rise to parametric amplification of the lattice response as discussed in Sec.~\ref{sec:param}.

\section{Summary of numerical methods \label{app:numerics}}
In this appendix, we provide a summary of numerical methods for solving the quantum Floquet-Boltzmann kinetic equations for the lattice and electronic degrees of freedom. In reality, the two systems are coupled and must propagated forward in time self-consistently. The perturbative framework adopted in this work (when physically permissible), allows us to study the two systems in iterations: the dynamics of the lattice is worked out assuming unperturbed equilibrium electron propagators, the nonequilibrium correction to electron propagators electrons are worked out on the backdrop of the driven lattice, and so on; see Fig.~\ref{fig:flowchart}. This iterative procedure is expected to converge to the self-consistent solution of the fully coupled system in the weak-coupling limit.

\subsection{Solving the Quantum Floquet-Boltzmann kinetic equation for lattice displacement and phonon propagators}\label{sec:QFB_phonon_numerics}
The quantum Floquet-Boltzmann kinetic equations for the lattice displacement and phonon propagators were worked out in Sec.~\ref{sec:QFB_lattice}. The final result is the coupled system of equations given in Eq.~\eqref{eq:EOM_phi}, Eqs.~\eqref{eq:FBDR}-\eqref{eq:FBDRadj}, and Eqs.~\eqref{eq:FBDK}-\eqref{eq:FBDKadj}. Coupling to electrons only appear in the bath term, $\Pi^{(\ell); R/A/K}_{n,m}(\omega; t)$, which we assume is given to us in this section. We take a further simplifying step and neglect the COM time dependence of $\Pi^{(\ell)}$ which in indeed the case if the bath is approximately calculated using equilibrium electron propagators (see Sec.~\ref{sec:landau_damping}). The following analysis can be easily generalized for time-dependent baths, e.g. as required for the next iterations if one were to follow the perturbative decoupling recipe mentioned above.

The major difficulty in time-stepping Eqs.~\eqref{eq:FBDR}-\eqref{eq:FBDRadj} and Eqs.~\eqref{eq:FBDK}-\eqref{eq:FBDKadj} using ODE solvers is three-fold:
\begin{itemize}
\begin{item}[1.]
Time-derivatives appears on both sides of equations, and $\partial_t$ of different Floquet components $\partial_t \DD^{R/K}_{n,m}(\omega;t)$ are coupled due to the lattice nonlinearity and the bath. In other words, $\partial_t \DD^{R/K}_{n,m}(\omega;t)$ is only {\em implicitly} given by Eqs.~\eqref{eq:FBDR}-\eqref{eq:FBDRadj} and Eqs.~\eqref{eq:FBDK}-\eqref{eq:FBDKadj}.
\end{item}
\begin{item}[2.]
We have two sets of evolution equations for the retarded and Keldysh propagators: one obtained from the forward KB equation (Eqs.~\ref{eq:FBDR} and \ref{eq:FBDK}), and another from the backward KB equation (Eqs.~\ref{eq:FBDRadj} and \ref{eq:FBDKadj}). In general, these two are complementary. For instance, the direct numerical solution of two-time propagators requires the forward and backward equations to step the propagators forward in the first and second times, respectively, e.g. see Ref.~\cite{semkat1999kadanoff}. In the kinetic approximation, however, only the COM time is stepped forward while the relative time is transformed to the frequency domain and is carried as a label. In theory, one may choose to work with either of the forward or backward equations for time stepping as both are correct to $\mathcal{O}(\partial_t)$. However, the mixing of large non-gradient terms and small gradient terms leads to undesirable numerical errors.
\end{item}
\begin{item}[3.]
We found the system to be marginally stiff, requiring a robust ODE solver with adaptive time-stepping and local error control. This leads to unavoidably long run times. The application of stiff solvers is challenging as the Jacobian of the system is dense and is difficult to calculate.
\end{item}
\end{itemize}

Let us note that we do not need to calculate $\DD^{A}$ as a separate quantity since the identity $\DD^R(t_1, t_2) = \DD^A(t_2, t_1)$ implies $\DD^A_n(\omega; t) = \DD^R_n(-\omega; t)$. Furthermore, the exact identities $\DD^R_{-n}(\omega) = [\DD^R_{n}(-\omega)]^*$, $i\DD^K_{-n}(\omega) = [i\DD^K_{n}(-\omega)]^*$, and $i\DD^K_n(-\omega) = i\DD^K_n(\omega)$ allow us to restrict the numerical calculation to non-negative Floquet indices.\\

\noindent {\em Setting up the linear system and calculating the explicit $\partial_t \DD^{R/K}_{n,m}(\omega;t)$---} The second issue mentioned above can be circumvented using anti-symmetric and symmetric combinations of Eqs.~\eqref{eq:FBDR}-\eqref{eq:FBDRadj} and Eqs.~\eqref{eq:FBDK}-\eqref{eq:FBDKadj}, respectively. The issue of implicitness, however, remains challenging. In particular, the $\omega$-integral appearing in $\chi_n(t)$ and the appearance of $\partial_t U_n(t)$ in the kinetic equations implies that neither Floquet indinces, nor $\omega$ are ``good'' numbers. In other words, the kinetic equations of the lattice displacement and phonon propagators pose a dense linear system for $\partial_t \DD^{R/K}_{n,m}(\omega;t)$ and $\partial_t \phii_n(t)$. In order to find $\partial_t \phii_n(t)$ and $\partial_t \DD^{R/K}_{n,m}(\omega;t)$ explicitly, at each time $t$, we carefully index $\partial_t \DD^{R/K}_{n,m}(\pm \omega;t)$ and $\partial_t \phii_n(t)$ for all $(\omega, n,m)$, cast the coupled kinetic equations into a linear system and solve it via LU decomposition.

We perform the calculations on a regular frequency grid $\omega \in [-\omega_M, \omega_M]$ where $\omega_M$ is a high frequency cutoff. The grid spacing is chosen as rational fraction of $\Omega/2$ close to $0.1 \gamma_\ell$ in order to ensure that $\omega \pm n\Omega/2$ belongs to the grid. This allows us to identify a large fraction of unknown time-derivatives and matrix elements with one another and greatly reduce the dimension of the linear system. The $\omega$-derivatives are calculated using the 5-point finite difference approximation, and the $\omega$-integral appearing in Eq.~\eqref{eq:chin} is approximated using the trapezoid rule. We choose the Floquet cutoff $n_D = 2$, and the frequency cutoff $\omega_M = 2\omega_0 + 5\gamma_\ell + (n_D+1)\Omega$. This choice ensures that all involved propagators remain small and negligible for $|\omega| > \omega_M$. We carefully checked that increasing $\omega_M$ and $n_D$ had a negligible and controllably small effect on the results. For an $\omega$-grid with $\sim 500$ points, one needs to solve a linear system of size $\sim 6000 \times 6000$ for each  calculation of the explicit time derivatives.\\

\noindent {\em Initial thermal state and renormalized phonon frequency---} The lattice is in a thermal equilibrium state at the bath temperature before the drive ramped up. To find conditions describing the equilibrium state, we set the external drive and time derivatives to zero in the described evolution equations, assume $\phii_n(t) \rightarrow \phii_0\,\delta_{n,0}$, $U_n(t) \rightarrow U_0\,\delta_{n,0}$, $\chi_n(t) \rightarrow \chi_0 \, \delta_{n,0}$, and $\DD_n^{R/A/K}(\omega) \rightarrow \delta_{n,0}\,\DD^{R/A/K}_0(\omega)$. This leads to the following set of coupled equations:
\begin{align}\label{eq:phonon_equilibrium_sc_eqs}
&\left(\omega_0^2 - \frac{1}{3}\, \omega_0\kappa_4 \, \phii_0^2 - \omega_0\kappa_4\,\chi_0 - \omega\kappa_3\,\phii_0\right)\phii_0 - \omega_0 \kappa_3 \, \chi_0 = 0,\nonumber\\
&\DD_0^{R/A}(\omega) = \frac{2\omega_0}{\omega^2 - \omega_0^2 - 2\omega_0\omega_L - 2\omega_0 U_0 \pm i\gamma_\ell\omega},\nonumber\\
&i\DD_0^K(\omega) = \frac{4\omega_0\gamma_\ell \, \omega \, \mathrm{coth}(\beta\omega/2)}{(\omega^2 - \omega_0^2 - 2\omega_0\omega_L - 2 \omega_0 U_0)^2 + \gamma_\ell^2\omega^2},\nonumber\\
&U_0 = -\frac{\kappa_4}{2}\,\phii_0^2 - \frac{\kappa_4}{2}\,\chi_0 - \kappa_3\phii_0,\nonumber\\
&\chi_0 = \frac{1}{2}\int_{-\infty}^{+\infty} \frac{\dd \omega}{2\pi}\, i\DD_0^K(\omega).
\end{align}
The first and last two equations must be solved self-consistently, leading to a renormalized phonon frequency:
\begin{equation}\label{eq:renorm_phonon_freq}
\Omega_0 \equiv \sqrt{\omega_0^2 + 2\omega_0 \bar{\omega}_{L} + 2\omega_0 U_0}.
\end{equation}

\noindent{\em Numerical time stepping---} Provided that $\{\phii_n(t)\}$, $\{U_n(t)\}$, $\{\chi_n(t)\}$, and $\{i\DD^{R,K}_n(\omega;t)\}$ are known for all $\omega$ on a regular grid, we obtain the explicit time-derivatives of these quantities using by solving a linear system as described earlier. We can then invoke an explicit ODE solver to perform time-stepping. Here, we integrated the ODE using the adaptive Runge-Kutta-Fehlberg(4,5) method with local relative error tolerance of $10^{-6}$.

\subsection{Solving the Quantum Floquet-Boltzmann kinetic equation for electrons}\label{sec:QFB_electron_numerics}
The quantum Floquet-Boltzmann kinetic equation for $\{\psi_n(\omega;t)\}$ (see Eq.~\ref{eq:psidef}) was derived in Sec.~\ref{sec:QFB_el}. Similar to the kinetic equation for the phonons, this kinetic equation is also a formidably dense implicit integral equation for $\{\partial_t\psi_n(\omega;t)\}$ in which all frequencies and Floquet indices are coupled and defies the immediate application of an explicit ODE solver. In this section, we describe a numerical strategy for solving this equation.\\

\noindent {\em Preliminaries---} As a first step, we use the exact identities $\psi^*_{n,m}(\omega) = \psi_{-n,m}(\omega)$ and $\Sigma^A_{n,m}(\omega) = [\Sigma^R_{-n,m}(\omega)]^*$ to cast Eq.~\eqref{eq:FBE} into a more useful form:
\begin{multline}\label{eq:FBE2}
\partial_t \psi_n = in\Omega \, \psi_{n} + i\Sigma^K_{n}\\
- i\Sigma^R_{n',n'-n} \, \psi_{n'-n,n'}^* + i \Sigma^{R,*}_{n',n'+n} \, \psi_{n'+n,n'}\\
+\frac{1}{2} \, \partial_\omega\Sigma^R_{n',n'-n} \, \partial_t \psi_{n'-n,n'}^* +\frac{1}{2} \, \partial_\omega\Sigma^{R, *}_{n',n'+n} \, \partial_t \psi_{n'+n,n'}\\
-\frac{1}{2} \, \partial_t \Sigma^R_{n',n'-n} \, \partial_\omega \psi_{n'-n,n'}^* - \frac{1}{2} \, \partial_t\Sigma^{R, *}_{n',n'+n} \, \partial_\omega \psi_{n'+n,n'}.
\end{multline}
We have dropped the common $(\omega;t)$ argument from all quantities for brevity. Summation over repeated indices is implied everywhere in this section. The numerical integration of this equation is complicated by the fact that the self-energies are functionals of $\psi$, so that $\partial_t \Sigma^R$ terms implicitly involve $\partial_t\psi$. This functional dependence can be made explicit using Eqs.~\eqref{eq:senRN_floquet}-\eqref{eq:senKN_floquet}:
\begin{align}
\Sigma^R_n(\omega;t) &= \FFF[i\DD^K]_n(t)\nonumber\\
&\quad+ \int_0^\infty\dd\omega'\, \mathsf{K}[\rho]_{n-n'}(\omega,\omega';t)\,\psi_n(\omega';t),\nonumber\\
\label{eq:app_Sig_K}
i\Sigma^K_n(\omega;t) &= \pi \int_{-\infty}^{+\infty} \dd\nu \,iF^K_{n-n'}(\nu;t)\,\psi_{n}(\omega-\nu;t).
\end{align}
where:
\begin{align}
\FFF[i\DD^K]_n(t) &= -i\pi \int_0^\infty \dd \nu\,iF^K_n(\nu;t),\nonumber\\
\mathsf{K}[\rho]_{n}(\omega,\omega';t) &= \KKK[\rho]^{PV,+}_n(\omega,\omega';t) + \KKK[\rho]^{\delta,+}_n(\omega,\omega';t)\nonumber\\
&\quad+ \KKK[\rho]^{PV,-}_n(\omega,\omega';t) + \KKK[\rho]^{\delta,-}_n(\omega,\omega';t),\nonumber\\
\KKK[\rho]^{PV,\pm}_n(\omega,\omega';t) &= PV\int_{-\infty}^{+\infty}\dd\nu\,\frac{\rho_n(\nu;t)}{\omega \mp \omega' - \nu}\nonumber\\
\KKK[\rho]^{\delta,\pm}_n(\omega,\omega';t) &\equiv -i\pi \int_{-\infty}^{+\infty} \dd\nu \, \rho_n(\nu;t)\,\delta(\omega \mp \omega' - \nu).
\end{align}
The following useful identities can be established using the symmetries of $\psi$, $F^{K/\rho}$, and the properties of Kramers-Kronig transforms:
\begin{align}\label{eq:app_FBE_useful_idents}
\mathsf{K}[\rho]^{PV,\pm}_{n}(-\omega,\omega';T) &= \mathsf{K}[\rho]^{PV,\mp}_{n}(\omega,\omega';T)\\
&= \left[\mathsf{K}[\rho]^{PV,\pm}_{-n}(-\omega,\omega';T)\right]^*,\\
\mathsf{K}[\rho]^{\delta,\pm}_{n}(-\omega,\omega';T) &= -\mathsf{K}[\rho]^{\delta,\mp}_{n}(\omega,\omega';T)\\
&= -\left[\mathsf{K}[\rho]^{\delta,\pm}_{-n}(-\omega,\omega';T)\right]^*,\\
\Sigma^R_n(-\omega;T) &= -[\Sigma^R_{-n}(\omega;T)]^*,\\
i\Sigma^K_n(-\omega;T) &= -i\Sigma^K_n(\omega;T) = [i\Sigma^K_{-n}(\omega;T)]^*
\end{align}
As a result, we only need to calculate each quantity only for $\omega>0$. Also, save for $\Sigma^R_n$, all other quantities can be calculated for $n \geq 0$.\\

\noindent {\em The frequency grid---} We proceed by generating a grid $X_\omega$ in the interval $[0, \omega_c]$. Here, $\omega_c$ is an appropriate cutoff $\omega_c \gg 1/\beta, \omega_0, \Omega$. We generate the grid $X_\omega$ such that for all $\omega \in X_\omega$, if $\omega_m \equiv \omega + m\Omega/2 < \omega_c$, then $\omega_m \in X_\omega$. We call such a grid $X_\omega$ as a {\em Floquet-closed} grid. We will shortly see that a Floquet-closed grid leads to a significant reduction in computational complexity by allowing us to reuse previously calculated integrals. In practice, it is necessary to generate a non-uniform grid that emphasizes on the $\omega \lesssim 1/\beta$ region. To this end, we create two uniform grids, $X^\mathrm{th}_\omega \in [0, c/\beta]$, and $X^\mathrm{>}_\omega \in [c/\beta, \omega_c]$, and concatenate them. Crucially, we choose the grid spacings $\delta\omega^\mathrm{th}$ and $\delta\omega^>$ such that both are integer multiples of $\Omega/(2N)$ for some $N$. Once we have this basic two-scale grid, we pool together $|X^\mathrm{>}_\omega \cup X^\mathrm{th}_\omega + m\Omega/2|$ for $|m|<m_c$ and keep the unique points to find $X_\omega$.\\

\noindent{\em Calculating the required matrix elements---} We assume that $\{\psi_n(\omega;t)\}$ are known for $\omega \in X_\omega$ and $0 \leq n \leq N_\psi$ for some cutoff $N_\psi \geq N_D$. It is trivial to calculate $i\Sigma^K$ numerically based on Eq.~\eqref{eq:app_Sig_K} using a quadrature formula. To find $\Sigma^R$ and $\partial_t\Sigma^R$, we first calculate $\FFF[i\DD^K]_n$ and $\partial_t\,\FFF[i\DD^K]_n$, both of which are trivial. To calculate the contribution from $\{\rho_n\}$, we calculate the following quantities:
\begin{align}
\KKK_{n,jk}[\rho](t) &\equiv \int_{\omega_j}^{\omega_{j+1}}\dd\omega'\,\KKK_{n}[\rho](\omega_j,\omega_k;t),\nonumber\\
\partial_t\,\KKK[\rho]_{n,jk}(t) &\equiv \int_{\omega_j}^{\omega_{j+1}}\dd\omega'\,\KKK_{n}[\partial_t\rho](\omega_j,\omega_k;t),\nonumber\\
\tilde{\KKK}[\rho]_{n,jk}(t) &\equiv \int_{\omega_c}^{\infty}\dd\omega'\,\KKK_{n}[\rho](\omega_j,\omega_k;t),\nonumber\\
\partial_t\,\tilde{\KKK}[\rho]_{n,jk}(t) &\equiv \int_{\omega_c}^{\infty}\dd\omega'\,\KKK_{n}[\partial_t\rho](\omega_j,\omega_k;t),
\end{align}
for $\omega_j, \omega_k \in X_\omega$. The $\tilde{\KKK}$ terms stem from $\int_{\omega_c}^\infty \dd\omega'$ assuming $\psi_{n}(\omega') \approx \delta_{n,0}$ for $\omega'>\omega_c$. The proper $\omega'$-integrals must be approximated with quadratures much finer that $X_\omega$ grid spacing, and this is necessary since the integrands can vary on shorter scales than the grid spacing of a practically sized $X_\omega$. The improper $\omega'$-integrals can be calculated using M\"obius transformation and then using standard proper quadratures. Calculating $\KKK_{n,jk}[\rho](t)$ and $\KKK_{n,jk}[\partial_t \rho](t)$ is quite expensive as the integrand is given by a Kramers-Kronig integral and must be obtained numerically for every integration point. Having calculated these quantities, we may compose the full expression for $\Sigma^R_n(\omega_j;t)$ and $\partial_t\Sigma^R_n(\omega_j;t)$ approximately as:
\begin{widetext}
\begin{align}\label{eq:app_FBE_Sig_R}
\Sigma^R(\omega_j;t) &\simeq \FFF[i\DD^K]_n(t) + \sum_{k=0}^{N_\omega-1}\KKK[\rho]_{n-n',jk}(t)\,\frac{\psi_{n'}(\omega_j;t) + \psi_{n'}(\omega_{j+1};t)}{2} + \tilde{\KKK}_{n,j}[\rho](t),\nonumber\\
\partial_t\Sigma^R(\omega_j;t) &\simeq \FFF[\partial_t i\DD^K]_n(t) + \sum_{k=0}^{N_\omega-1}\partial_t\KKK[\rho]_{n-n',jk}(t)\,\frac{\psi_{n'}(\omega_j;t) + \psi_{n'}(\omega_{j+1};t)}{2}\nonumber\\
&+ \sum_{k=0}^{N_\omega-1}\KKK[\rho]_{n-n',jk}(t)\,\frac{\partial_t\psi_{n'}(\omega_j;t) + \partial_t\psi_{n'}(\omega_{j+1};t)}{2} + \partial_t\tilde{\KKK}_{n,j}[\rho](t).
\end{align}
\end{widetext}
Here, $N_\omega$ is the number of grid points in $X_\omega$. We have also used linear interpolation for the values of $\psi$ between consecutive grid points.  If $\omega_c \gg 1/\beta$, the high energy tail of Fermi distribution indeed remains intact (i.e. we assume $\omega_c$ is large enough so that no particles will be excited to energies above $\omega_c$). The important point about using the Floquet closed $X_\omega$ is that once we calculate $\KKK_n(\omega_j, \omega_k)$ for $\omega_j \in X_\omega$, we immediately get $\KKK_n(\omega_j + m\Omega/2, \omega_k)$ for all $m$ using a combination of shifts, inversions, and conjugation (see Eq.~\ref{eq:app_FBE_useful_idents}). In other words, we do not need to perform the expensive calculation of $\KKK$ for all $m$-shifted $\omega$ points.\\

\noindent{\em Setting up the linear system and time-stepping---} Plugging Eq.~\eqref{eq:app_FBE_Sig_R} expression into Eq.~\eqref{eq:FBE2}, we find an explicit linear system for $\partial_t \psi_n(\omega_j; t)$ for $\omega_j \in X_\omega$. The $\omega$-derivatives are found using 5-point finite difference approximation on the $X_\omega$ grid. This system can be mapped to a matrix equation by (1) indexing $\{\mathrm{Re} \, \psi_n(\omega_j;t), \mathrm{Im} \, \psi_n(\omega_j;t)\}$ for all $n$ and $\omega_j \in X_\omega$, and (2) setting up a mapping from $\{\mathrm{Re}\,\psi_n(\omega_j + m\Omega/2;t), \mathrm{Im}\,\psi_n(\omega_j + m\Omega/2;t)\}$ to the corresponding indexed values for all $m$ using the symmetries of $\psi$. Having a recipe to calculate $\partial_t \psi_n(\omega; t)$, we proceed and integrate the ODE using the adaptive Runge-Kutta-Fehlberg(4,5) method with local relative error tolerance of $10^{-6}$. 

\subsection{Numerical analysis of the spectrum of Floquet-Migdal-Eliashberg gap functional}
Calculating the spectrum of the Floquet-Migdal-Eliashberg (FME) gap functional, given in Eq.~\eqref{eq:scfin}, involves three steps: (1) calculating $\Sigma^{R/K}_{n,m}(\omega;t)$ in the normal state, (2) calculating the Floquet matrix elements of the anomalous response $\QQ^{n,m}_{n',m'}(\omega;t)$, and (3) calculating the spectrum of the FME gap functional.

The first step is identical to the procedure described in Sec.~\ref{sec:QFB_electron_numerics}. In the equilibrium-electron approximation, we use $\psi_{n,m}(\omega;t) \rightarrow \delta_{n,0}\,\tanh[\beta(\omega - m\Omega/2)/2]$ in calculating the self-energies rather than using $\psi_{n,m}(\omega;t)$ found from solving the Floquet-Boltzmann equation for electrons.

The second step involves inverting the coefficient matrix of $\delta \FF^R_{n,m}(\omega;t)$ which can be read from Eq.~\eqref{eq:dFFloq2}. To this end, we truncate the intermediate $n'$ Floquet band index summation to $|n'| \leq N_\phi$ and Floquet quasi-momentum indices to $|m| \leq N_m$. The truncated system of equations is then carefully mapped to a proper linear system, $\sum_{n'=-N_\phi}^{+N_\phi}\sum_{m'=-N_m}^{+N_m}\mathcal{C}^{n,m}_{n',m'}(\omega;t)\,\delta \FF^R_{n,m}(\omega;t) = -2\pi i\,\phi_{n,m}(\omega;t)$. The Floquet matrix elements of the anomalous response is readily found by inverting $\mathcal{C}^{n,m}_{n',m'}$ in the space of paired Floquet indices $(n,m)$:
\begin{equation}
\QQ^{n,m}_{n',m'}(\omega;t) = -2\pi i\, [\mathcal{C}^{-1}]^{n,m}_{n',m'}.
\end{equation}
In practice, we found the final results to be accurate to $10^{-4}$ with the choice $N_\phi = N_m = N_D + 2$ where $N_\Sigma$ is the previously chosen Floquet band cutoff in calculating the retarded self-energy.

The last step is slightly more involved. The overall strategy is to formally interpolate $\Delta_{n}(\omega;t)$ over a finite grid $G_\omega$, plug the interpolation formula in Eq.~\eqref{eq:scfin}, read off the coefficients of $\Delta_{n}(\omega \in G_\omega;t)$, and calculate its spectrum. Even though a brute-force discretization is equally applicable in principle, the uniform grid must be very dense in order to obtain accurate results, leading to calculating the spectrum of intractably large matrices. The interpolation procedure allows us to obtain accurate results using much coarser grids.\\

\noindent{\em Setting up the grid---} We generate $G_\omega$ by concatenating three grids $G_\omega = G^{(1)}_\omega \cup G^{(2)}_\omega \cup G^{(3)}_\omega$ where $G^{(1)}_\omega$ is a uniform grid for $\omega \in [0, 10/\beta)$ where $\beta^{-1} \sim 0.05\,\Omega_0$ is the typical effective temperature of electrons, $G^{(2)}_\omega$ in another uniform grid for $\omega \in [10/\beta, \omega_c)$ where $\omega_c \sim 10\,\Omega_0$ is a typical scale beyond which variations of $\Delta_n(\omega;t)$ becomes negligible, and finally, $G^{(3)}_\omega$ is a log-scaled grid for $\omega \in [\omega_c, \infty)$. In practice, we found allocating 100 points for each sub-grid produced results accurate to $10^{-4}$.\\

\noindent{\em Setting up the coefficient matrix---} We approximate $\Delta_n(\omega;t)$ over $G_\omega$ using a linear interpolant:
\begin{multline}
\Delta_n(\omega;t) \approx \frac{\omega_{j_\omega+1} - \omega}{\omega_{j_\omega+1} - \omega_{j_\omega}}\,\Delta_n(\omega_{j_\omega};t)\\
+ \frac{\omega - \omega_{j_\omega}}{\omega_{j_\omega+1} - \omega_{j_\omega}}\,\Delta_n(\omega_{j_\omega+1};t),
\end{multline}
where $j_\omega$ is the nearest grid point to the left of $\omega$. Plugging this ansatz into Eq.~\eqref{eq:scfin}, we find:
\begin{multline}\label{eq:scfin_grid}
\Delta_n(\omega_j;t) = \frac{i\omega_j}{2\pi}\sum_{n',n'',m'}\sum_{k=1}^{|G_\omega|}\bigg\{\QQ^{n,0}_{n',m'}(\omega_j;t)\\
\times\mathsf{K}^\Delta_{n'',m'}(\omega_j,\omega_k;t)\,\Delta_{n'-n''}(\omega';t) \\
- \left[\QQ^{-n,0}_{n',m'}(\omega;t)\right]^*\,\mathsf{K}^{\Delta*}_{n'',m'}(\omega_j,\omega_k;t)\, \Delta^*_{n'-n''}(\omega';t)\bigg\},
\end{multline}
where:
\begin{multline}
\mathsf{K}^\Delta_{n,m}(\omega_j,\omega_k;t) = \int_{\omega_{k-1}}^{\omega_k} \frac{\dd\omega'}{\omega'} \, K_{n,m}(\omega_j,\omega';t)\,\frac{\omega' - \omega_{k-1}}{\omega_k - \omega_{k-1}}\\
+ \int_{\omega_k}^{\omega_{k+1}} \frac{\dd\omega'}{\omega'} \, K_{n,m}(\omega_j,\omega';t)\,\frac{\omega_{k+1} - \omega'}{\omega_{k+1} - \omega_{k}}.
\end{multline}
The end points, $k=|\GG_\omega|$ and $k=1$, only get contributions from the first and second integrals, respectively. The $\omega'$-integrals are performed via an adaptive Gauss-Kronrod quadrature and is refined until a tolerance of $10^{-6}$ is achieved. The kernel $K_{n,m}(\omega,\omega';t)$ is given by Eq.~\eqref{eq:K_gap_def} and each evaluation requires performing a numerical Kramers-Kronig transform. Calculating the matrix elements $\mathsf{K}^\Delta_{n,m}(\omega_j,\omega_k;t)$ is the most computationally expensive part of this section. Finally, we decompose $\Delta_n(\omega;t)$ into real and imaginary parts and use the relations $\mathrm{Re}[\Delta_n(\omega;t)] = \mathrm{Re}[\Delta_{-n}(\omega;t)]$, $\mathrm{Im}[\Delta_n(\omega;t)] = -\mathrm{Im}[\Delta_{-n}(\omega;t)]$ to cast Eq.~\eqref{eq:scfin_grid} as a matrix equation. The coefficient matrix acts on space of bundled labels $(n, j, o)$ where $n$ is the Floquet index, $j$ is the grid point index, and $o=0, 1$ indicates real and imaginary component of $\Delta_n$.

Finally, we impose cutoffs $N_\Delta$ and $N_m$ over the Floquet band index of $\Delta$ and the internal $m'$ quasi-momentum summation. The Floquet cutoff for $K$ is $N_K = N_D + N_\psi$ where $N_D$ and $N_\psi$ are the previously chosen Floquet cutoffs for phonon propagators and electron energy statistics, respectively. We found $N_\Delta = N_m = N_K + 2$ to produce results accurate to $10^{-4}$. Assuming $N_\psi = N_D = 2$ and $|G_\omega| = 300$, the final coefficient matrix is has a dimension $3900 \times 3900$ and its spectrum can be easily found numerically. 

\bibliography{library}



\end{document}